\documentclass[preprint,aps,nofootinbib]{revtex4-1}	
\usepackage{amsmath}
\usepackage{amsfonts}
\usepackage{amssymb}
\usepackage{color}
\usepackage{bm}		
\usepackage{graphicx}
\usepackage{feynmp}
\usepackage{slashed}	
\usepackage{bbm}		
\usepackage{xspace}	
\usepackage{empheq}
\usepackage[toc,page]{appendix}
\usepackage{hyperref}	

\providecommand{\be}{\begin{equation}}
\providecommand{\ee}{\end{equation}}
\providecommand{\ba}[1]{\begin{align}{#1}\end{align}}
\providecommand{\nl}{\nonumber \\ }

\providecommand{\mean}[1]{\langle #1 \rangle}

\def\Tr{\mathrm{Tr}} 
\def\hc{\text{h.c.}} 

\def\GeV{\text{ GeV}} 
\providecommand{\oln[1]}{\overline{#1}} 
\providecommand{\wh[1]}{\widehat{#1}} 
\def\MSbar{\overline{\text{MS}}} 
\def\mb{\overline{m}} 
\def\tt{\tilde{t}} 
\def\muMS{\hat{\mu}} 
\def\XtMS{\widehat{X}_t}
\def\AtMS{\widehat{A}_t}
\def\AbMS{\widehat{A}_b}
\def\AtauMS{\widehat{A}_\t}
\def\YtMS{\widehat{Y}_t}%

\def\a{\alpha}
\def\b{\beta}
\def\c{\chi}
\def\d{\delta}
\def\e{\epsilon}
\def\f{\phi}
\def\g{\gamma}

\def\k{\kappa}
\def\l{\lambda}
\def\m{\mu}

\def\q{\theta}

\def\s{\sigma}
\def\t{\tau}

\def\D{\Delta}
\def\F{\Phi}

\begin{document}
\vspace{-0.2cm}
\begin{flushright}
EFI Preprint 15-24 \\
\end{flushright}

\begin{center}
\Large\bf Higgs bosons in heavy supersymmetry with an intermediate $m_A$ 
\end{center}

\vspace{0.2cm}
\begin{center}
{\sc Gabriel Lee$^{(a,b)}$ and Carlos E.~M.~Wagner$^{(b,c,d)}$ } \\
\vspace{0.5cm}
{\it
$^{(a)}$ Physics Department, Technion -- Israel Institute of Technology, \\
Haifa 32000, Israel
}\\
{\it
$^{(b)}$ Enrico Fermi Institute and Department of Physics, \\
University of Chicago, Chicago, Illinois 60637, USA
}\\
{\it
$^{(c)}$ Kavli Institute for Cosmological Physics, \\
University of Chicago, Chicago, Illinois 60637, USA
}\\
{\it 
$^{(d)}$ HEP Division, Argonne National Laboratory, \\
9700 Cass Avenue, Argonne, Illinois 60439, USA
}
\end{center}
\vspace{0.5cm}

\begin{abstract}
The minimal supersymmetric standard model leads to precise predictions of the properties of the light Higgs boson degrees of freedom 
that depend on only a few relevant supersymmetry-breaking parameters. In particular, there is an upper bound on the mass of the lightest
neutral Higgs boson, which for a supersymmetric spectrum of the order of a TeV is barely above the one of the Higgs resonance recently observed  at
the LHC.  This bound can be raised by considering a heavier supersymmetric spectrum, relaxing the tension between 
theory and experiment. In a previous article, we studied the predictions for the lightest $CP$-even Higgs mass for large values of
the scalar-top and heavy Higgs boson masses. In this article we perform a similar analysis, considering also the case of  a
$CP$-odd Higgs boson mass $m_A$ of the order of the weak scale. We perform the calculation using effective theory techniques, considering
a two-Higgs doublet model and a Standard Model-like theory and resumming the large logarithmic corrections that appear at scales above and below $m_A$, respectively. 
We calculate the mass and couplings of the lightest $CP$-even Higgs boson and compare our results with the ones obtained by other methods. 
\end{abstract}

\maketitle


\section{Introduction} \label{sect:intro}

Since the discovery of the Higgs boson at the LHC in 2012 \cite{Chatrchyan:2012ufa, Aad:2012tfa}, both the ATLAS and CMS experiments have made increasingly precise measurements of its mass $M_h$, mainly in the $h \rightarrow ZZ, \gamma\gamma$ decay channels. Using $\simeq$ 5 fb${}^{-1}$ of data at $\sqrt{s} = 7$ TeV and $\simeq$ 20 fb${}^{-1}$ of data at $\sqrt{s} = 8$ TeV,  the ATLAS and CMS experiments have measured \cite{Aad:2014aba,Khachatryan:2014jba}
\ba{
\text{ATLAS: } & M_h = 125.36 \pm 0.37 \pm 0.18 \GeV \,, \\
\text{CMS: } & M_h = 125.02  \ {}^{+0.26}_{-0.27}  \ {}^{+0.14}_{-0.15} \GeV \,,
}
where the quoted uncertainties are statistical and systematic, respectively. 
The result from a recent combination of the measurements from ATLAS and CMS is $M_h = 125.09 \pm 0.21 \pm 0.11$ GeV \cite{Aad:2015zhl}.

Low-energy supersymmetry (SUSY) is a highly predictive framework that can accommodate the observed Higgs mass and Standard Model (SM)-like properties in a variety of models~\cite{Heinemeyer:2011aa}. These models contain at least an extra Higgs doublet and the observed Higgs boson is usually identified with the lightest $CP$-even state $h$, with properties that deviate from the SM one depending on the mixing with the other neutral scalar states in the theory. In the Minimal Supersymmetric Standard Model (MSSM), 
the Higgs sector reduces at tree level to a type II two-Higgs doublet model (THDM), with the mass of the lightest $CP$-even Higgs boson bounded to be smaller than the neutral gauge boson mass $M_Z$.

This tree-level result, however, is modified by SUSY-breaking effects, receiving large radiative corrections from heavy top squarks. In the case of heavy supersymmetric particles and nonstandard Higgs bosons, the Higgs boson mass may be determined as a function of the stop masses and their mixings, depending only weakly on 
other SUSY-breaking parameters.  Models with heavy supersymmetric particles are motivated by the absence of any significant deviations of flavor or precision measurement
observables with respect to the SM predictions.  Hence, a precise computation of the Higgs mass as a function of the stop mass parameters is of significant interest. 

There has been much activity in the computation of the Higgs mass in the MSSM in the past.  The Higgs mass has been calculated by performing fixed-order perturbative calculations in the MSSM, as well as in effective theory analyses, in which the dominant logarithmic dependence has been resummed by renormalization group (RG) methods.  For supersymmetric particle masses of the order of the weak scale, an accurate prediction of the Higgs mass may be obtained by computing the radiative effects diagrammatically up to a fixed order in perturbation theory~\cite{Haber:1990aw,Heinemeyer:1998jw, Martin:2004kr,Degrassi:2001yf,Brignole:2002bz,Dedes:2003km}. 
Alternatively, the dominant radiative corrections at a given order in perturbation theory may be obtained from effective potential methods, 
using derivatives of the effective potential $V(H_1, H_2)$, for values of the Higgs field equal to their vacuum expectation values $\mean{H_1} = v_1, \, \mean{H_2} = v_2$~
\cite{Espinosa:1999zm, Espinosa:2000df, Martin:2001vx, Martin:2002iu}. 
These fixed-order calculations have been now carried out up to partial three-loop order~\cite{Martin:2007pg,Harlander:2008ju,Kant:2010tf,Feng:2013tvd}.

On the other hand, for heavy supersymmetric particles, the effective field theory approach
may be implemented by integrating out MSSM particles, considering the induced thresholds
to the relevant couplings and running them down to the electroweak scale, evaluating the
effective potential approximation of the Higgs mass, and, after appropriate corrections, 
the pole mass~\cite{Casas:1994us, Carena:1995bx, Carena:1995wu, Haber:1996fp, Carena:2000dp}. 
It is clear that for low values of the supersymmetric particle masses, where the logarithmic 
corrections are similar in size to the nonlogarithmic ones, the fixed-order calculations are
expected to lead to the most accurate values. For very heavy supersymmetric particles, 
the logarithmic corrections become very large, the fixed-order perturbation theory breaks down, 
and the RG approach leads to an appropriate resummation of the leading logarithmic corrections. 
In this case, the effective field theory methods may lead to a more accurate determination of the Higgs mass.

In a previous work~\cite{Draper:2013oza}, we used effective field theory (EFT) calculations 
to compute the mass of the lightest $CP$-even Higgs boson in the MSSM, in the case of heavy
stops and nonstandard Higgs bosons. A similar approach was also taken recently in Refs.~\cite{Bagnaschi:2014rsa, Vega:2015fna, Cheung:2014hya}. 
We studied the cases of light and heavy charginos and neutralinos, which can lead to
relevant radiative corrections to the lightest $CP$-even Higgs mass. I
Furthermore, we provided an analytical approximation for the relevant three- and four-loop
corrections to the Higgs mass that revealed a large cancellation between the dominant and subdominant leading-log contributions, 
leading to a large difference between our computations and the previous partial three-loop calculations discussed above.

\begin{figure}[tb]
\begin{center}
\includegraphics[width=0.85\linewidth]{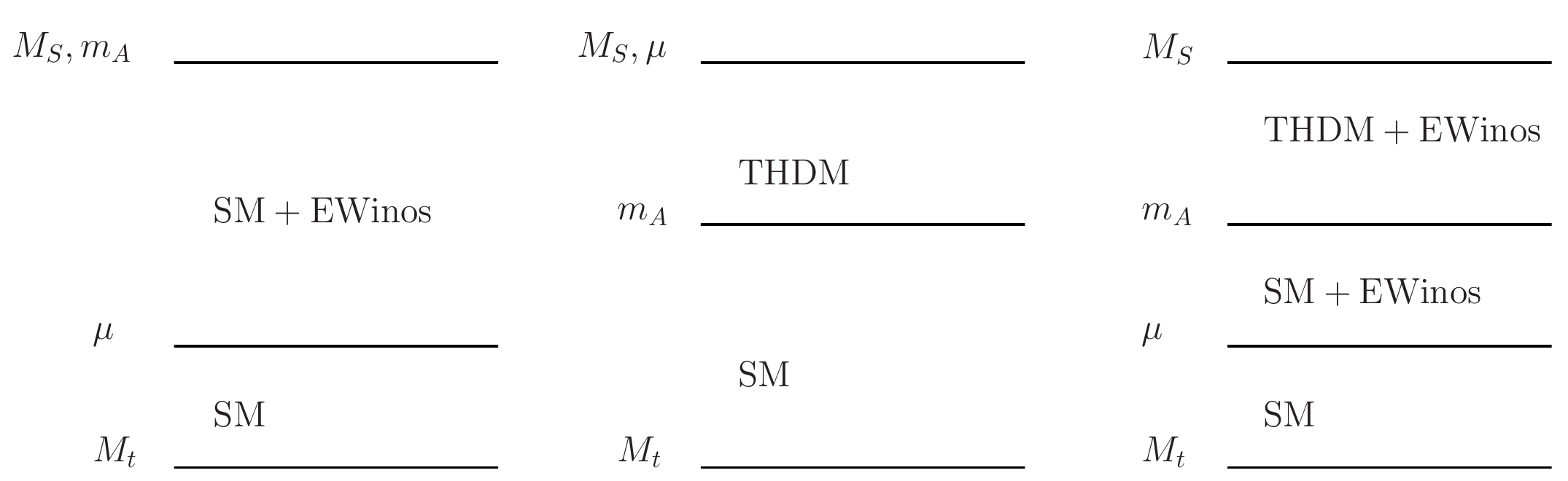}
\end{center}
\vspace{-1.5em}
\caption{Examples of hierarchies of scales examined in Ref.~\cite{Draper:2013oza} (left), and in this paper (middle, right).}
\label{fig:scales}
\end{figure}

In this article, we perform a similar study for the case of a small $CP$-odd Higgs mass, characterizing a light nonstandard Higgs boson spectrum. 
In this case, the theory below the stop mass scale is a THDM, with the possibility of additional charginos and neutralinos, 
depending on the choice of the gaugino and Higgsino mass parameters; see Fig.~\ref{fig:scales}.
This approach was first detailed in Ref.~\cite{Haber:1993an}.
The presence of two $CP$-even Higgs bosons at low energies leads to mixing effects in the $CP$-even Higgs sector that affect the predicted lightest $CP$-even Higgs boson mass and couplings, and therefore modify the Higgs physics at high energy colliders. 

This article is organized as follows: in Sec.~\ref{sect:THDM} we describe the properties of the low energy effective theory, the THDM. 
In Sec.~\ref{sect:MSSM} we describe the constraints on this generic framework when we assume the presence of a softly broken supersymmetric theory.
In Sec.~\ref{sect:rge} we study the numerical predictions for the Higgs boson masses and mixing angles. 
In Sec.~\ref{sect:comparison} we describe the approach to the alignment limit and the comparison with the values predicted in the hMSSM approach. 
We reserve Sec.~\ref{sect:concl} for our conclusions.


\section{Two-Higgs Doublet Model} \label{sect:THDM}

The most general scalar potential with two complex $SU(2)_L$ doublet Higgs fields $\F_1, \F_2$, each carrying hypercharge $Y = 1$, is \cite{Haber:1993an}
\begin{equation} 
\begin{split}
V &= m_1^2 \F_1^\dag \F_1 + m_2^2 \F_2^\dag \F_2 - (m_{12}^2 \F_1^\dag \F_2 + \hc) + \frac{\l_1}2 (\F_1^\dag \F_1)^2 + \frac{\l_2}2 (\F_2^\dag \F_2)^2 \\
& \quad + \l_3 (\F_1^\dag \F_1) (\F_2^\dag \F_2) + \l_4 (\F_1^\dag \F_2) (\F_2^\dag \F_1) \\
& \quad + \Bigg\{ \frac{\l_5}2 (\F_1^\dag \F_2)^2 + \Big[ \l_6 (\F_1^\dag \F_1) + \l_7 (\F_2^\dag \F_2) \Big] \F_1^\dag \F_2 + \hc \Bigg\} \,. \label{eqn:thdmpotn}
\end{split} 
\end{equation}
We assume $CP$-conservation and for simplicity, will take the coefficients $m_{12}^2, \l_5, \l_6,$ and $\l_7$ to be real. At the minimum of the scalar potential, the Higgs fields acquire vacuum expectation values
\be
\mean{\F_i} = \frac1{\sqrt2} \begin{pmatrix} 0 \\ v_i \end{pmatrix} \,,
\ee
and we can parametrize them by writing
\be
\F_i = \begin{pmatrix} \f_i^+ \\ \frac1{\sqrt2} (v_i + \f_i^0 + i a_i^0) \end{pmatrix} \,,
\ee
where $\f_i^+$ is complex and $\f_i^0, a_i^0$ are real. We choose the $v_i$ to be real and non-negative, with the usual relations
\be
v = \sqrt{v_1^2 + v_2^2} \simeq 246 \GeV \,, \qquad t_\b \equiv \tan\beta = v_2/v_1 \,.
\ee

After electroweak symmetry breaking (EWSB), five physical Higgs bosons remain in the spectrum: two $CP$-even, charged $H^\pm$; two $CP$-even, neutral $h, H$; and one $CP$-odd, neutral $A$. Minimizing the scalar potential, we can eliminate $m_1^2, m_2^2$, and we have the following expressions for the squared masses of $A$ and $H^\pm$:
\ba{
m_A^2 &= m_{12}^2 - \frac12 v^2 (2\l_5 + \l_6 t_\b^{-1} + \l_7 t_\b) \,, \\
m_{H^\pm}^2 &= m_A^2 + \frac12 v^2 (\l_5 - \l_4) \,.
}
The squared mass matrix for the $CP$-even, neutral Higgs bosons in the $\{\F_1, \F_2\}$ basis is
\be
\mathcal{M}^2 = 
\begin{pmatrix} \mathcal{M}_{11}^2 & \mathcal{M}_{12}^2 \\ \mathcal{M}_{12}^2 & \mathcal{M}_{22}^2 \end{pmatrix} =
m_A^2 \begin{pmatrix} s_\b^2 & -s_\b c_\b \\ -s_\b c_\b & c_\b^2 \end{pmatrix}
+ v^2 \begin{pmatrix} f_{11} & f_{12} \\ f_{12} & f_{22} \end{pmatrix} \,,
\label{CPevenmass}
\ee
where $s_\b = \sin\b, c_\b = \cos\b$. Throughout this paper, we will employ similar shorthand $s_\q = \sin\q, c_\q = \cos\q, t_\q = \tan\q$ for a generic angle $\q$. The $f_{ij}$ are
\ba{
f_{11} &= \l_1 c_\b^2 + 2\l_6 c_\b s_\b + \l_5 s_\b^2 \,, \\
f_{12} &= (\l_3 + \l_4) c_\b s_\b + \l_6 c_\b^2 + \l_7 s_\b^2 \,, \\
f_{22} &= \l_2 s_\b^2 + 2\l_7 c_\b s_\b + \l_5 c_\b^2 \,.
}
Diagonalizing this matrix, the masses of the physical $CP$-even, neutral Higgs bosons are given by
\be
\begin{split}
m_{H, h}^2 = \frac12 \Big( \Tr \mathcal{M}^2 & \pm \sqrt{(\Tr \mathcal{M}^2)^2 - 4 \det \mathcal{M}^2} \Big) 
= \frac12 \Big( \mathcal{M}_{11}^2 + \mathcal{M}_{22}^2 \pm \d m^2 \Big) \,, \\
\d m^2 &= \sqrt{ \big( \mathcal{M}_{11}^2 - \mathcal{M}_{12}^2 \big)^2 + 4 \big( \mathcal{M}_{12}^2 \big)^2} \,,
\end{split}
\ee
where the mixing angle $\a$ for the neutral $CP$-even states is
\be
\begin{split}
c_\a = \sqrt{\frac{\d m^2 + \mathcal{M}_{11}^2 - \mathcal{M}_{22}^2}{2 \d m^2}} \,, & \qquad
s_\a = \frac{\sqrt2 \mathcal{M}_{12}^2}{\sqrt{\d m^2 (\d m^2 + \mathcal{M}_{11}^2 - \mathcal{M}_{22}^2)}} \,, \\
\begin{pmatrix} H \\ h \end{pmatrix} &= \begin{pmatrix} c_\a & s_\a \\ -s_\a & c_\a \end{pmatrix} \begin{pmatrix} \F_1 \\ \F_2 \end{pmatrix} \,,
\end{split}
\ee
and the mixing angle is defined in the range $-\pi/2 \leq \a \leq 0$.

We can also rotate to the \textit{Higgs basis} $\{H_1, H_2\}$ \cite{Gunion:2002zf},
\be
\begin{pmatrix} H_1 \\ H_2 \end{pmatrix} = \begin{pmatrix} c_\b & s_\b \\ -s_\b & c_\b \end{pmatrix} \begin{pmatrix} \F_1 \\ \F_2 \end{pmatrix} \,,
\ee
where only one of the scalars receives a vacuum expectation value, $\mean{H_1} = v/\sqrt2$. In the Higgs basis, the $CP$-even mass matrix takes a similar form, 
\be
\mathcal{M}_H^2 = 
m_A^2 \begin{pmatrix} 0 \ & \ 0 \\ 0 \ & \ 1 \end{pmatrix}
+ v^2 \begin{pmatrix} g_{11} & g_{12} \\ g_{12} & g_{22} \end{pmatrix} \,, \label{eq:MHHiggs}
\ee
where
\ba{
g_{11} &= \l_1 c_\b^4 + \l_2 s_\b^4 + 2 (\l_3 + \l_4 + \l_5) s_\b^2 c_\b^2 + 4 \l_6 c_\b^3 s_\b + 4 \l_7 s_\b^3 c_\b \,, 
\label{eqn:g11} \\
g_{12} &= c_\b s_\b (\l_2 s_\b^2 - \l_1 c_\b^2 + ( \l_3 + \l_4 + \l_5) c_{2\beta} ) + 3 (\l_7 - \l_6) s_\b^2 c_\b^2  + \l_6 c_\b^4 - \l_7 s_\b^4 \,, \label{eqn:g12} \\
g_{22} &= (\l_1 + \l_2 ) c_\b^2 s_\b^2  - 2 (\l_3 + \l_4) s_\b^2 c_\b^2 + \l_5 (s_\b^4 + c_\b^4) + (\l_7-\l_6) s_{2\beta} c_{2\beta}  \,, \label{eqn:g22}
}
and it follows that in this basis the mixing angle is $\beta - \alpha$, namely
\be
\begin{pmatrix} H \\ h \end{pmatrix} = \begin{pmatrix} c_{\beta-\alpha} & -s_{\beta-\alpha} \\ s_{\beta-\alpha} & c_{\beta-\alpha} \end{pmatrix} \begin{pmatrix} H_1 \\ H_2 \end{pmatrix} \,.
\ee

When the mixing $c_{\beta-\alpha}$ is small, this basis is convenient since the lightest $CP$-even Higgs tree-level couplings are identified with the SM Higgs ones. More generally, the Lagrangian describing the coupling of the  Higgs bosons to the top and bottom quarks at scales below $M_S$ may be parametrized in the following way:
\be
{\cal{L}} = \left( h_b + \delta h_b \right) \bar{b}_R \Phi_1^{i,*} Q_L^i+ \epsilon_{ij} \left( h_t + \delta h_t \right) \bar{t}_R Q_L^i \Phi_2^j  
+ \Delta h_b \bar{b}_R Q_L^i \Phi_2^{i *} + \epsilon_{ij}\Delta h_t \bar{t}_R Q_L^i \Phi_1^{j} + \text{H.c.}
\ee

From here it follows that the bottom and quark running masses are given by
\begin{eqnarray}
m_b & = & \frac{h_b v}{\sqrt{2}} c_\beta \left( 1 + \frac{ \delta h_b}{h_b} + \frac{ \Delta h_b t_\beta}{h_b} \right) \,, \label{eq:bloopcouplings} \\
m_t & = & \frac{h_t v}{\sqrt{2}} s_\beta \left( 1 + \frac{\delta h_t}{h_t} + \frac{ \Delta h_t}{h_t t_\beta} \right) \,. 
\label{eq:tloopcouplings}
\end{eqnarray}
with the relevant couplings evaluated at the weak scale. 
Observe that while the corrections to the bottom coupling are loop suppressed, 
they are enhanced at moderate or large values of $t_\b$ and therefore they may take values of order 1 in this regime. 
On the contrary, the corrections to the top coupling are suppressed by both loop and $t_\b$ factors and therefore tend to be small.

At tree level, the MSSM Yukawa couplings are related to the SM Yukawa couplings by
\be
y_t = h_t s_\b , \qquad y_b = h_b c_\b, \qquad y_\tau = h_\tau c_\b \,.
\ee
From Eqs.~(\ref{eq:bloopcouplings})--(\ref{eq:tloopcouplings}), it follows that these couplings are modified at one-loop order at $M_S$ in the following forms \cite{Guasch:2001wv, Carena:2000yi},
\ba{
h_t &= \frac{y_t}{s_\b} \frac1{1 + \d_t + \D_t} \,, \label{eqn:hytMSbar} \\
h_b &= \frac{y_b}{c_\b} \frac1{1 + \d_b + \D_b} \,, \label{eqn:hybMSbar}  \\
h_\tau &= \frac{y_\tau}{c_\b} \frac1{1 + \d_\tau + \D_\tau} \,, \label{eqn:hytauMSbar}
}
where $\d_i = \d h_i/h_i$ are the terms without factors of $t_\b$, and $\D_t = (\D h_t \, t_b^{-1})/h_t$ $[ \D_b = (\D h_b \, t_b)/h_b, \D_\t =  (\D h_\t \, t_b)/h_\t ]$ are $t_\b$ suppressed [enhanced] terms:
\ba{
\frac{\d_t}{\k} &= - \frac83 g_3^2 m_{\tilde{g}} A_t \, I(m_{\tt_1}, m_{\tt_2}, m_{\tilde{g}})
- h_b^2 \mu^2 \, I(m_{\tilde{b}_1}, m_{\tilde{b}_2}, \mu) - \frac29 g_Y^2 M_1 A_t I( m_{\tt_1}, m_{\tt_2}, M_1) \,, \label{eqn:dt} \\
\frac{t_\b \D_t}{\k} &= \frac83 g_3^2 m_{\tilde{g}} \mu \, I(m_{\tt_1}, m_{\tt_2}, m_{\tilde{g}})
+ h_b^2 \mu A_b \, I(m_{\tilde{b}_1}, m_{\tilde{b}_2}, \mu) \nl
& \quad - g_2^2 M_2 \mu \bigg\{ \Big[ c_b^2 I(m_{\tilde{b}_1},  M_2, \mu) + s_b^2 I(m_{\tilde{b}_2},  M_2, \mu) \Big]
+ \frac12 \Big[ c_t^2 I(m_{\tt_1}, M_2, \mu) +  s_t^2 I(m_{\tt_2}, M_2, \mu) \Big] \bigg\} \nl
& \quad + \frac13 g_Y^2 M_1 \mu \bigg\{ \frac23 I( m_{\tt_1}, m_{\tt_2}, M_1) 
+ \frac12 \Big[ c_t^2 I(m_{\tt_1}, M_1, \mu) + s_t^2 I(m_{\tt_2}, M_1, \mu) \Big] \nl
& \quad - 2 \Big[ s_t^2 I(m_{\tt_1}, M_1, \mu) + c_t^2 I(m_{\tt_2}, M_1, \mu) \Big] \bigg\} \,, \label{eqn:Dt}
}
\ba{
\frac{\d_b}{\k} &= - \frac83 g_3^2 m_{\tilde{g}} A_b \, I(m_{\tilde{b}_1}, m_{\tilde{b}_2}, m_{\tilde{g}})
- h_t^2 \mu^2 \, I(m_{\tt_1}, m_{\tt_2},  \mu) 
+ \frac19 g_Y^2 M_1 A_b \, I( m_{\tilde{b}_1}, m_{\tilde{b}_2}, M_1) \,, \label{eqn:db} \\
\frac{\D_b}{\k \, t_\b}  &= \frac83 g_3^2 m_{\tilde{g}} \mu \, I(m_{\tilde{b}_1}, m_{\tilde{b}_2}, m_{\tilde{g}})
+ h_t^2 \mu A_t \, I(m_{\tt_1}, m_{\tt_2},  \mu) \nl
& \quad - g_2^2 M_2 \mu \bigg\{ \Big[ c_t^2 I(m_{\tt_1}, M_2, \mu) +  s_t^2 I(m_{\tt_2}, M_2, \mu) \Big]
+ \frac12 \Big[ c_b^2 I(m_{\tilde{b}_1},  M_2, \mu) + s_b^2 I(m_{\tilde{b}_2},  M_2, \mu) \Big] \bigg\} \nl
& \quad - \frac13 g_Y^2 M_1 \mu \bigg\{ \frac13 I( m_{\tilde{b}_1}, m_{\tilde{b}_2}, M_1) 
+ \frac12 \Big[ c_b^2 I(m_{\tilde{b}_1}, M_1, \mu) + s_b^2 I(m_{\tilde{b}_2}, M_1, \mu) \Big] \nl
& \quad + \Big[ s_b^2 I(m_{\tilde{b}_1}, M_1, \mu) + c_b^2 I(m_{\tilde{b}_2}, M_1, \mu) \Big] \bigg\} \,, \label{eqn:Db}
}
\ba{
\frac{\d_\tau}{\k} &= g_Y^2 M_1 A_\tau I( m_{\tilde{\tau}_1}, m_{\tilde{\tau}_2}, M_1) \,, \label{eqn:dtau} \\
\frac{\D_\tau}{\k \, t_\b} &= - g_2^2 M_2 \mu \bigg\{ I(m_{\tilde{\nu}_\tau}, M_2, \mu) 
+ \frac12 \Big[ c_\tau^2 I(m_{\tilde{\tau}_1},  M_2, \mu) + s_\tau^2 I(m_{\tilde{\tau}_2},  M_2, \mu) \Big] \bigg\} \nl
& \quad - g_Y^2 M_1 \mu \bigg\{ I( m_{\tilde{\tau}_1}, m_{\tilde{\tau}_2}, M_1) 
- \frac12 \Big[ c_\tau^2 I(m_{\tilde{\tau}_1}, M_1, \mu) + s_\tau^2 I(m_{\tilde{\tau}_2}, M_1, \mu) \Big] \nl
& \quad + \Big[ s_\tau^2 I(m_{\tilde{\tau}_1}, M_1, \mu) + c_\tau^2 I(m_{\tilde{\tau}_2}, M_1, \mu) \Big] \bigg\} \,. \label{eqn:Dtau}
}
In the above expressions, $\k = (1/16\pi^2)$ is a loop factor; $A_{t}$ ($A_{b,\tau}$) are the trilinear couplings of the stops to the Higgs field $\Phi_2$ ($\Phi_1$); $m_{\tilde{f}}$ are the sfermion eigenstate masses; $M_{1,2}$ are the hypercharge and weak gaugino masses; $m_{\tilde{g}}$ is the gluino mass; and $\mu$ is the Higgsino mass parameter.  
The parameters $s_t, s_b, s_\tau$ $(c_t, c_b, c_\tau)$ are the sines (cosines) of the stop, sbottom, and stau mixing angles, and the function $I(a,b,c)$ is defined as
\be
I(a,b,c) = \frac{a^2 b^2 \log(a^2/b^2) + b^2 c^2 \log(b^2/c^2) + a^2 c^2 \log(c^2/a^2)}{(a^2 - b^2)(b^2 - c^2)(a^2 - c^2)} .
\ee
We will assume that the masses $m_{\tilde{g}} = m_{\tilde{b}_i} = m_{\tilde{t}_i} = m_{\tilde{\tau}_i} = m_{\tilde{\nu}_i} = M_S$ (such that $s_X^2 = c_X^2 = 1/2$ with $X = t, b, \tau$). 
We will consider the two scenarios $M_2 = M_1 = \mu = M_S$ and $M_2 = M_1 = \mu = 200$ GeV.
With these choices, the above expressions contain the dominant contributions to the threshold corrections, 
which also include all terms necessary for consistency with our threshold corrections to the quartic couplings.~\cite{Bagnaschi:2014rsa}

Strictly speaking, below $M_S$, the couplings $\Delta h_{t,b}$ and $h_{t,b} + \delta h_{t,b}$ evolve in slightly different ways. 
However, since the dominant contribution from QCD in the RG evolution of the couplings is the same, 
and the couplings $\Delta h_{t, b}$ are already loop suppressed, we shall approximate the ratios $\Delta_{t,b}$ 
as constants below $M_S$ and concentrate only on the RG evolution of the top- and bottom-quark couplings to the fields $H_u$ and $H_d$, respectively. 
We expect this approximation to have a negligible impact on the Higgs boson masses.

Using the above expressions, one can easily prove that the couplings of the light physical Higgs boson $h$ to top and bottom quarks and vector gauge bosons are given by (see, e.g.~Ref.~\cite{Carena:2014nza})
\begin{equation}
\begin{split}
g_{htt} &= \left[\left( s_{\beta - \alpha}  + \frac{c_{\beta - \alpha}}{t_\beta} \right) - \frac{\Delta_t}{1 + \delta_t+ \Delta_t} \left( \frac{ t_\beta  \ c_{\beta-\alpha} }{s_\beta^2}\right) \right] \ g_{htt}^{\rm SM} \,, \\
g_{hbb} &= \left[\left( s_{\beta - \alpha}  - c_{\beta - \alpha} \ t_\beta \right)  + \frac{\Delta_b}{1 + \delta_b + \Delta_b} \left( \frac{t_\beta \ c_{\beta-\alpha}}{s_\beta^2} \right) \right]\ g_{hbb}^{\rm SM} \,, \\
g_{hVV} &= s_{\beta-\alpha}  \ g_{hVV}^{\rm SM} \,, \label{eqn:ghffVV}
\end{split}
\end{equation}
where $g_{htt}^{\rm SM}, g_{hbb}^{\rm SM}$ and $g_{hVV}^{\rm SM}$ denote the SM couplings of top quarks, bottom quarks and weak gauge bosons to the Higgs.  One observes that in the regime of moderate or large values of $t_\b$ and small values of $c_{\b - \a}$, the bottom coupling can get sizable departures from the SM value, while the top and vector gauge boson couplings tend to be close to their SM values.

Similarly, for the heavy Higgs boson $H$, one obtains
\begin{equation}
\begin{split}
g_{Htt} &= \left[\left( -\frac{s_{\beta - \alpha}}{t_\beta} + c_{\beta - \alpha} \right) + \frac{\Delta_t}{1+ \delta_t + \Delta_t}\left(\frac{t_\beta \ s_{\beta-\alpha}}{s_\beta^2}\right) \right] \ g_{htt}^{\rm SM} \,, \\
g_{Hbb} &= \left[ \left( s_{\beta - \alpha} \ t_\beta + c_{\beta - \alpha} \right)
- \frac{\Delta_b}{1 + \delta_b+ \Delta_b} \left( \frac{t_\beta \ s_{\beta-\alpha}}{s_\beta^2} \right) \right] \ g_{hbb}^{\rm SM} \,, \\
g_{HVV} &= c_{\beta-\alpha}  \ g_{hVV}^{\rm SM} \,. \label{eqn:gheavyHffVV}
\end{split}
\end{equation}
Hence, one observes that for small values of $c_{\beta-\alpha}$ the coupling $g_{Hbb}$ of the heavy $CP$-even Higgs to bottom quarks is affected by loop corrections and can become sizable at large values of $t_\b$.  The top-quark coupling to the heavy Higgs instead remains suppressed by either loop or $t_\b$ factors. 
 
For completeness, we stress that there is a close connection between the coupling of the heavy $CP$-even Higgs and of the $CP$-odd Higgs to top and bottom quarks. These $CP$-odd Higgs boson couplings are given by
\begin{equation}
\begin{split}
g_{Att} &= \left[ \frac{1}{t_\beta} - \frac{t_\beta \ \Delta_t}{( 1 + \delta_t+ \Delta_t) s_\beta^2 } \right] g_{htt}^{\rm SM} \\
g_{Abb} &= \left[ t_\beta - \frac{t_\beta  \ \Delta_b }{(1 + \delta_b+ \Delta_b) s_\beta^2} \right] g_{hbb}^{\rm SM} \label{eqn:gAffVV}
\end{split}
\end{equation}


\section{The MSSM Higgs Sector} \label{sect:MSSM}

The MSSM Higgs potential for the two Higgs doublets $H_D, H_U$ with respective hypercharges $Y = -1, 1$ is
\ba{
V_H &= \frac18 (g_2^2 + g_Y^2) (|H_D|^2 - |H_U|^2)^2 + \frac12 g_2^2 |H_D^\dag H_U|^2 + |\mu|^2 (|H_D|^2 + |H_U|^2) \nl
& \quad + m_{11}^2 |H_D|^2 + m_{22}^2 |H_U|^2 + m_{12}^2 (H_D \cdot H_U + \hc) \,, \label{eqn:mssmpotn}
}
where $H_D \cdot H_U = \e_{ab} H_D^a H_U^b$. These originate from the D-terms in the superpotential and the soft supersymmetry-breaking terms.
To recover the form of the THDM potential in Eq. (\ref{eqn:thdmpotn}), let $m_{k}^2 = m_{kk}^2 + |\mu|^2$ for $k \in \{1, 2\}$ and $m_{12}^2 = B \mu$, 
with the following relations between the fields,
\be
\F_1 = -i\s_2 H^{*}_D \,, \qquad \F_2 = H_U \, .
\ee
The terms in Eq. (\ref{eqn:mssmpotn}) become
\be
|H_D^\dag H_U|^2 \rightarrow |\F_1|^2 |\F_2|^2 - (\F_1^\dag \F_2) (\F_2^\dag \F_1) \,, \qquad H_D \cdot H_U \rightarrow -\F_1^\dag \F_2 \,,
\ee
and we have the following tree-level relations for the quartic couplings:
\ba{
\l_1 = \l_2 &= \frac14 (g_2^2 + g_Y^2) \,, \label{eqn:l12tree} \\
\l_3 &= \frac14 (g_2^2 - g_Y^2) \,, \\
\l_4 &= -\frac12 g_2^2 \,, \\
\l _5 = \l_6 = \l_7 &= 0 \,,
}
where the notation for the above couplings is shorthand for $\l_i^{\text{MSSM}} (M_S)$.

The one-loop threshold corrections to $\l_k (M_S)$ in the MSSM from box and triangle diagrams are tabulated in, e.g. Ref.~\cite{Haber:1993an}:
\ba{
\D^{(1)}_{\text{th}} \l_1 &= - \frac{\k}2 h_t^4 \muMS^4 + 6\k h_b^4 \AbMS^2 \Big(1 - \frac{\AbMS^2}{12} \Big) + 2\k h_\t^4 \AtauMS^2 \Big(1 - \frac{\AtauMS^2}{12} \Big) \nl
&\quad + \k \frac{g_2^2 + g_Y^2}4 \Big[ 3h_t^2 \muMS^2 - 3h_b^2 \AbMS^2 - h_\t^2 \AtauMS^2 \Big] \,, \label{eqn:1loopL1} \\
\D^{(1)}_{\text{th}} \l_2 &= 6 \k h_t^4 \AtMS^2 \Big( 1 - \frac{\AtMS^2}{12} \Big) - \frac{\k}2 h_b^4 \muMS^4 - \frac{\k}6 h_\t^4 \muMS^4 \nl
&\quad - \k \frac{g_2^2 + g_Y^2}4 \Big[ 3h_t^2 \AtMS^2 - 3h_b^2 \muMS^2 - h_\t^2 \muMS^2 \Big] \,, \label{eqn:1loopL2} \\
\D^{(1)}_{\text{th}} \l_3 &= \frac{\k}6 \muMS^2 \Big[ 3h_t^4 (3 - \AtMS^2) + 3h_b^4 (3 - \AbMS^2) + h_\t^4 (3 - \AtauMS^2) \Big] \nl
&\quad + \frac{\k}2 h_t^2 h_b^2 \Big[ 3(\AtMS + \AbMS)^2 - (\muMS^2 - \AtMS \AbMS)^2 - 6\muMS^2 \Big] \\
&\quad - \frac{\k}2 \frac{g_2^2 - g_Y^2}4 \Big[ 3h_t^2 (\AtMS^2 - \muMS^2) + 3h_b^2 (\AbMS^2 - \muMS^2) + h_\t^2 (\AtauMS^2 - \muMS^2) \Big] \,, \label{eqn:1loopL3} \\
\D^{(1)}_{\text{th}} \l_4 &= \frac{\k}6 \muMS^2 \Big[ 3h_t^4 (3 - \AtMS^2) + 3h_b^4 (3 - \AbMS^2) + h_\t^4 (3 - \AtauMS^2) \Big] \nl
&\quad - \frac{\k}2 h_t^2 h_b^2 \Big[ 3(\AtMS + \AbMS)^2 - (\muMS^2 - \AtMS \AbMS)^2 - 6\muMS^2 \Big] \nl
&\quad + \frac{\k}2 \frac{g_2^2}2 \Big[ 3h_t^2 (\AtMS^2 - \muMS^2) + 3h_b^2 (\AbMS^2 - \muMS^2) + h_\t^2 (\AtauMS^2 - \muMS^2) \Big] \,, \label{eqn:1loopL4} \\
\D^{(1)}_{\text{th}} \l_5 &= -\frac{\k}6 \muMS^2 \Big[ 3h_t^4 \AtMS^2 + 3h_b^4 \AbMS^2 + h_\tau^4 \AtauMS^2 \Big] \,, \label{eqn:1loopL5}
}
\ba{
\D^{(1)}_{\text{th}} \l_6 &= \frac{\k}6 \muMS \Big[ 3h_t^4 \muMS^2 \AtMS + 3h_b^4 \AbMS (\AbMS^2 - 6) + h_\t^4 \AtauMS (\AtauMS^2 - 6) \Big] \,, \label{eqn:1loopL6} \\
\D^{(1)}_{\text{th}} \l_7 &= \frac{\k}6 \muMS \Big[ 3h_t^4 \AtMS (\AtMS^2 - 6) + 3h_b^4 \muMS^2 \AbMS + h_\t^4 \muMS^2 \AtauMS \Big] \,, \label{eqn:1loopL7} 
}
where $\wh{A}_i = A_i/M_S$, $\hat{\mu} = \mu/M_S$, the Yukawas $h_{t,b,\tau}$ are given in Eqs.~(\ref{eqn:hytMSbar}$-$\ref{eqn:Dtau}), and all parameters are in the $\MSbar$ scheme.%
\footnote{We have not included the small threshold corrections to the quartic couplings from electroweakinos, which involve only $g_Y, g_2, \l$. 
They are listed in Ref.~\cite{Ibrahim:2000qj}, and we estimate they lower $m_h$ by about $0.5$ GeV.}

In addition, there are self-energy corrections to the Higgs bosons which, after redefinition of the Higgs fields, give rise to one-loop corrections to the quartic couplings:
\ba{
\D^{(1)}_{\F} \l_1 &= -\frac{\k}6 \frac{g_2^2 + g_Y^2}2 \Big[ 3h_t^2 \muMS^2 + 3h_b^2 \AbMS^2 + h_\t^2 \AtauMS^2 \Big] \,, \\
\D^{(1)}_{\F} \l_2 &= -\frac{\k}6 \frac{g_2^2 + g_Y^2}2 \Big[ 3h_t^2 \AtMS^2 + 3h_b^2 \muMS^2 + h_\t^2 \muMS^2 \Big] \,, \\
\D^{(1)}_{\F} \l_3 &= -\frac{\k}6 \frac{g_2^2 - g_Y^2}4 \Big[ 3h_t^2 (\AtMS^2 + \muMS^2) + 3h_b^2 (\AbMS^2 + \muMS^2) + h_\t^2 (\AtauMS^2 + \muMS^2) \Big] \,, \\
\D^{(1)}_{\F} \l_4 &= \frac{\k}6 \frac{g_2^2}2 \Big[ 3h_t^2 (\AtMS^2 + \muMS^2) + 3h_b^2 (\AbMS^2 + \muMS^2) + h_\t^2 (\AtauMS^2 + \muMS^2) \Big] \,, \\
\D^{(1)}_{\F} \l_5 &= \D^{(1)}_{\F} \l_6 = \D^{(1)}_{\F} \l_7 = 0 \,. \label{eqn:l5671Lse}
}

We extend these corrections with additional two-loop $h_t^4 g_3^2$ terms, which can be extracted from the corrections to $\l$ in the $m_A \sim M_S$ case \cite{Draper:2013oza},
\be
\D^{(h_t^4 g_3^2)}_{\text{th}} \l = 16\k^2 h_t^4 s_\b^4 g_3^2 \Bigg\{ -2\XtMS + \frac13 \XtMS^3 - \frac1{12} \XtMS^4 \Bigg\} \,,
\ee
and these are matched to the quartic couplings in Eq. (\ref{eqn:g11}) by picking out the terms proportional to $(c_\b^4, s_\b^4, c_\b^2 s_\b^2, c_\b^3s_\b, s_\b^3 c_\b)$ for $(\l_1$, $\l_2, \l_{345} \equiv \l_3 + \l_4 + \l_5, \l_6, \l_7)$, respectively:
\ba{
\D^{(h_t^4 g_3^2)}_{\text{th}} \l_1 & = -\frac43 \k^2 h_t^4 g_3^2 \muMS^4 \,, \label{eqn:l12L} \\
\D^{(h_t^4 g_3^2)}_{\text{th}} \l_2 &= 16\k^2 h_t^4 g_3^2 \Big( -2\AtMS + \frac13 \AtMS^3 - \frac1{12} \AtMS^4 \Big) \,, \\
\D^{(h_t^4 g_3^2)}_{\text{th}} \l_{345} & = 4\k^2 h_t^4 g_3^2 \AtMS \muMS^2 \Big( 1 - \frac12 \AtMS \Big) \,, \\
\D^{(h_t^4 g_3^2)}_{\text{th}} \l_6 & = \frac43 \k^2 h_t^4 g_3^2 \muMS^3 \Big(-1 + \AtMS \Big) \,, \\
\D^{(h_t^4 g_3^2)}_{\text{th}} \l_7 & = 4\k^2 h_t^4 g_3^2 \muMS \Big( 2 - \AtMS^2 +  \frac13 \AtMS^3 \Big) \,. \label{eqn:l72L}
}
Note that there is an asymmetry $\D^{(h_t^4 g_3^2)}_{\text{th}} \l$ when $\XtMS \rightarrow -\XtMS$; 
however, this is subdominant to the asymmetric contribution from the $h_t$ threshold in Eq.~(28), 
which leads to log-enhanced corrections to the quartic couplings at the two-loop level.


\section{RG Evolution and Diagonalizing the Higgs Mass Matrix} \label{sect:rge}

From the low energy effective theory point of view, the tree-level values of the quartic couplings, as well as the threshold
corrections enumerated above, should be defined with the gauge and Yukawa couplings at the stop mass scale. 
More precisely, these values define the boundary conditions for the RG evolution of the quartic couplings as well as the Yukawa couplings from the
scale $M_S$ down to the scale of the $CP$-odd Higgs boson $m_A$.

It is clear that if the scale $m_A$ is of the order of the weak scale, the $CP$-even mass matrix elements, Eq.~(\ref{CPevenmass}), may be computed by evolving all quartic couplings down to the weak scale. The $CP$-even Higgs masses may then be calculated by diagonalizing the mass matrix at the weak scale and adding the proper corrections converting the running masses at the weak scale to the pole masses.  

On the other hand, if the $CP$-odd Higgs boson mass is much larger than the weak scale, 
decoupling of the heavy Higgs bosons should be achieved, 
and the mixing between the nonstandard and standard $CP$-even Higgs boson, $c_{\a-\b}$, should go to zero as $1/m_A^2$.  
Therefore, for a heavy supersymmetric spectrum, the effective theory is just the SM below the scale $m_A$. 
The lightest $CP$-even Higgs mass computed in this case should reduce to the one previously computed in Ref.~\cite{Draper:2013oza}.

Both limits may be appropriately recovered by evolving the quartic couplings up to the scale $m_A$ and computing the matrix elements at that scale in the Higgs basis. 
At scales below $m_A$ we simply evolve the $(1,1)$ matrix element by the full two-loop SM RG evolution of the quartic coupling up to the weak scale. 
On the other hand, the corrections of the off-diagonal elements coming from the evolution from $m_A$ to the weak scale become relevant only for large values of $m_A$, 
for which $c_{\a-\b}$ becomes small and therefore irrelevant from the phenomenological point of view. 
Consequently, we consider the evolution of this matrix element from $m_A$ to the weak scale by resumming the dominant top-induced corrections at the one-loop level. 
For similar reasons, for large values of $m_A$, the radiative corrections to the $(2,2)$ matrix element coming from the running between $m_A$ and the weak scale, 
which depend logarithmically on the $CP$-odd Higgs mass, become small compared with the tree-level value, 
which depend quadratically on this mass. 
For low values of $m_A < 500$~GeV and $t_\b > 4$, we have checked that our results for the Higgs boson masses (mixing angles)
differ by less than 0.5 GeV (1\%) from the ones that would be obtained by evolving all quartic couplings until the weak scale. 
For larger values of $m_A$ and lower values of $t_\b$, however, the effects of decoupling of the heavy Higgs bosons become relevant 
and for $m_A \gg 1$~TeV, the lightest $CP$-even Higgs boson mass can become significantly different from the one that would be obtained without decoupling the heavy Higgs bosons.

As a starting point for the evolution of the quartic couplings, one specifies the SM values of the gauge couplings and Yukawa couplings at some scale.
We work in the third-generation approximation, so the six couplings $g_3, g_2, g_1, y_t, y_b, y_\tau$ are relevant.
We use the low energy parameters $g_i, y_j$ at the scale of the top-quark pole mass $M_t$, which are extracted from the SM observables in Table~\ref{tab:SMobs}, 
and have values given in Table~\ref{tab:SMparamsMt}.%
\footnote{Unlike in Refs.~\cite{Draper:2013oza, Vega:2015fna}, we use the NNLO value of $y_t(M_t)$ 
instead of the NNLO+N${}^3$LO QCD value because we use three-loop SM RG equations below $m_A$, 
but only two-loop THDM RG equations above $m_A$.}
These couplings, along with an initial value%
\footnote{We have checked that the final values for $M_h$ do not have a strong dependence on the initial condition for $\l(M_t)$ if it is chosen to correspond to a value of $m_h(M_t) = \l v^2 \sim$ 100--150 GeV.}~\cite{Buttazzo:2013uya, Machacek:1983tz, Machacek:1983fi, Machacek:1984zw, Luo:2002ey, Mihaila:2012fm, Mihaila:2012pz, Chetyrkin:2012rz, Bednyakov:2012en, Chetyrkin:2013wya, Bednyakov:2013eba} 
of $\l \sim 0.25$, are evolved to the intermediate scale $m_A$ using three-loop SM RG equations for $g_3, g_2, g_1, y_t, \l$ and two-loop SM RG equations for $y_b, y_\t$. 
There are additional loop contributions to $g_1, g_2, y_t, \l$ from electroweakinos if $\mu, M_1, M_2 < m_A$.~\cite{Giudice:2011cg}
Due to their weak couplings, we have only included the dominant one-loop log-enhanced contributions from RG running 
using tree-level gauge couplings of the electroweakinos to the Higgs bosons. 

\begin{table}[tb!]
\begin{center}
\begin{tabular}{ll}
Observable \qquad & \qquad Value \qquad \\ 
\hline
$SU(3)_c$ $\MSbar$ gauge coupling (5 flavours) & $\a_s(M_Z) = 0.1184 \pm 0.0007$ \\
Fermi constant from muon decay & $V = (\sqrt2 G_F)^{-1/2} = 246.21971 \pm 0.00006 \GeV$ \\
Top-quark pole mass & $M_t = 173.34 \pm 0.76 \pm 0.3 \GeV$ \\
$W$ boson pole mass & $M_W = 80.384 \pm 0.014 \GeV$ \\
$Z$ boson pole mass & $M_Z = 91.1876 \pm 0.0021 \GeV$ \\
Higgs pole mass & $M_h = 125.09 \pm 0.21 \pm 0.11 \GeV$
\end{tabular}
\end{center}
\vspace{-1em}
\caption{SM observables, collected in Table 2 of \cite{Buttazzo:2013uya}.}
\label{tab:SMobs}
\end{table}

Above the scale $m_A$, the effective theory is the THDM, 
and two-loop type II THDM RG equations are employed in the running between $m_A$ and $M_S$.
These are listed in Appendix~\ref{app:THDM2LRGEs}, and can also be found in Ref.~\cite{Dev:2014yca}.
As above, we have included one-loop contributions to the running of $g_1, g_2, h_t, \l_k$ from electroweakinos if $m_A < \mu, M_1, M_2 < M_S$.
We note that for perturbative consistency of the RG running, three-loop RG equations should be used; however, these are not known for the THDM. 
Also, inclusion of the three-loop order RG equations in the SM running has a small effect, and we expect the same holds for the THDM.
To determine the approximate values of the MSSM gauge and Yukawa couplings at the high scale $M_S$, 
we run the couplings up to $M_S$ setting $\l_k = 0$ $(k = 1, \ldots, 7)$ in the running; 
this has subpercent level effects on the running of the Yukawas and the gauge couplings.
At $M_S$, we calculate the threshold corrections to the Yukawas according to Eqs.~(\ref{eqn:hytMSbar})--(\ref{eqn:Dtau}),
and use these results in the expressions for the MSSM values of the $\l_k$ in Eqs.~(\ref{eqn:l12tree})--(\ref{eqn:l5671Lse}, (\ref{eqn:l12L})--(\ref{eqn:l72L}).%
\footnote{We use $t_\b(m_A)$ as an input, and run $v_1$ and $v_2$ to $M_S$ using two-loop RG equations for the anomalous dimensions. When calculating the MSSM values at $M_S$, we use $t_\b(M_S) = v_2(M_S)/v_1(M_S)$.}
With these values of $\l_k(M_S)$, we use the full type II THDM RG equations in the running back down to $m_A$.
The matrix elements of $\mathcal{M}_H^2$ in Eqs.~(\ref{eq:MHHiggs}, \ref{eqn:g11}$-$\ref{eqn:g22}) are computed, 
and the value of $g_{11}$ is used as the boundary value for $\l(m_A)$ for the SM RG running from $m_A$ to $M_t$. 
$\mathcal{M}_H^2$ is then diagonalized at $M_t$, and the running masses $m_h^2, m_H^2$ and the mixing angle $\b - \a$ are computed.

The running mass $m_h$ is converted to the pole mass $M_h$ using the SM one-loop formula as in Ref.~\cite{Draper:2013oza}, in which SM $\MSbar$ running couplings are used,
\be
\begin{split}
M_h^2 &= \l (M_t) v^2 (M_t) + \k \Bigg\{ 
3y_t^2 (4\mb_t^2 - m_h^2) B_0(\mb_t, \mb_t, m_h) - \frac92 \l m_h^2 \Big[2 - \frac{\pi}{\sqrt3} - \log \frac{m_h^2}{Q^2} \Big] \\
& \qquad - \frac{v^2}4 \Big[ 3g_2^4 - 4\l g_2^2 + 4\l^2 \Big] B_0(m_W, m_W, m_h) 
+ \frac12 g_2^2 v^2 \Big[ g_2^2 - \l \Big( \log\frac{m_W^2}{Q^2} - 1 \Big) \Big] \\
& \qquad - \frac{v^2}8 \Big[ 3(g_2^2 + g_Y^2)^2 - 4\l (g_2^2 + g_Y^2) + 4\l^2 \Big] B_0(m_Z, m_Z, m_h) \\
& \qquad + \frac14 (g_2^2 + g_Y^2) v^2 \Big[ (g_2^2 + g_Y^2) - \l \Big( \log\frac{m_Z^2}{Q^2} - 1 \Big) \Big]
\Bigg\} \Bigg|_{Q^2 = M_t^2} \,,
\label{eqn:Mh1L}
\end{split}
\ee
where $B_0$ is the one-loop Passarino-Veltman integral
\be
B_0(m_1, m_2, m_3) = -\int_0^1 \log \Big[ \frac{(1-x)m_1^2 + x m_2^2 - x(1-x) m_3^2}{Q^2} \Big] \,.
\ee
We have checked that this gives similar results as the two-loop conversion using parameter values in the on-shell (OS) scheme, as in Ref.~\cite{Buttazzo:2013uya}.
We have not included contributions from light electroweakinos in the conversion formula, but we expect these contributions to be subdominant to those from the SM.
For low values of $m_A$ and $t_\b$ and large values of $M_S$, the top Yukawa $y_t$ can deviate sizably 
from the coupling of the physical Higgs to the top, $g_{htt}$ in Eq.~(\ref{eqn:ghffVV}); 
however, we checked that the shift in $M_h$ when substituting $g_{htt}$ for $y_t$ in Eq.~(\ref{eqn:Mh1L}) is less than 0.5 GeV.
Similarly, in the scenarios we investigated, we expect the difference between the running mass and the pole mass of the heavy Higgs to be small 
due to its larger mass and its reduced coupling of the heavy Higgs to the top quark in Eq.~(\ref{eqn:gheavyHffVV}).

\begin{table}[tb!]
\begin{center}
\begin{tabular}{cc|cc}
$g_i$ \qquad & \qquad $g_i(M_t)$ \qquad & \qquad $y_j$ & \qquad $y_j(M_t)$ \qquad \\ 
\hline
$g_3$ \qquad & \qquad 1.1666 \qquad & \qquad $y_t$ \qquad & \qquad 0.94018 \\
$g_2$ \qquad & \qquad 0.64779 \qquad & \qquad $y_b$ \qquad & \qquad 0.0156 \\
$g_Y = \sqrt{3/5} g_1$ \qquad & \qquad 0.35830 \qquad & \qquad $y_\tau$ \qquad & \qquad 0.0100 
\end{tabular}
\end{center}
\vspace{-1em}
\caption{Values of SM parameters at $Q = M_t$ computed in the $\MSbar$ scheme. $g_2, g_Y$, and $y_t$ are computed at NNLO, and the $SU(5)$ normalization relates $g_1$ to the SM hypercharge coupling $g_Y$. The value of $g_3$ is obtained using three-loop QCD matching to the SM. We have used the two-loop, five-flavour $\MSbar$ RG equations in the broken phase from \cite{Arason:1991ic} to run $m_b, m_\tau$ from their initial values $m_b (m_b) = 4.18 \GeV, M_\tau = 1.777 \GeV$ \cite{PDG}. For more details, see \cite{Buttazzo:2013uya}.}
\label{tab:SMparamsMt}
\end{table}

It is instructive to consider the dominant one-loop contributions to the $CP$-even Higgs matrix elements in the Higgs basis. 
This was discussed in detail in Ref.~\cite{Carena:2014nza} (for the $CP$-violating case see Ref.~\cite{Li:2015yla}), in which it was shown that
\begin{eqnarray}
g_{11} v^2 &=& m_Z^2 c^2_{2\beta}+\frac{3v^2 s_\beta^4 h_t^4}{8\pi^2}\left[\ln\left(\frac{M_S^2}{m_t^2}\right)+\XtMS^2 \left(1-\frac{\XtMS^2}{12}\right)\right]\,,\label{mhmax} \\[8pt]
g_{12} v^2&=& -s_{2\beta}\left\{m_Z^2 c_{2\beta}-\frac{3v^2 s_\beta^2  h_t^4}{16\pi^2}\biggl[\ln\left(\frac{M_S^2}{m_t^2}\right)+\frac{\XtMS(\XtMS+\YtMS)}{2}-\frac{\XtMS^3 \YtMS}{12}\biggr]\right\}, \label{zeesixcorr} 
\end{eqnarray}
where $\XtMS = X_t/M_S$, $X_t = A_t - \mu/t_\b$ is the stop mixing parameter associated with the coupling of the SM-like Higgs to the stops, $\YtMS = Y_t/M_S$ and $Y_t = A_t + \mu t_\b$.  

From the above, the mixing angle $c_{\beta-\alpha}$ may be computed,
\be
c_{\beta-\alpha}=\frac{-g_{12} v^2}{\sqrt{(m_H^2-m_h^2)(m_H^2-g_{11} v^2)}}\,.\label{cbmaf}
\ee 
For values of $m_H$ larger than the weak scale, one can show that \cite{Carena:2014nza}
\be \label{eqn:tbcbma}
t_\beta\; c_{\beta-\alpha}\simeq \frac{-1}{m_H^2-m_h^2}\left[m_h^2-m_Z^2 c_{2\beta} +
\frac{3m_t^4 \XtMS(\YtMS-\XtMS)}{4\pi^2 v^2}\left(1-\frac{\XtMS^2}{6}\right)\right] \,,
\ee
and therefore, all dominant radiative corrections to the mixing angle, which come from the renormalization of the quartic coupling $\lambda_2$ at scales above $m_A$, may be absorbed into the definition of the Higgs mass $m_h$. The remaining terms are proportional to $\muMS \XtMS \tan\beta$, vanish for maximal mixing $\XtMS^2= 6$, and cannot be absorbed into a redefinition of $m_h$.


\subsection{The hMSSM scenario} \label{sect:hMSSM}

The fact that the dominant corrections to the lightest $CP$-even Higgs mass and the mixing parameter have a common origin motivated the authors of Ref.~\cite{Djouadi:2015jea} to define
the ``hMSSM scenario", in which one assumes that for a given value of $t_\b$ and $m_A$ the proper Higgs boson mass may be obtained by choosing appropriate stop masses
and mixings. As discussed above, the dominant radiative corrections to the Higgs boson mass and the $CP$-even Higgs mixing may be absorbed into the
definition of the Higgs boson mass $m_h$.  More precisely, in the hMSSM scenario, only radiative corrections to the quartic coupling $\lambda_2$ are considered, namely $\lambda_1 = -\lambda_{345} = M_Z^2/v^2$,
$\lambda_2 = M_Z^2/v^2 + \Delta\mathcal{M}_{22}^2/(v^2 s_\beta^2)$ and $\lambda_6 = \lambda_7 = 0$. 
One can easily show that, in order to achieve the proper Higgs pole mass $M_h$, the radiative corrections must be given by
\be
\Delta\mathcal{M}_{22}^2 = \frac{M_h^2 (m_A^2 + M_Z^2 - M_h^2) - m_A^2 M_Z^2 c_{2\b}^2}{M_Z^2 c_\b^2 + m_A^2 s_\b^2 - M_h^2} \,.
\ee
The heavy $CP$-even Higgs mass is given by
\be
m_H^2 = \frac{(m_A^2 + M_Z^2 - M_h^2)(M_Z^2  c_\beta^2 + m_A^2 s_\beta^2) - m_A^2 M_Z^2 c_{2\beta}^2}{M_Z^2 c_\beta^2 + m_A^2 s_\beta^2 - M_h^2} \,. \label{eqn:hMSSMMH2} \\
\ee

Once these expressions are considered, the $CP$-even Higgs mixing angle is given by
\be
\a = -\arctan \Bigg( \frac{(M_Z^2 + m_A^2) c_\b s_\b}{M_Z^2 c_\b^2 + m_A^2 s_\b^2 - M_h^2} \Bigg) \,. \label{eqn:hMSSMalpha}
\ee
It is straightforward to show that, for values of $m_A$ larger than the weak scale, 
the mixing angle in this approximation  agrees with the one presented
in Eq.~(\ref{eqn:tbcbma}), in which the last term inside the bracket is neglected. 

From Eq.~(\ref{eqn:tbcbma}), one can then identify the main difference between our approach and the hMSSM approximation. 
For low values of $\muMS$, the main difference is associated with the radiative corrections to the quartic couplings $\lambda_1$ and $\lambda_{345}$. 
For moderate values of $t_\b$, the main logarithmic corrections to these couplings are governed by weak couplings, 
and hence are small compared to the dominant corrections absorbed into $m_h$. 

On the other hand, for sizable values of $\muMS$ and moderate values of $t_\b$,
the last term in Eq.~(\ref{eqn:tbcbma}) may become relevant and therefore we expect the hMSSM scenario to fail to accurately describe the Higgs phenomenology in this case. 
The difference will be maximal for sizable values of $\muMS$ and $\XtMS$ away from the 
maximal mixing value.  Beyond the  difference associated with the $\mu$-induced radiative
corrections, the hMSSM works under the assumption that one can always choose the supersymmetry-breaking 
parameters so that the proper Higgs mass is obtained. As we shall show, this is
not always possible: for low values of $m_A$ and $t_\b$, there are regions of parameter space 
for which obtaining the right Higgs mass would demand pushing the supersymmetric particle masses 
close or above the Planck scale, where the effective low energy supersymmetry description is no longer valid. 

We can also examine the $Hhh$ coupling. We define the Feynman rule for the vertex as $ig_{Hhh}$. This is given by \cite{Gunion:2002zf, Carena:2013ooa}
\be
\begin{split}
g_{Hhh} = -3v s_\b c_\b^3 \Big\{ &
-\l_6 s_{\a\b}^3 t_\b^{-1} 
+ s_{\a\b} c_{\a\b} \Big[ (\l_1 - \l_{345}) s_{\a\b}
+ t_\b \Big( \l_6 (2c_{\a\b} + s_{\a\b}) - \l_7 (2s_{\a\b} + c_{\a\b}) \Big) \\
& + c_{\a\b} t_\b^2 ( -\l_2 + \l_{345}) \Big] 
+ \l_7 c_{\a\b}^3 t_\b^3 \Big\}
- \l_{345} v c_{\a-\b} \,,
\label{eqn:gHhh}
\end{split}
\ee
where $s_{\a\b} \equiv (-s_\a/c_\b)$ and $c_{\a\b} \equiv c_\a/s_\b$, both of which tend to 1 in the alignment limit.

The above expressions can be compared to the expression given in the hMSSM approximation~\cite{Djouadi:2015jea},
\be
g_{Hhh} = -\frac{M_Z^2}{v} \bigg\{ 2s_{2\a} s_{\b+\a} - c_{2\a} c_{\b+\a} + 3 \frac{\Delta\mathcal{M}_{22}^2}{M_Z^2} \frac{s_\a}{s_\b} c_\a^2 \bigg\} \,, \label{eqn:hMSSMgHhh}
\ee
which can be recovered from Eq.~(\ref{eqn:gHhh}) when the radiative corrections to $\lambda_2$ alone are considered.  Hence, as with the mixing angle $\alpha$, we expect the hMSSM to provide a better approximation to the correct results provided $\muMS$ is small.


\section{Numerical Results} \label{sect:numresults}

\begin{figure}[tb!]
\begin{center}
\includegraphics[width=0.45\linewidth]{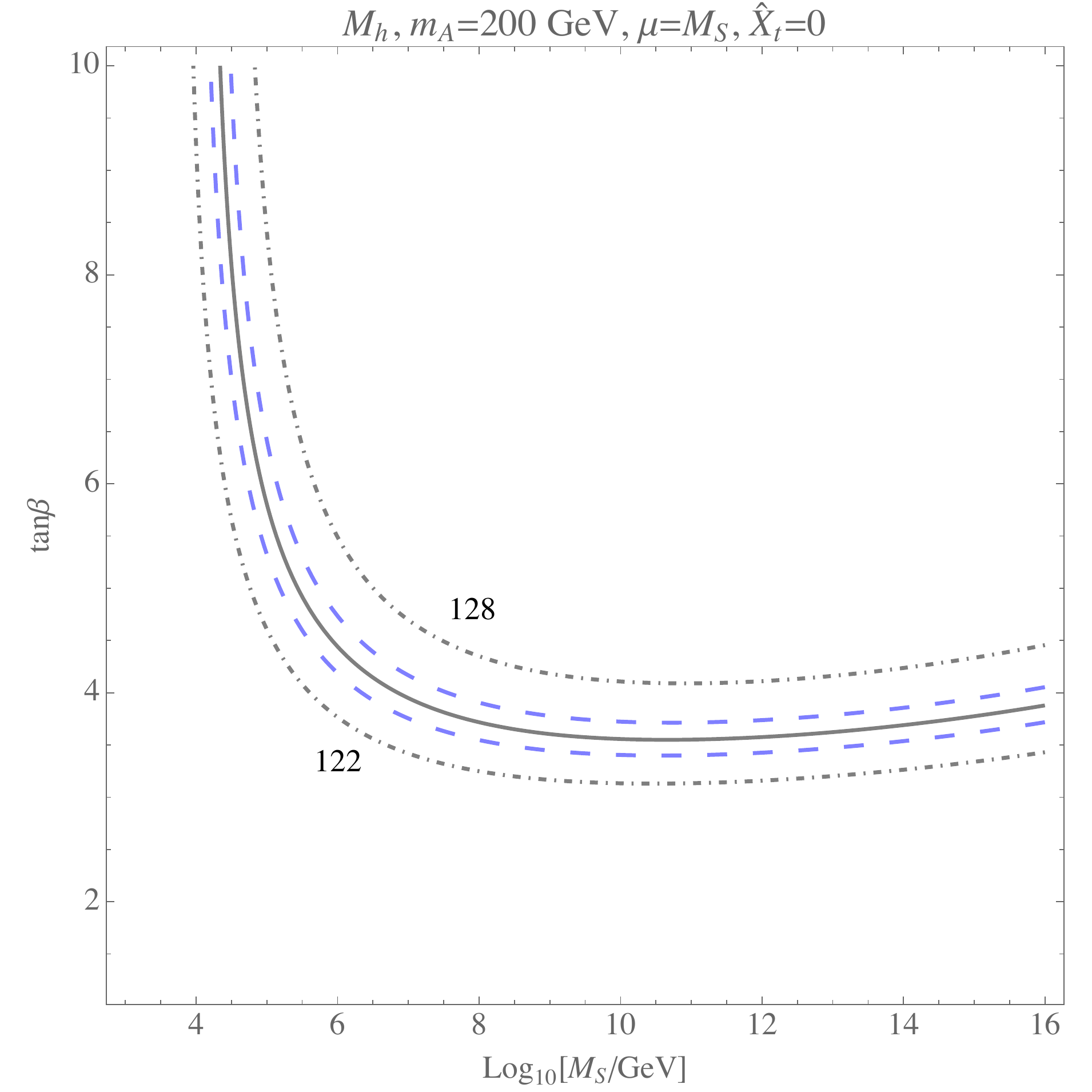} \quad
\includegraphics[width=0.45\linewidth]{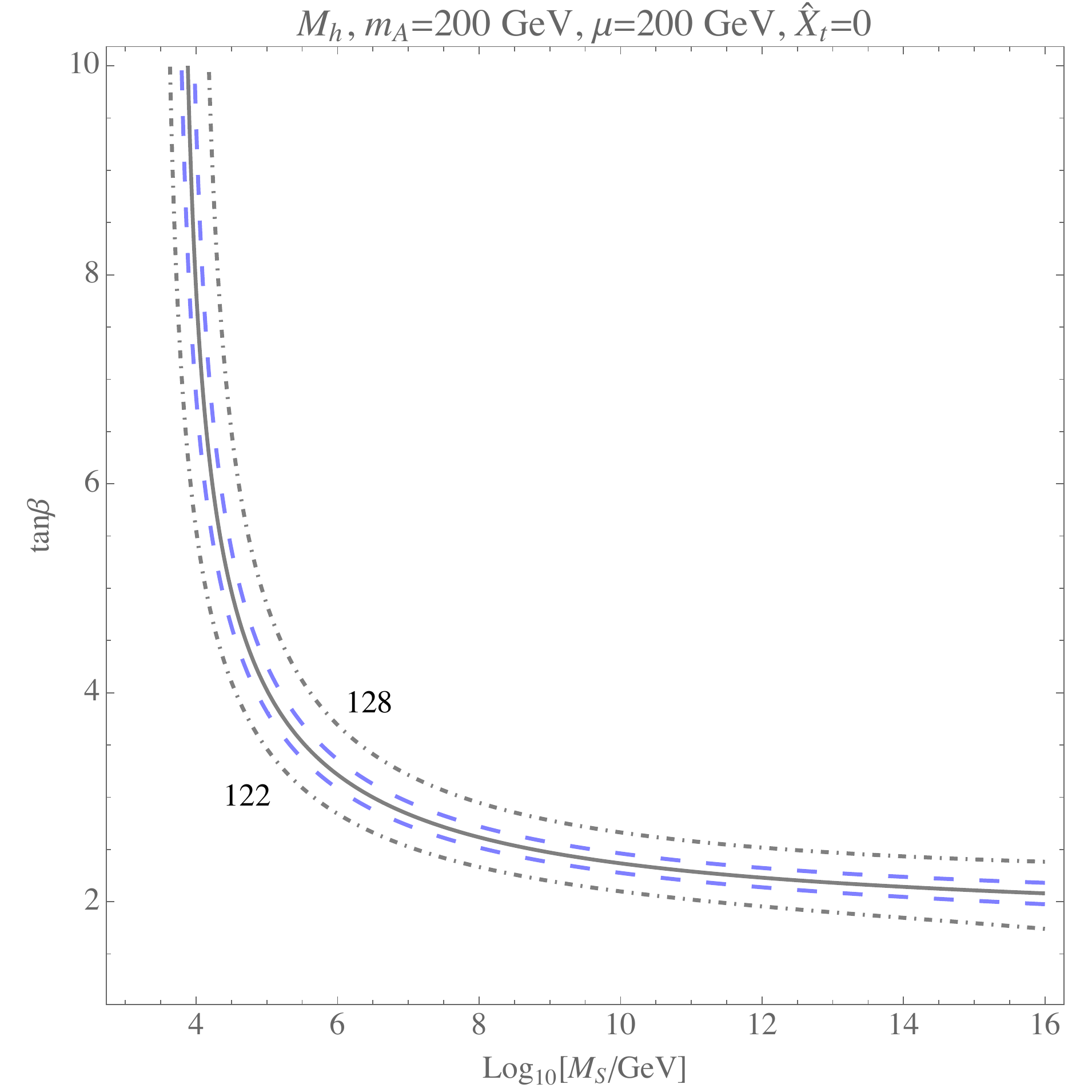}
\includegraphics[width=0.45\linewidth]{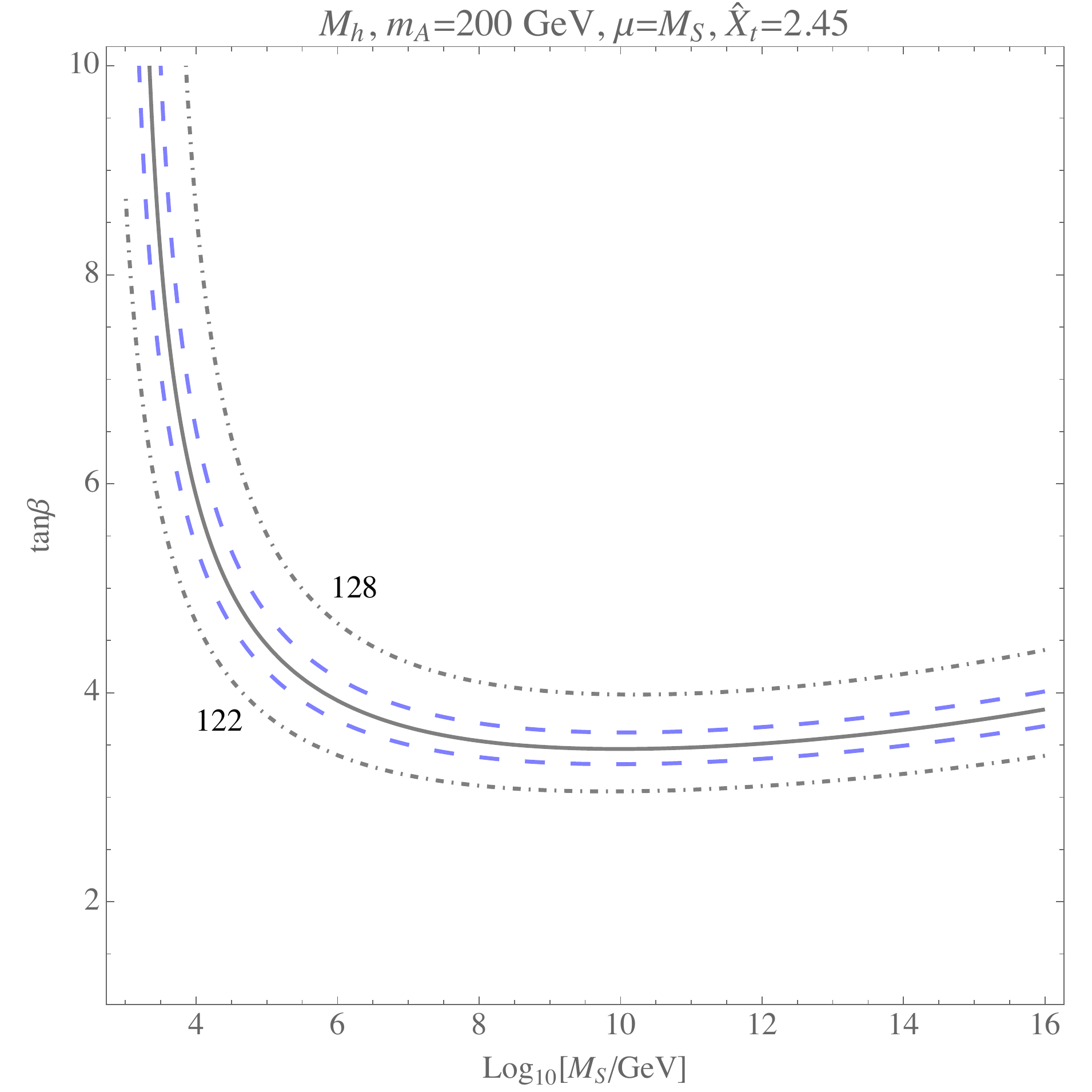} \quad
\includegraphics[width=0.45\linewidth]{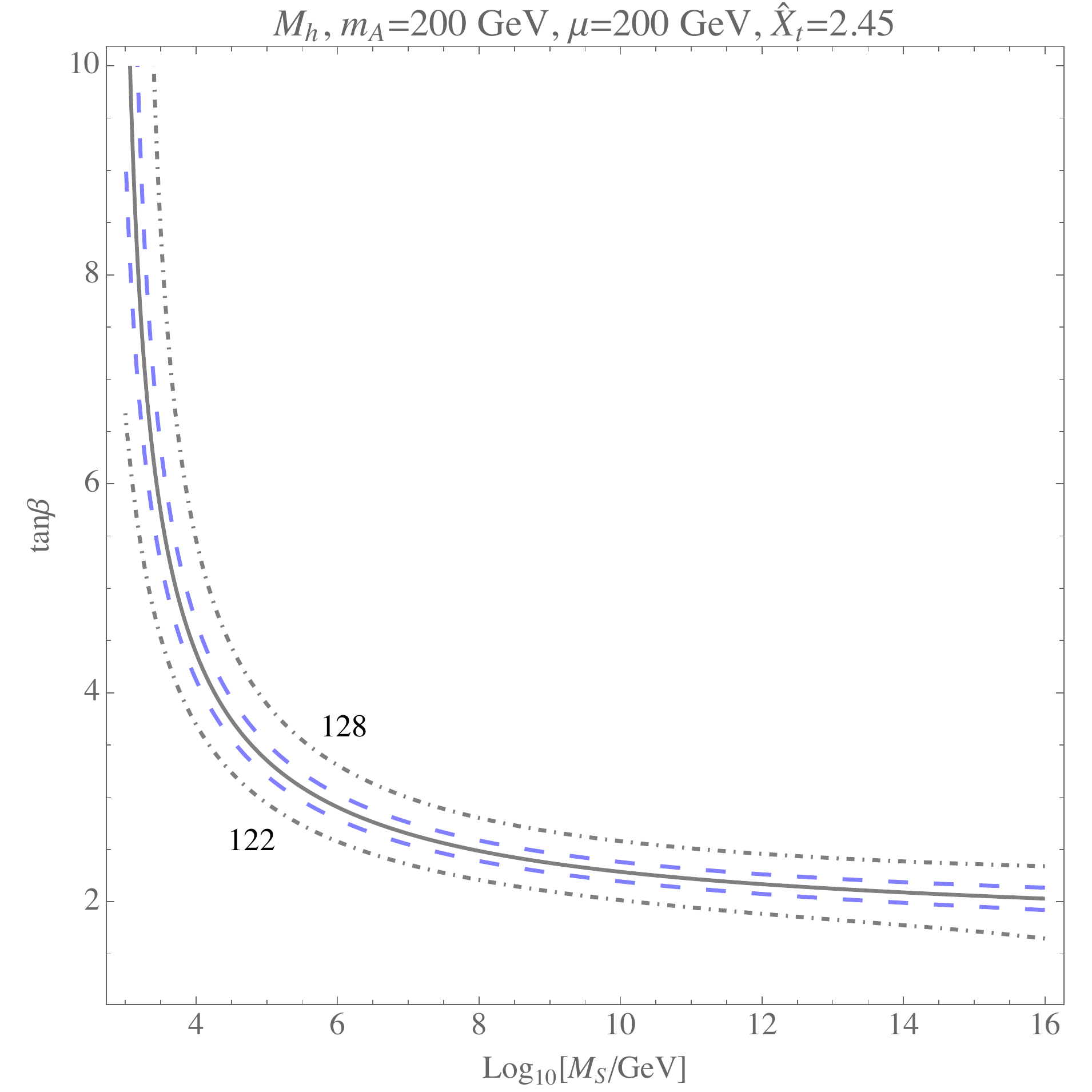}
\end{center}
\vspace{-1.5em}
\caption{Contour plots for $M_h$ in the plane $t_\b, M_S$ with $m_A = 200$ GeV, $M_1 = M_2 = \mu, A_b = A_\tau = A_t$. The outer black, dot-dashed lines are contours of $M_h = 122, 128$ GeV as labelled. The blue, dashed lines correspond to $M_h = 124, 126$ GeV, and the central black, solid line to $M_h = 125$ GeV. Plots in the top (bottom) row have $\XtMS = 0$ ($\sqrt{6}$), and plots in the left (right) column have $\mu = M_S$ ($200$ GeV).}
\label{fig:MhGUT_tanbvMS_mA200}
\end{figure}

\begin{figure}[tb!]
\begin{center}
\includegraphics[width=0.45\linewidth]{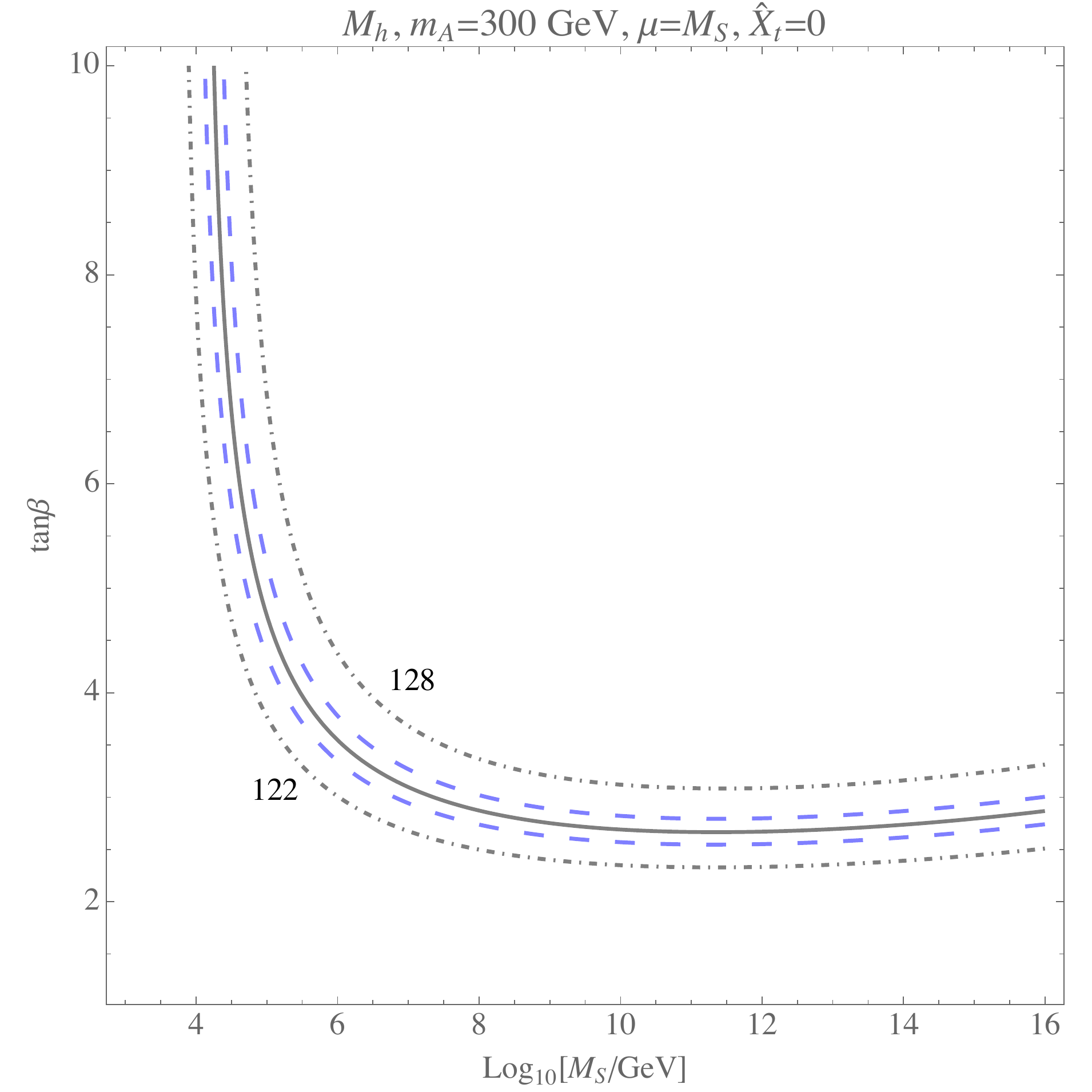} \quad
\includegraphics[width=0.45\linewidth]{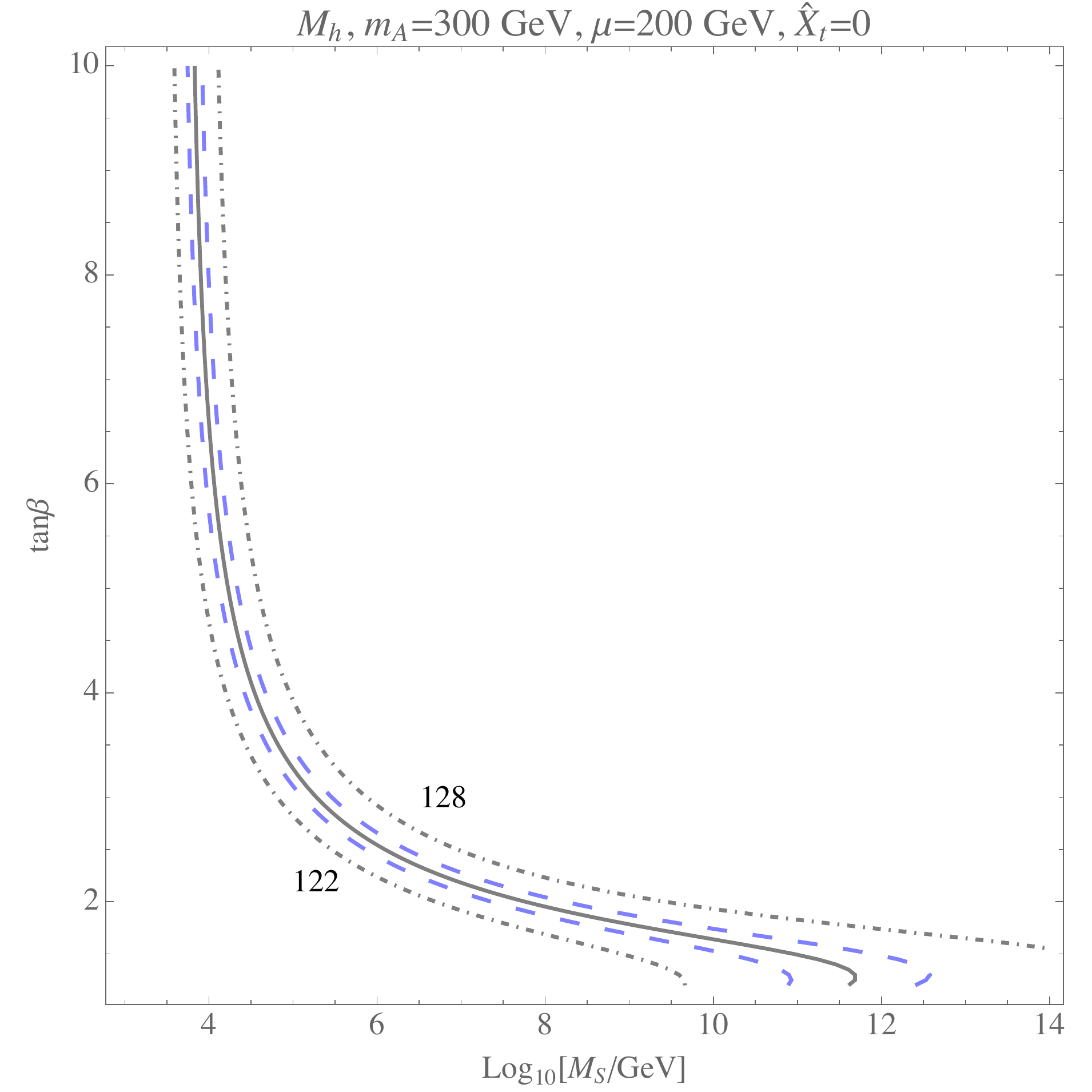}
\includegraphics[width=0.45\linewidth]{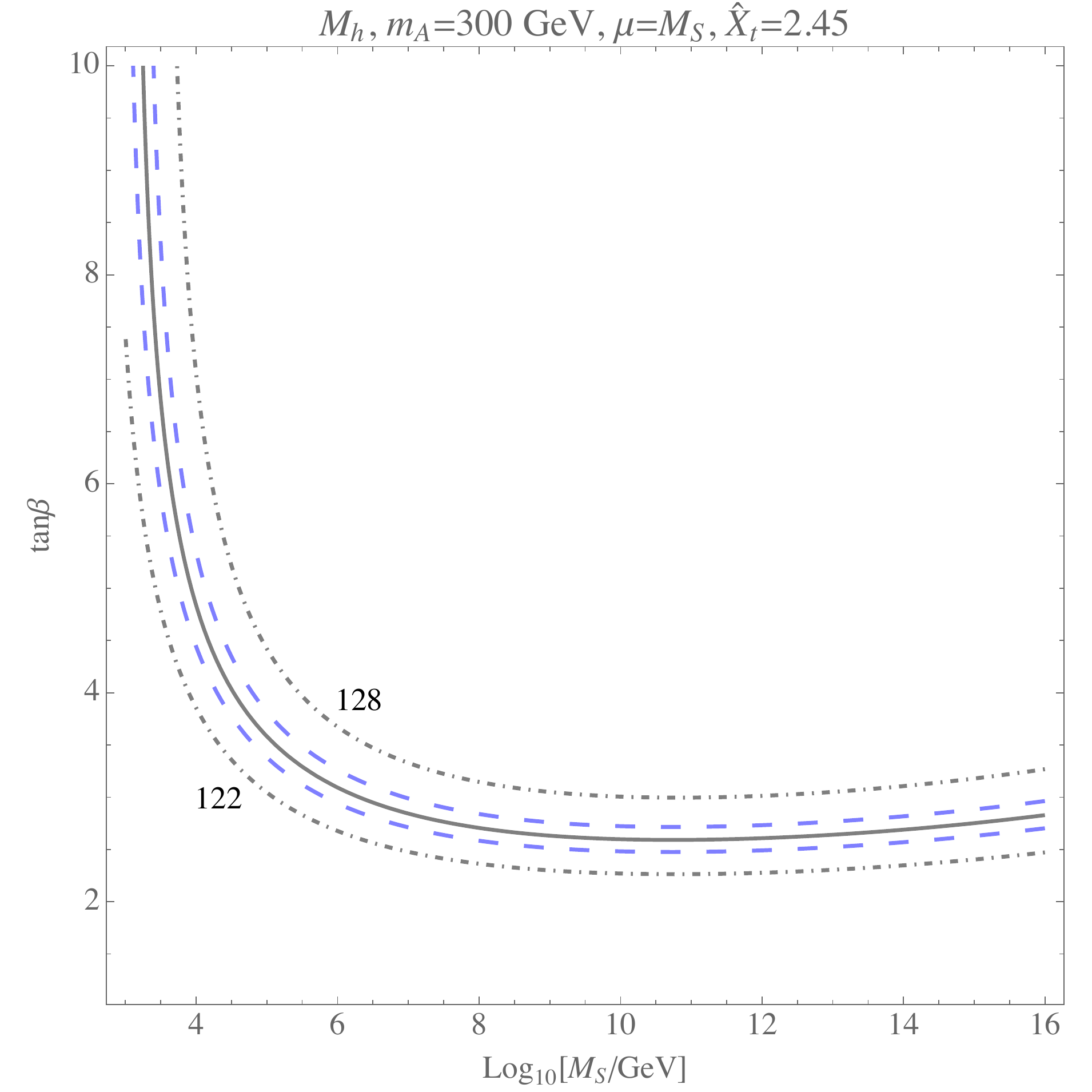} \quad
\includegraphics[width=0.45\linewidth]{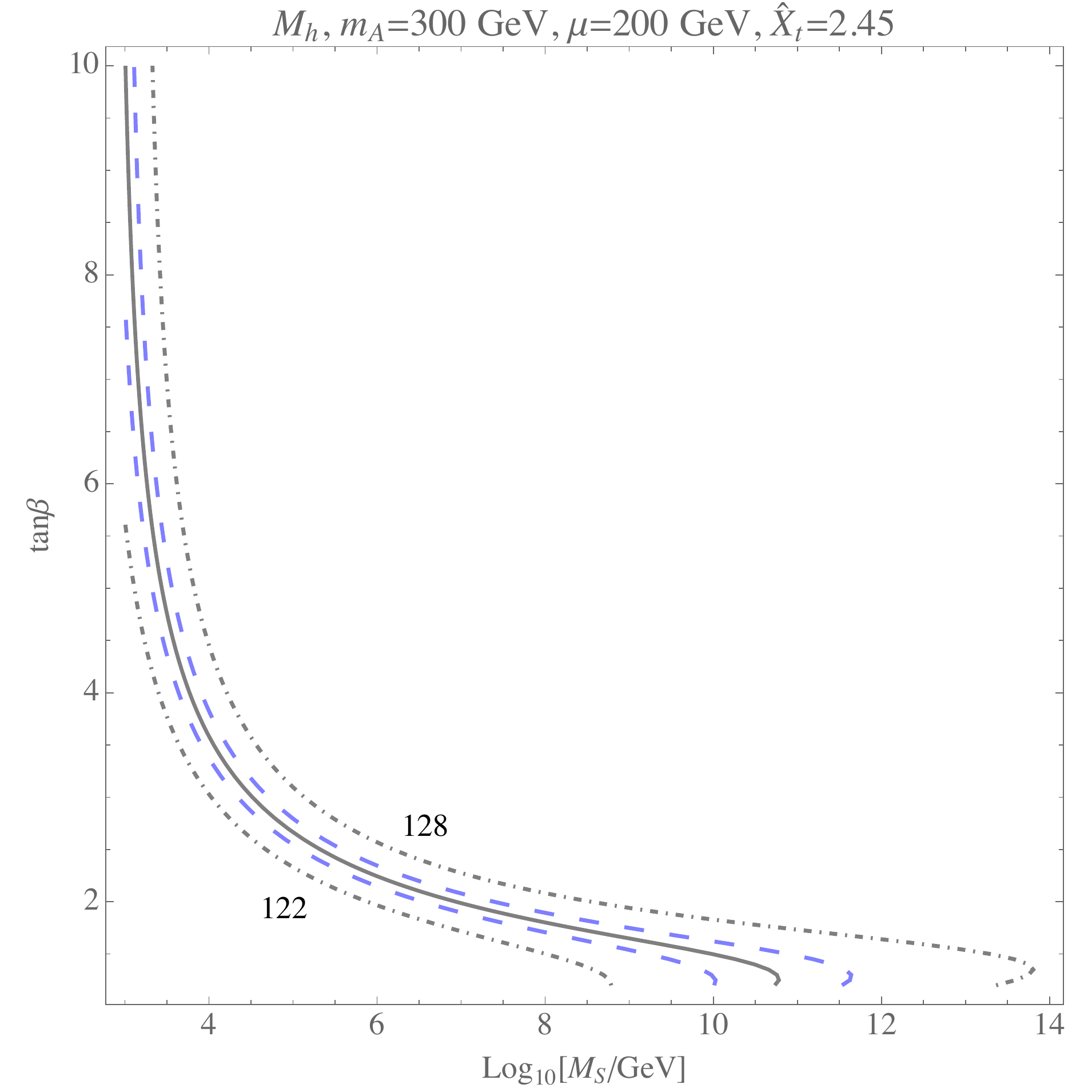}
\end{center}
\vspace{-1.5em}
\caption{As in Fig. \ref{fig:MhGUT_tanbvMS_mA200}, with $m_A = 300$ GeV.}
\label{fig:MhGUT_tanbvMS_mA300}
\end{figure}

\begin{figure}[tb]
\begin{center}
\includegraphics[width=0.45\linewidth]{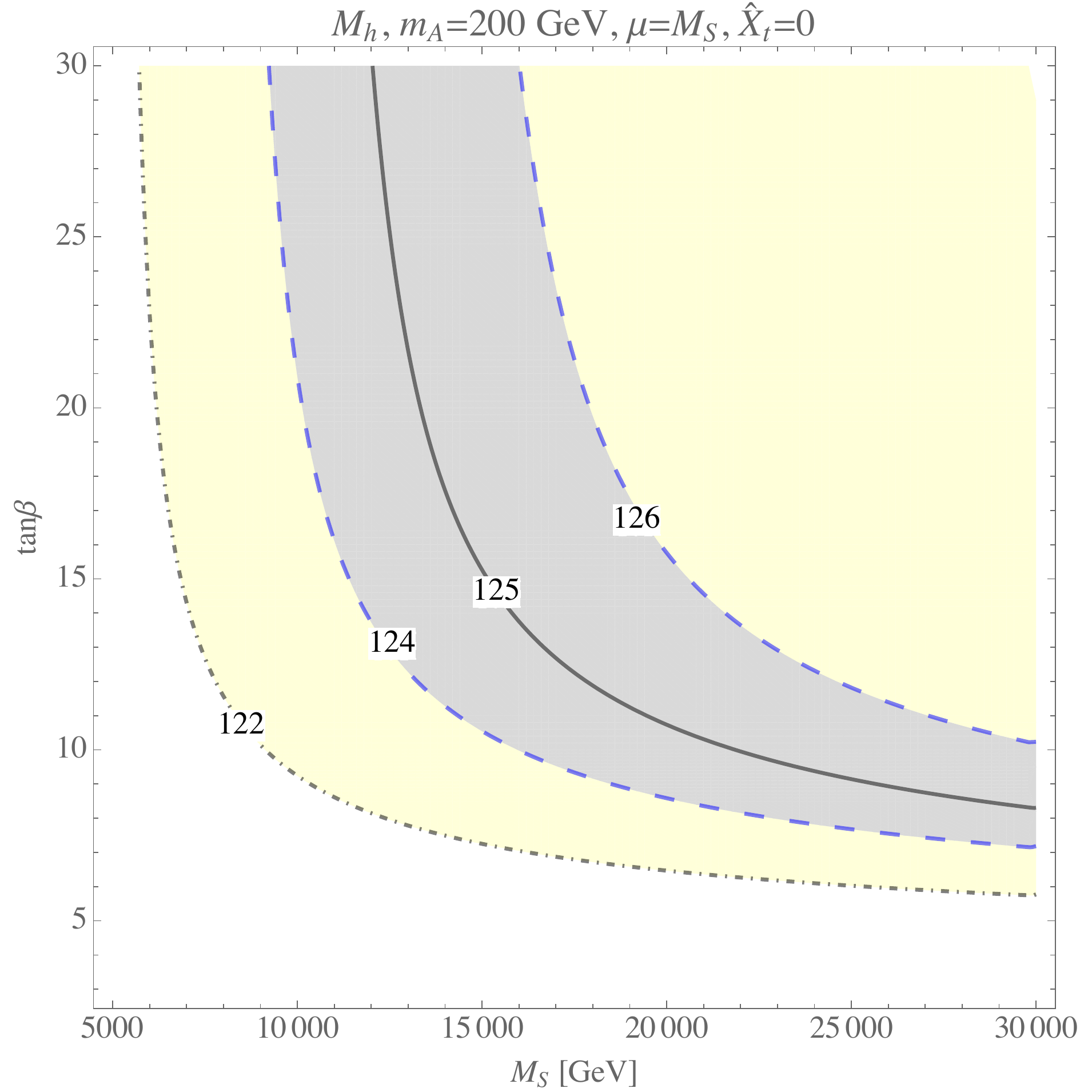} \quad
\includegraphics[width=0.45\linewidth]{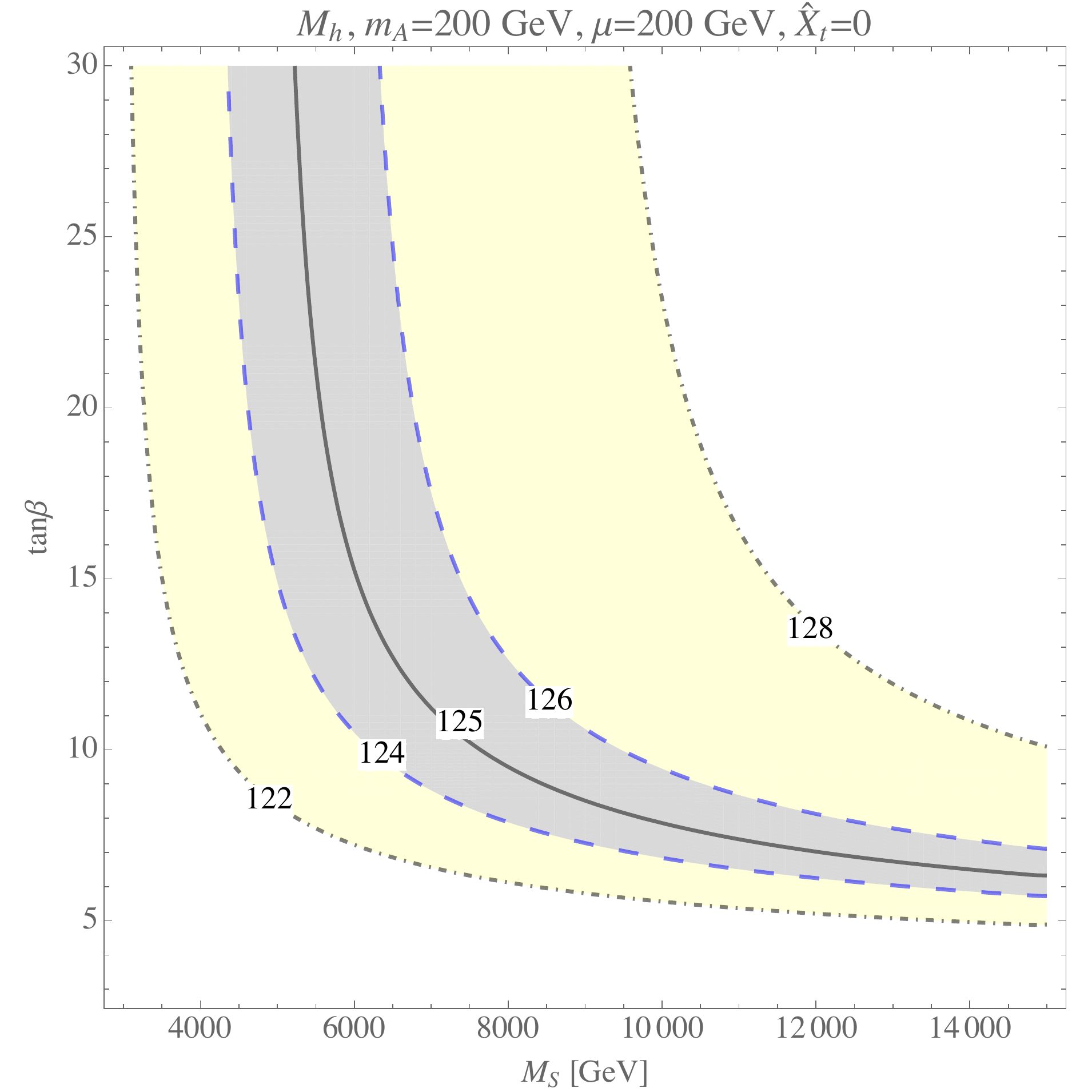}
\includegraphics[width=0.45\linewidth]{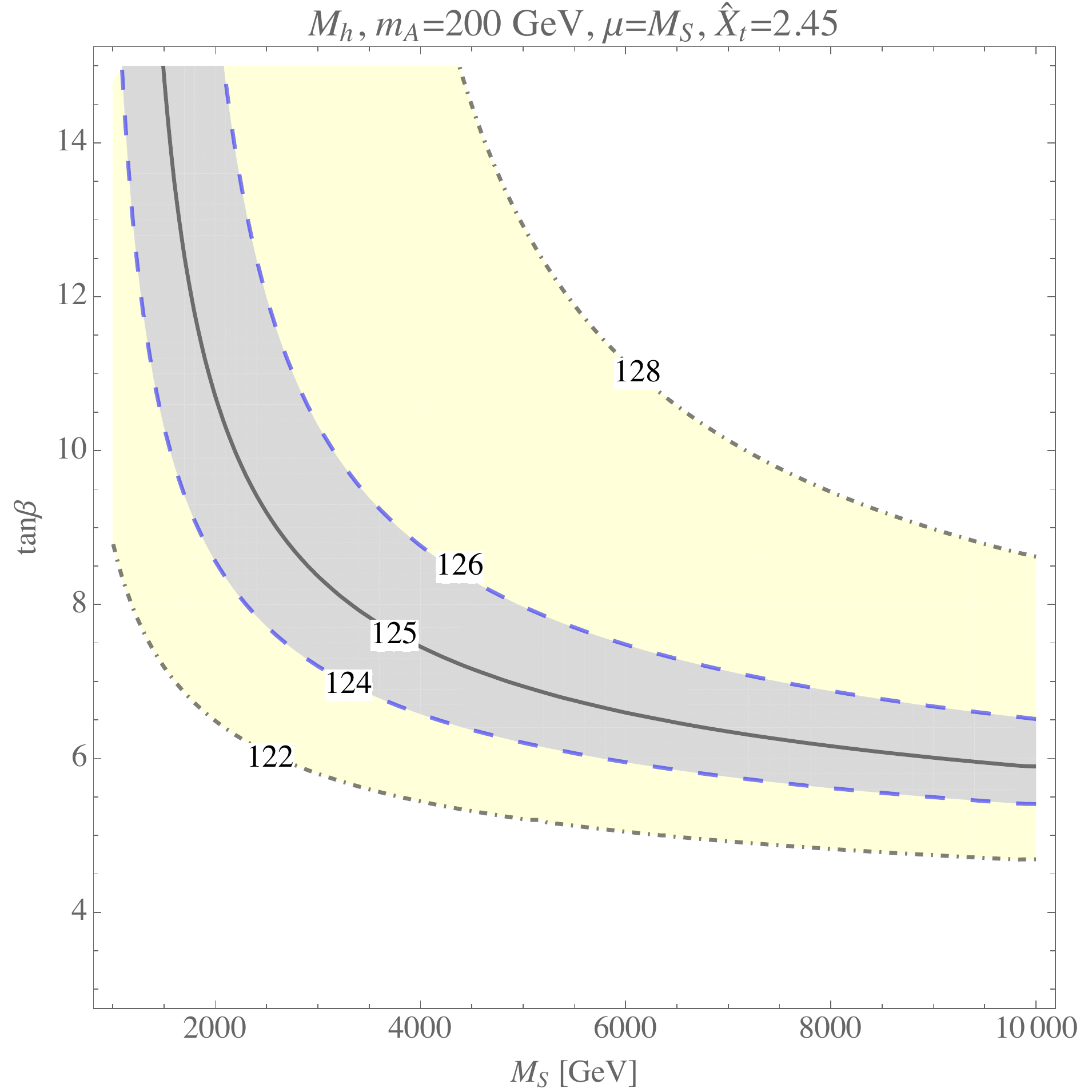} \quad
\includegraphics[width=0.45\linewidth]{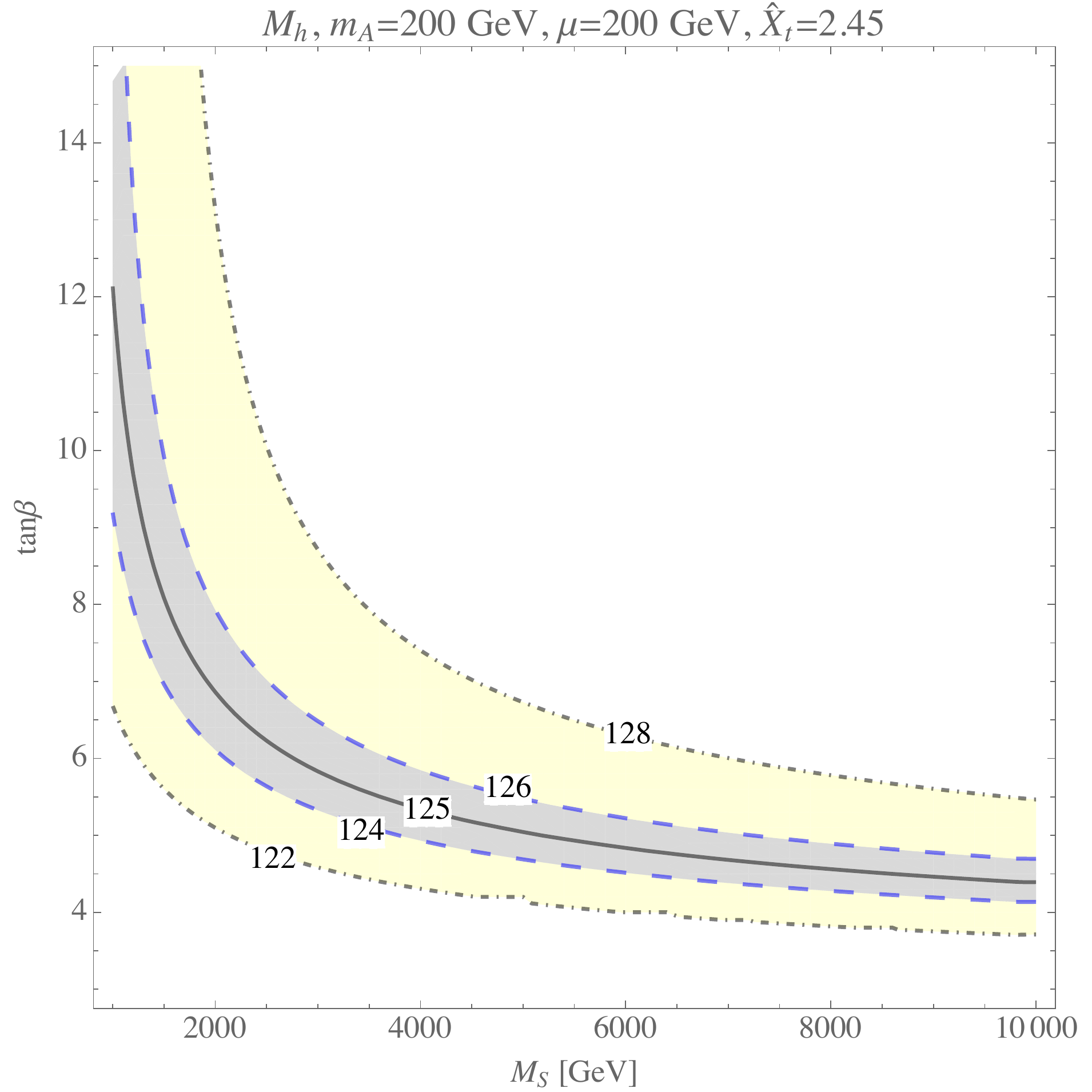}
\end{center}
\vspace{-1.5em}
\caption{As in Fig. \ref{fig:MhGUT_tanbvMS_mA200}, but with $M_S$ restricted to the 1--30 TeV range and modified ranges in $t_\b$. Shading has been added between the contours for visual clarity.}
\label{fig:MhTeV_tanbvMS_mA200}
\end{figure}

\begin{figure}[tb]
\begin{center}
\includegraphics[width=0.45\linewidth]{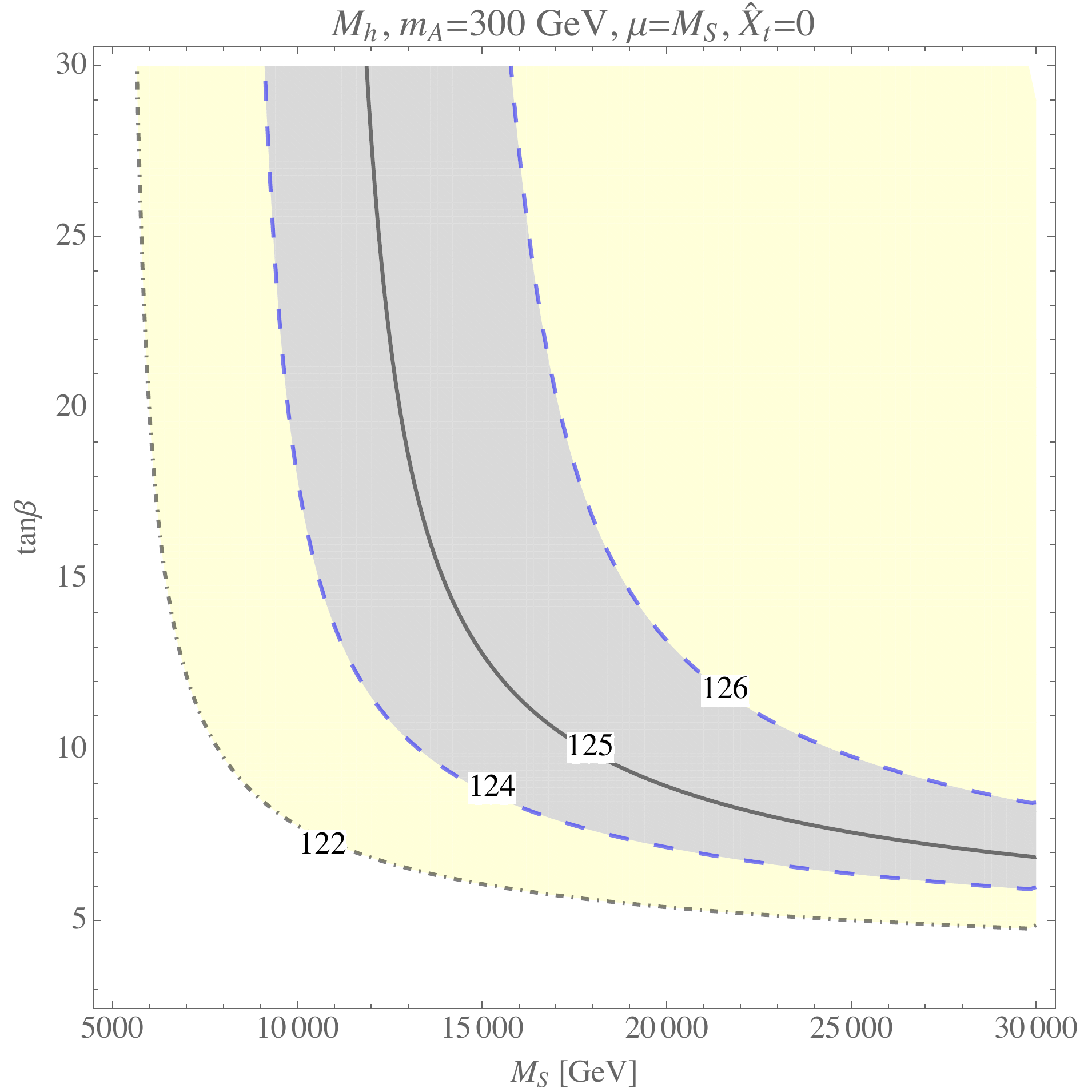} \quad
\includegraphics[width=0.45\linewidth]{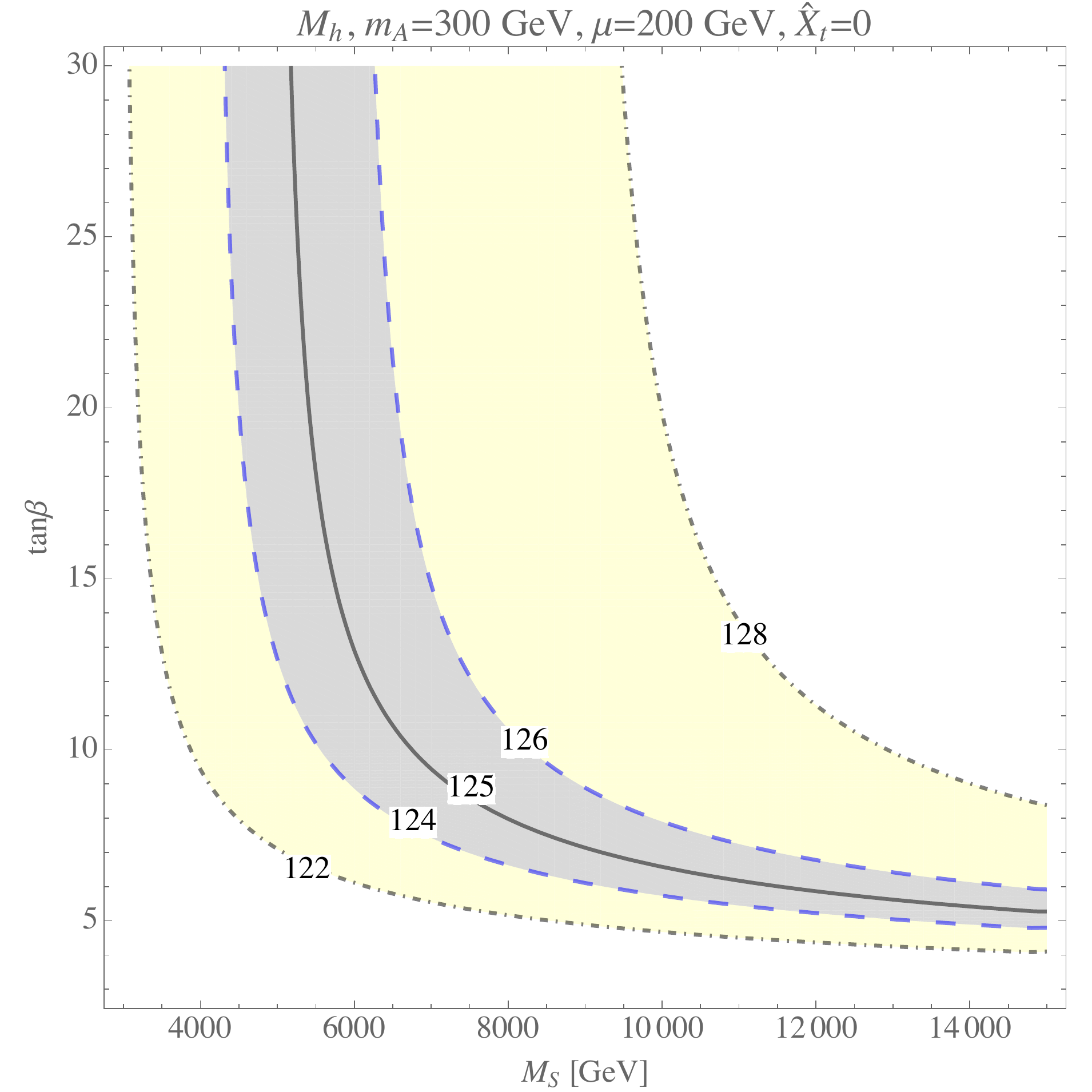}
\includegraphics[width=0.45\linewidth]{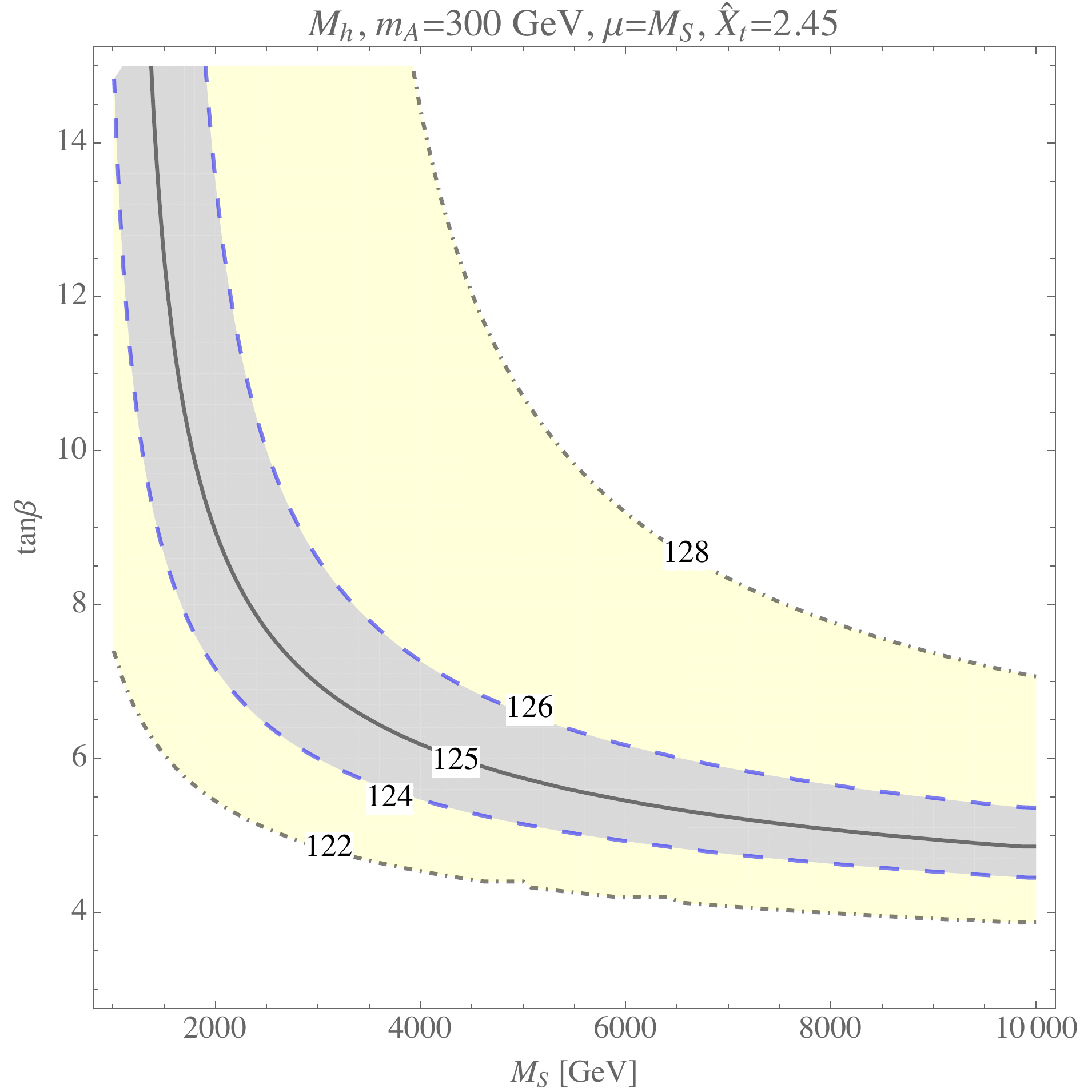} \quad
\includegraphics[width=0.45\linewidth]{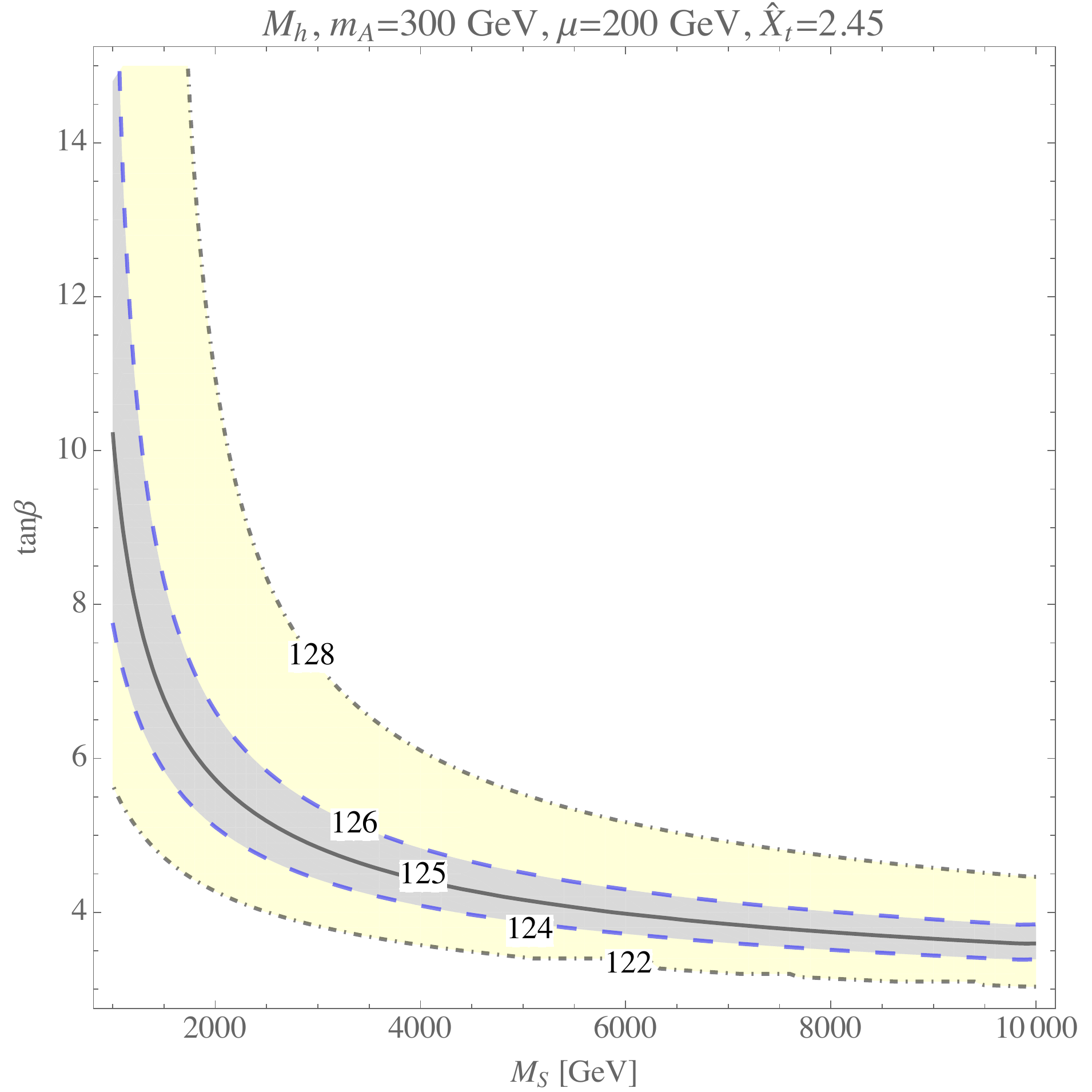}
\end{center}
\vspace{-1.5em}
\caption{As in Fig. \ref{fig:MhGUT_tanbvMS_mA300}, but with $M_S$ restricted to the 1--30 TeV range and modified ranges in $t_\b$. Shading has been added between the contours for visual clarity.}
\label{fig:MhTeV_tanbvMS_mA300}
\end{figure}

The results of our analysis are presented in Figs.~\ref{fig:MhGUT_tanbvMS_mA200},~\ref{fig:MhGUT_tanbvMS_mA300},~\ref{fig:MhTeV_tanbvMS_mA200} and~\ref{fig:MhTeV_tanbvMS_mA300}. 
In Fig.~\ref{fig:MhGUT_tanbvMS_mA200} we present contour plots of the lightest $CP$-even Higgs mass for $m_A = 200$~GeV in the $M_S$--$t_\b$ plane for small values of the stop mixing parameter, $\XtMS = 0$, and for values close to maximal mixing, $\XtMS = \sqrt{6}$. 
In addition, we compare values of $M_h$ obtained for $\mu= M_S$, for which the chargino and neutralino contributions to the Higgs mass decouple below the scale $M_S$, 
with the ones for low values of $\mu = 200$~GeV, for which the corrections to the Higgs mass induced by RG-evolution effects of charginos and neutralinos become relevant. 
We see that in order to obtain the proper value of the Higgs mass at low values of $t_\b \sim 2$, 
low values of $\mu$ of the order of the weak scale and large values of $M_S$ of the order of the GUT scale are necessary. 
We also note that for $t_\b \lesssim 1.5$, values of $M_h = 122$~GeV may not be obtained even if the supersymmetric spectrum is pushed to the GUT scale.  

The values of the Higgs mass at $m_A = 200$~GeV are heavily susceptible to Higgs mixing effects.
In contrast, we show in Fig.~\ref{fig:MhGUT_tanbvMS_mA300} contour plots of the
lightest $CP$-even Higgs mass for $m_A = 300$~GeV and similar supersymmetry
breaking parameters as in Fig.~~\ref{fig:MhGUT_tanbvMS_mA200}. 
The qualitative behavior is the same as in the previous case, but the proper Higgs mass is achieved
at lower values of $M_S$. In particular, for low values of $\mu$, values of $t_\b = 1$ no longer demand sparticles
above the GUT scale, a result that is independent of the stop mixing parameter.

It is relevant to show the previous results for values of $M_S$ of order of the TeV scale,
as expected if supersymmetry is related to the mechanism of electroweak symmetry
breaking. The results are presented in Figs.~\ref{fig:MhTeV_tanbvMS_mA200} and~\ref{fig:MhTeV_tanbvMS_mA300}, in which the values of $M_S$ are restricted to vary
between 1~TeV and 30~TeV.  For $m_A = 200$ GeV, it is clear that $M_h = 125$ GeV cannot be achieved with values of $t_\b \lesssim 4$; 
similarly, requiring values of $M_S$ of the order of 1~TeV demands large $t_\b$ and values of $X_t$ close to the maximal mixing values. 
As expected, the values of $t_\b$ necessary to achieve the proper Higgs mass increase for lower values of $m_A$ and large values of $\mu$, due to mixing and chargino and neutralino effects, respectively.

\begin{figure}[tb!]
\begin{center}
\includegraphics[width=0.48\linewidth]{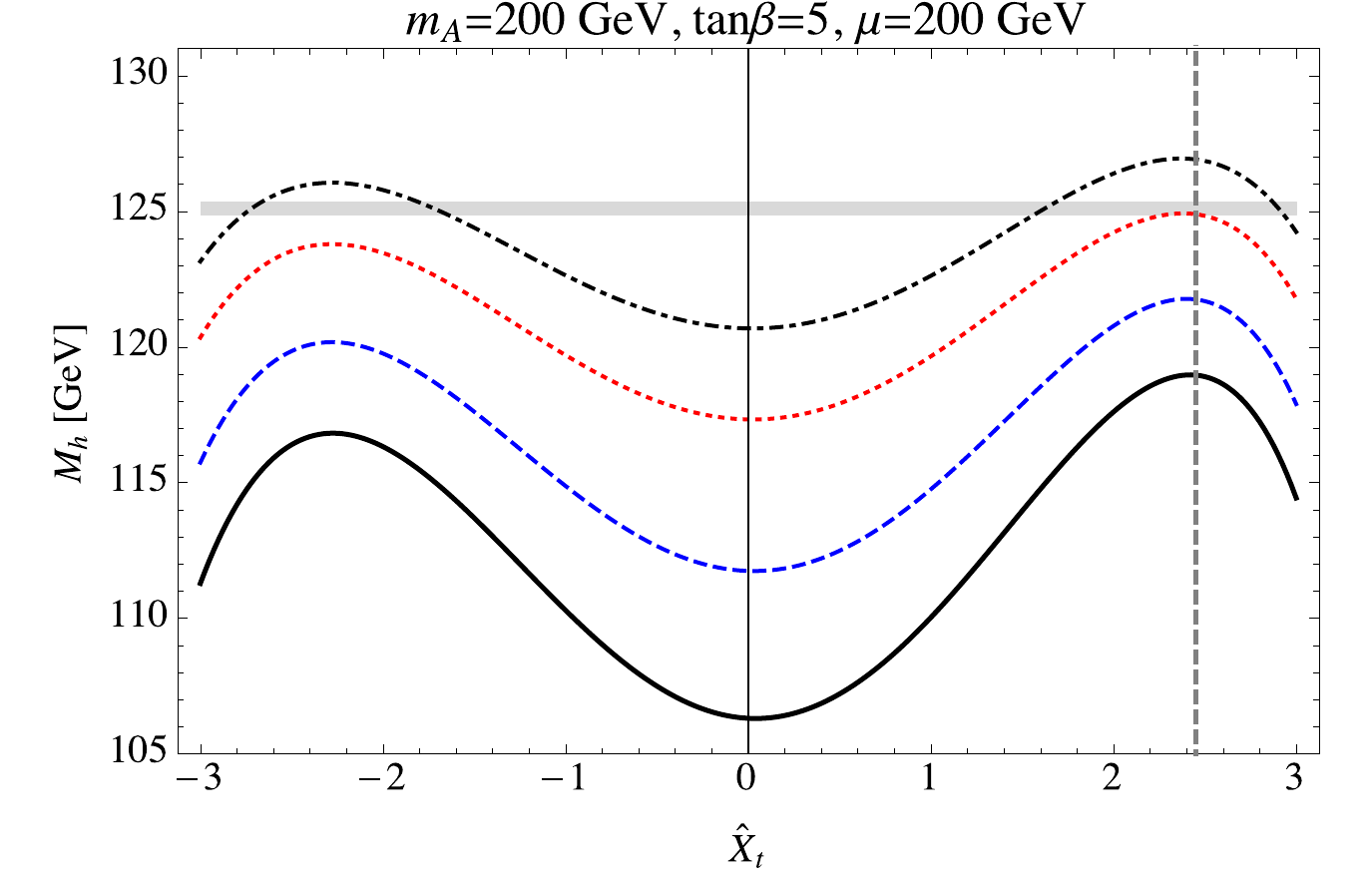} \quad
\includegraphics[width=0.48\linewidth]{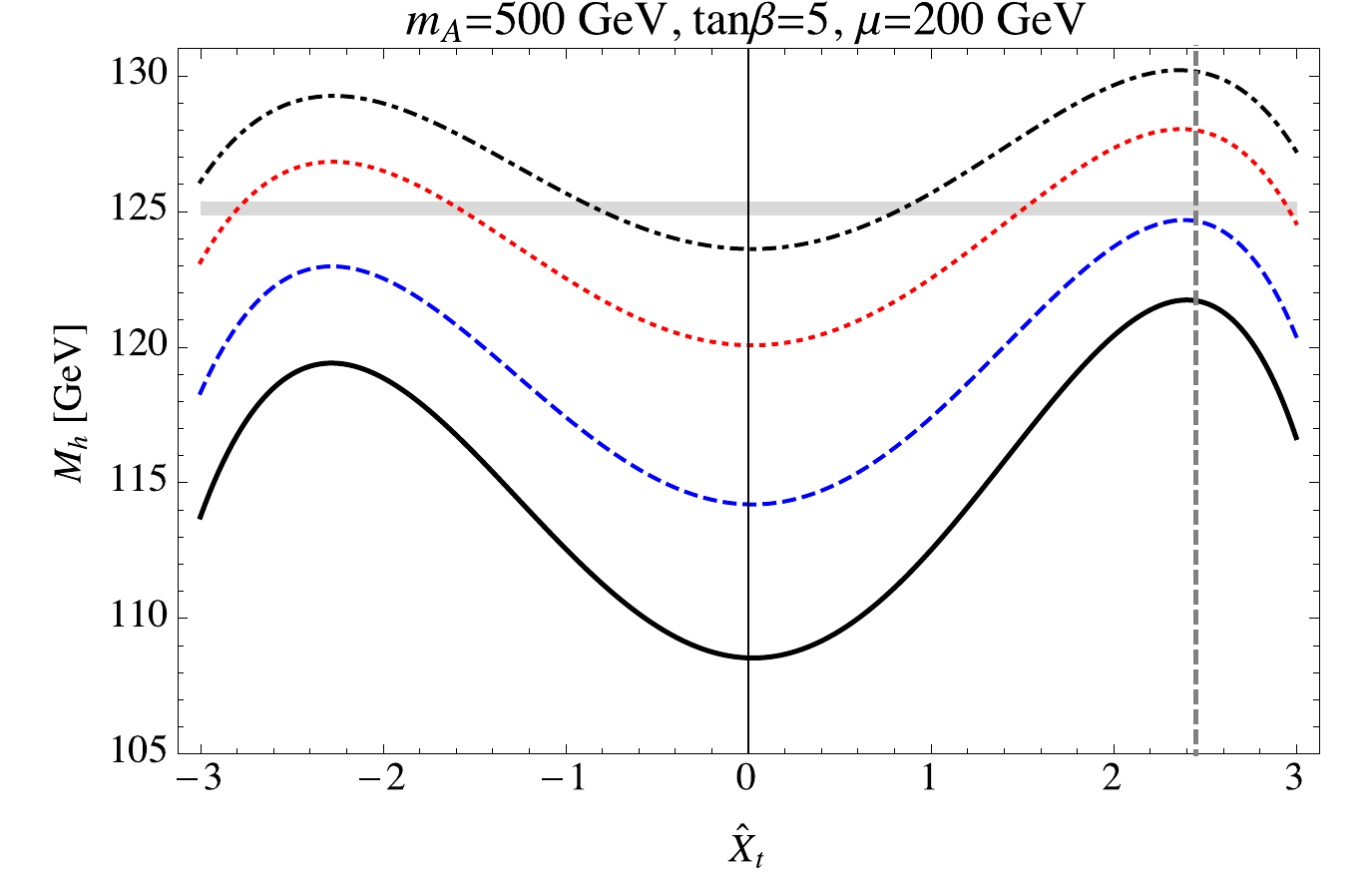}
\includegraphics[width=0.48\linewidth]{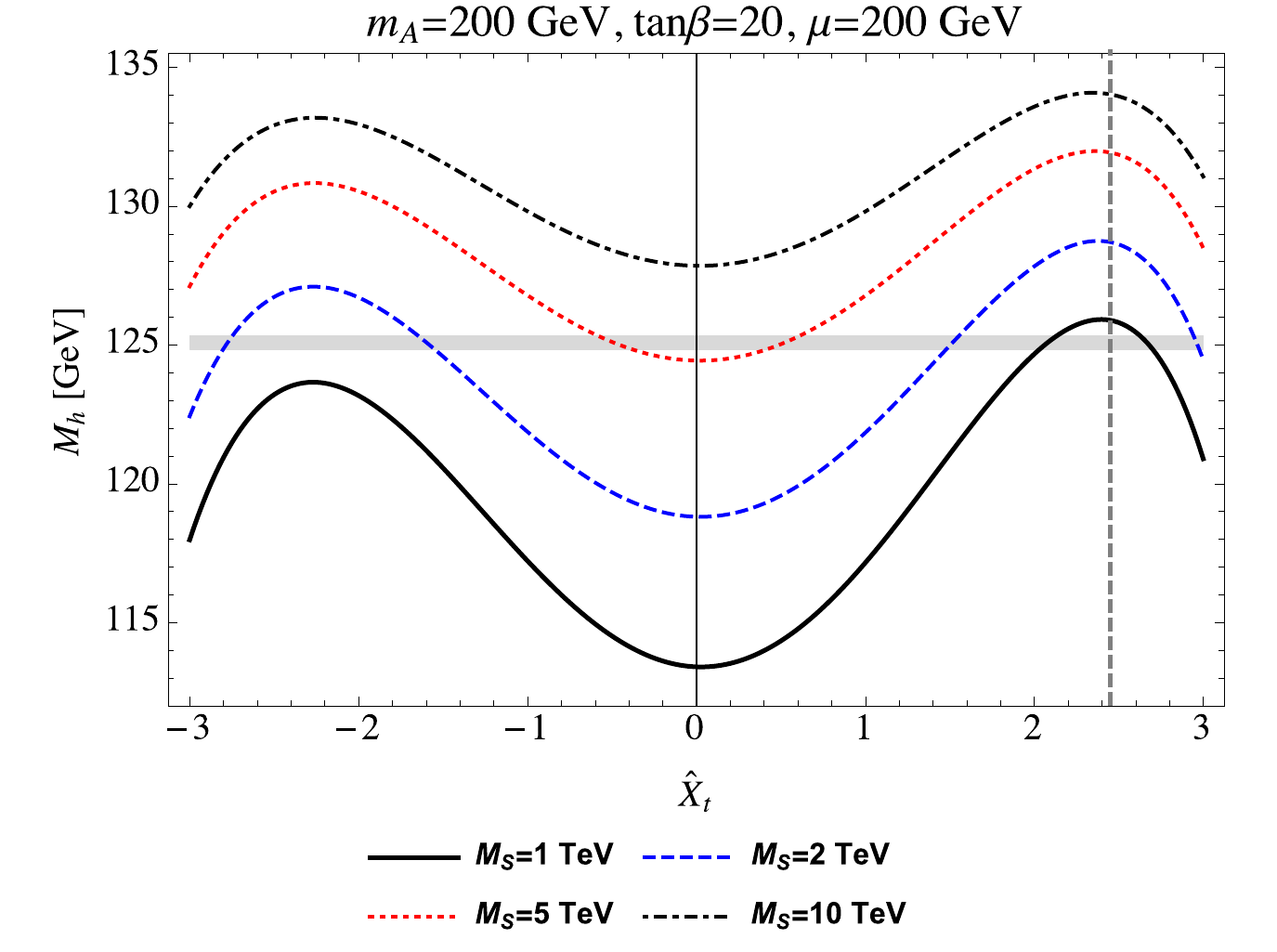} \quad
\includegraphics[width=0.48\linewidth]{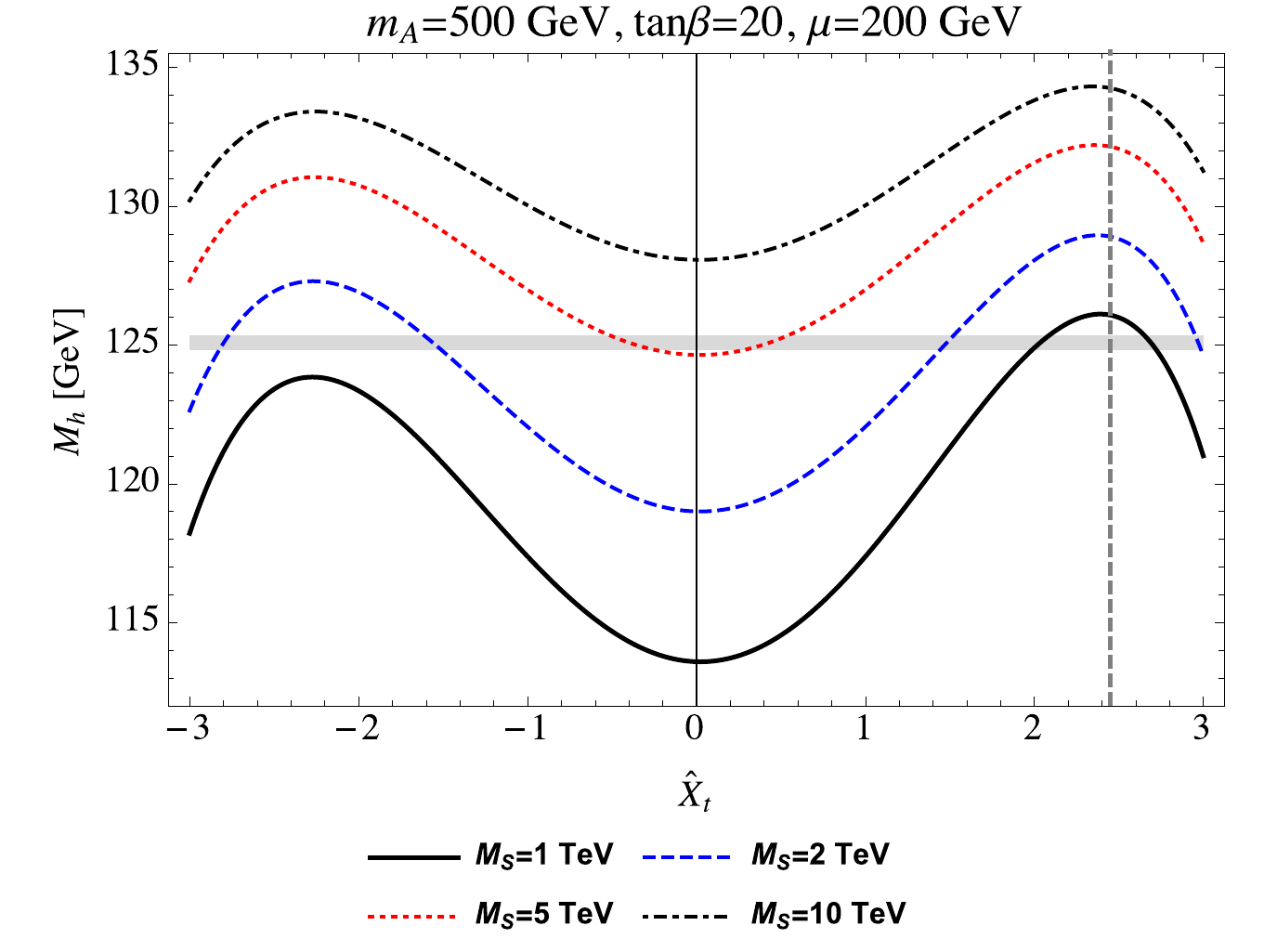}
\end{center}
\vspace{-1.5em}
\caption{$M_h$ vs $\XtMS$ for $m_A = (200, 500)$ GeV in the (left, right) columns, $t_\b = (2, 20)$ in the (top, bottom) rows, $A_b = A_\tau = M_S$, and $\mu = M_1 = M_2 = 200$ GeV. The four curves are for $M_S$ values of $1, 2, 5, 10$ TeV from bottom to top. The vertical grey dashed line indicates the value at the one-loop maximal mixing value $\XtMS = \sqrt6$. The horizontal light grey box is the $1\sigma$ band $M_h = 125.09 \pm 0.24$ GeV.}
\label{fig:MhvXt}
\end{figure}

In Fig. \ref{fig:MhvXt}, we plot $M_h$ as a function of $\XtMS$ to show the effect of mixing in the Higgs mass matrix at different values of $m_A$, $t_\b$, fixing $\mu = 200$ GeV. 
The different curves correspond to choice of $M_S$ between 1 and 10 TeV. At low $t_\b = 5$, the effect of mixing for $m_A = 200$ GeV is pronounced; 
the value of $M_h$ with $m_A = 200$ GeV is between 2--3 GeV lower than with $m_A = 500$ GeV, which approximates the decoupling limit. 
For higher values of $t_\b = 20$, the difference between the respective curves for the two values of $m_A$ falls to less than 0.5 GeV. 
The logarithmic dependence of $M_h$ on $M_S$ is evident in these plots: increasing $M_S$ from 1 to 5 TeV increases $M_h$ by approximately 12 GeV, 
while doubling $M_S$ from 5 to 10 TeV yields a more modest 3--4 GeV change. 
We also note that the maximum $M_h$ achieved for $M_S = 1$ TeV is $M_h \sim 126.1$ GeV in the lower right panel. 
Within uncertainties, this agrees with results previously found in the $m_A = M_S$ case in Ref.~\cite{Draper:2013oza}.

\begin{figure}[tb!]
\begin{center}
\includegraphics[width=0.48\linewidth]{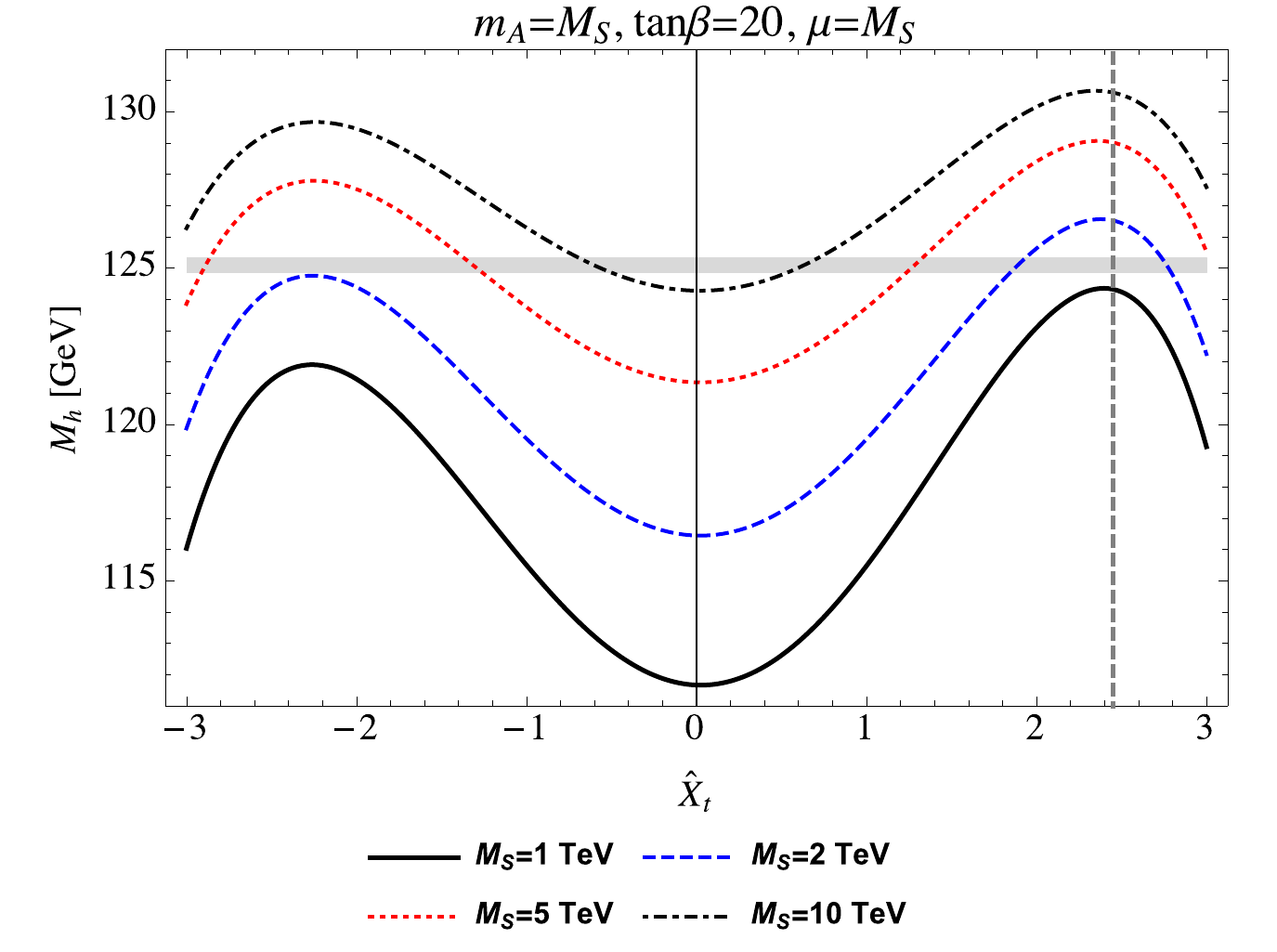}
\end{center}
\vspace{-1.5em}
\caption{As in Fig.~\ref{fig:MhvXt}, with $m_A = M_S, t_\b = 20, A_b = A_\tau = M_S$, and $\mu = M_1 = M_2 = M_S$.}
\label{fig:MhvXtmAMS}
\end{figure}

In Fig.~\ref{fig:MhvXtmAMS}, we have plotted $M_h$ in the high-scale SUSY scenario, with large $t_\b = 20$. 
For $M_S = 2$ TeV, the dashed blue curve, we obtain $M_h = 126.5$ GeV.
Also, in contrast to the bottom-right panel of Fig.~\ref{fig:MhvXt},
$M_h = 125$ GeV is no longer achieved for $M_S = 1$ TeV at maximal mixing without light electroweakinos. 
We can compare with the recent results produced by the \textsc{SusyHD} code of Ref.~\cite{Vega:2015fna}. 
Our values are $\lesssim 1$ GeV higher than the central result of Ref.~\cite{Vega:2015fna}. 
Part of this discrepancy is attributed to the use of the lower value of $y_t(M_t)$: 
if we instead use the NNLO + N${}^3$LO QCD value $y_{t, \text{N${}^3$LO QCD}}(M_t) = 0.93690$, $M_h$ is lowered by 0.5 GeV. 
The remaining small difference may be explained by the more complete calculation of thresholds in the $m_A \sim M_S$ case of Refs.~\cite{Draper:2013oza, Vega:2015fna}.

\section{Comparison to Previous Results} \label{sect:comparison}

In this section, we compare our results with the results obtained in the hMSSM scenario as well in the \textsc{FeynHiggs} version 2.10.2, 
in which relevant logarithmic effects to the SM quartic couplings are resummed in order to increase the accuracy of the results at large values of $M_S$ \cite{Degrassi:2002fi, Hahn:2013ria}.

\begin{figure}[htb!]
\begin{center}
\includegraphics[width=0.44\linewidth]{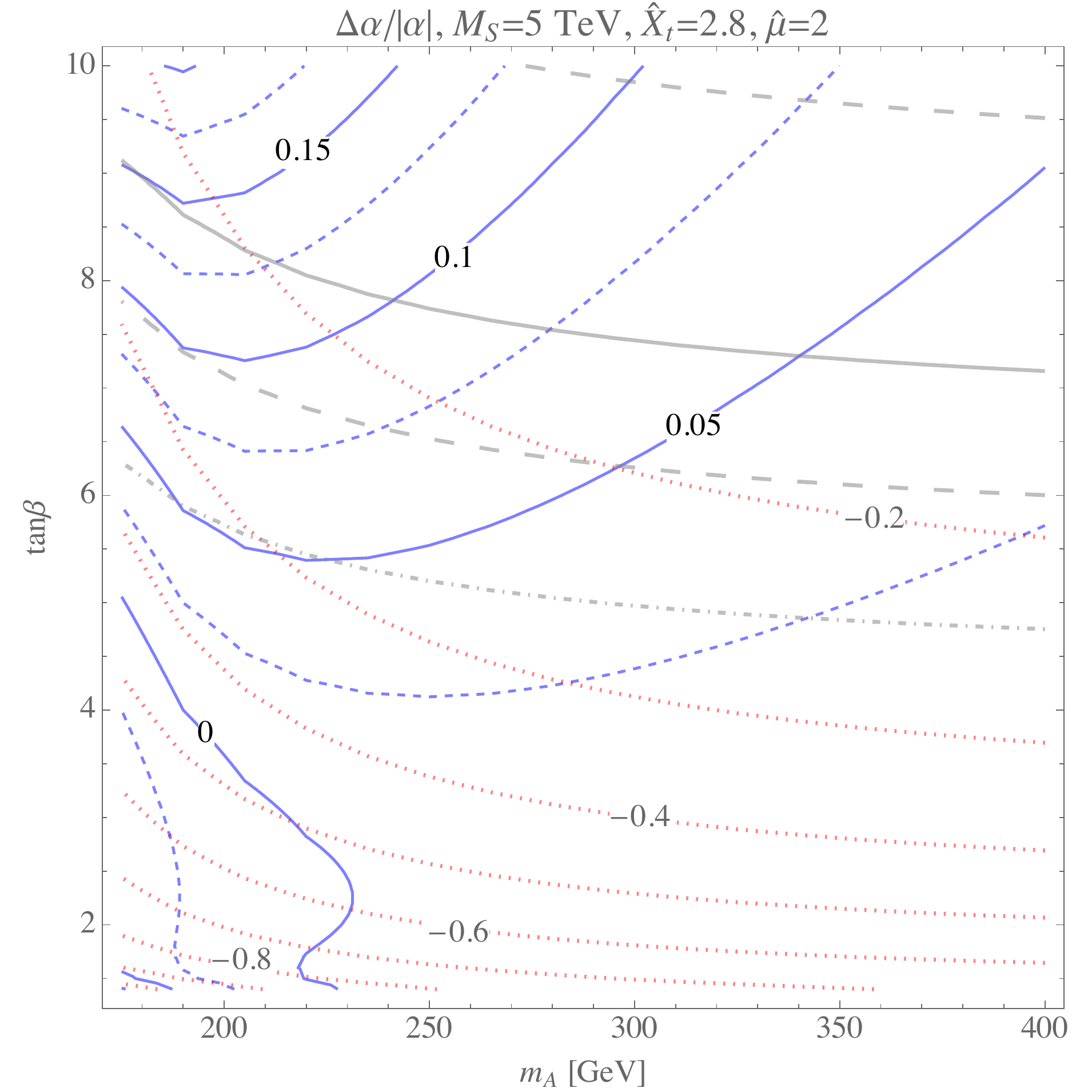} \quad
\includegraphics[width=0.44\linewidth]{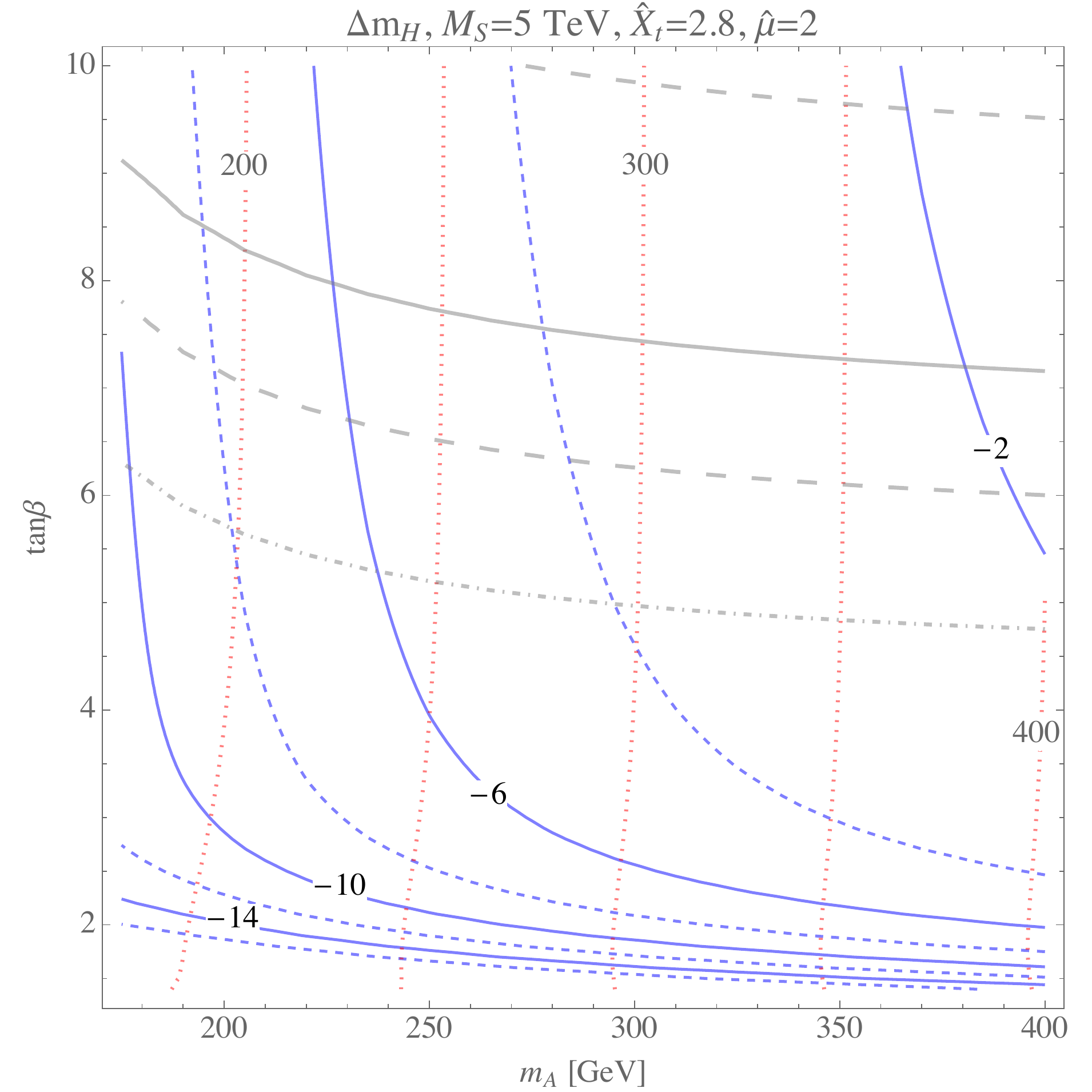}
\includegraphics[width=0.44\linewidth]{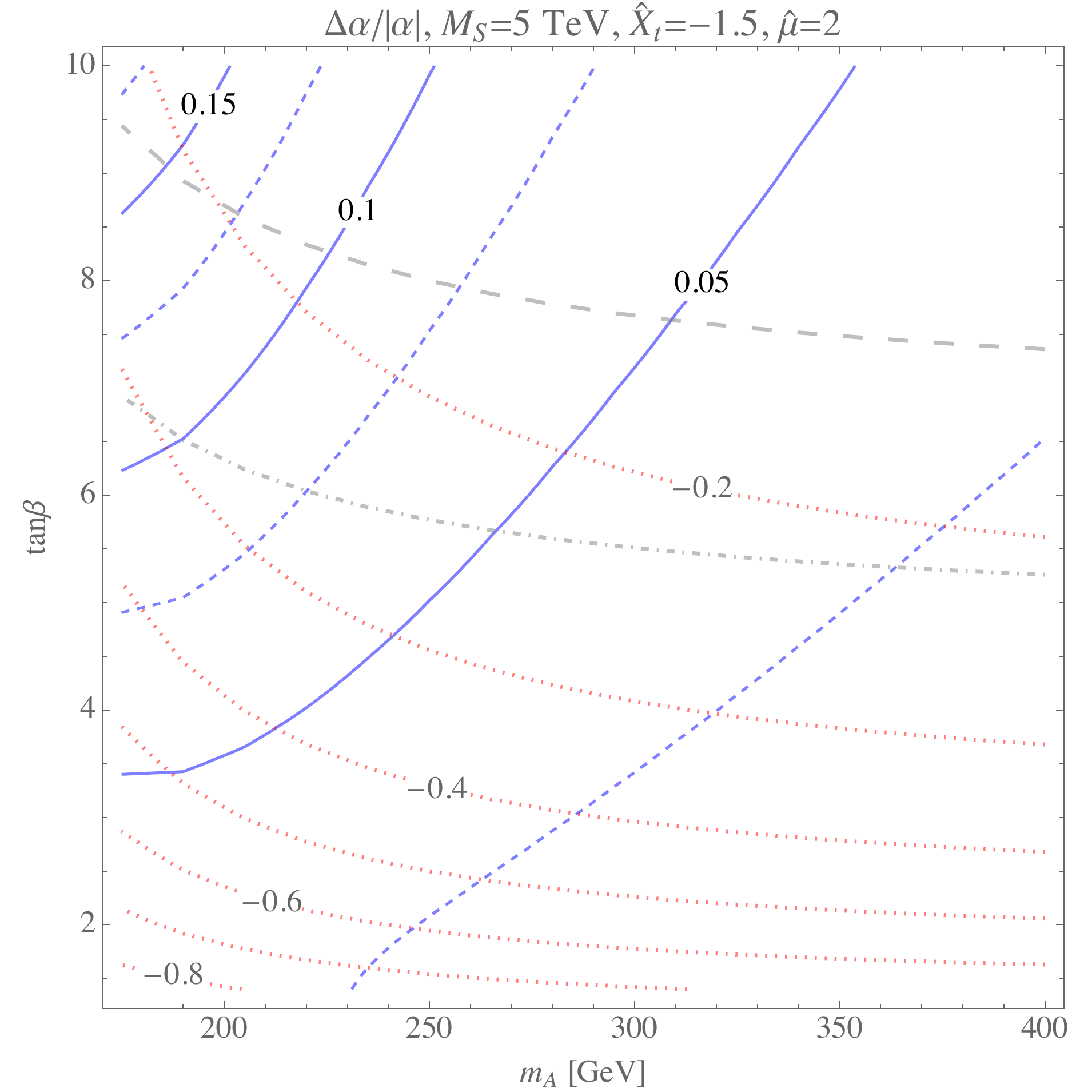} \quad
\includegraphics[width=0.44\linewidth]{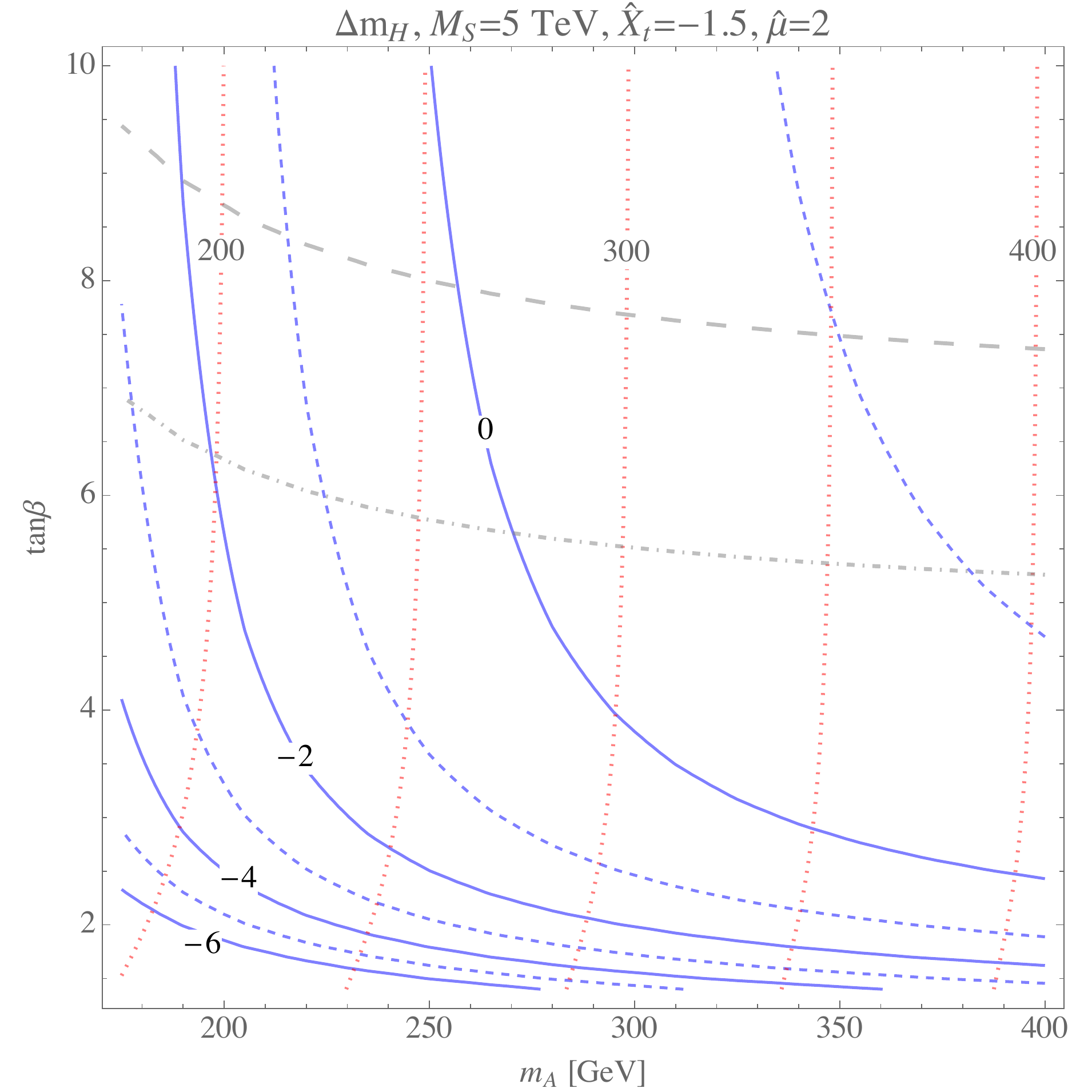}
\end{center}
\vspace{-1.5em}
\caption{Difference between hMSSM and effective THDM calculations of $\a$ and $M_H$ in the plane $t_\b, m_A$ for $M_S = 5$ TeV, $\XtMS = [2.8, -1.5]$ [top, bottom], $\mu = M_1 = M_2 = 2M_S$, and $A_b = A_\t = A_t$. From the bottom to the top of each plot, the light grey lines (dot-dashed, dashed, and solid) correspond to $M_h = (122, 124, 125, 126, 128)$ GeV. Red, dashed lines in the plots in the left (right) column are contours of $\alpha$ ($M_H$) computed using the effective THDM. Solid and dashed blue lines in the plots in the left (right) column are contours of $(\a - \a_\text{hMSSM})/|\a|$ ($M_H - M_{H, \text{hMSSM}}$, in GeV); dashed lines indicate values halfway between adjacent solid lines.}
\label{fig:THDMvhMSSM_MS5_muMS2}
\end{figure}

In Fig.~\ref{fig:THDMvhMSSM_MS5_muMS2}, we present the comparison of our results with the hMSSM approximation 
for sizable values of $\muMS = 2$ and values of $\XtMS = -1.5$ and $\XtMS = 2.8$, away from maximal mixing, 
for which the hMSSM results are expected to show a worse approximation to the correct results 
than for low values of $\mu$ at moderate or large values of $t_\b$. 
The results of our computation for the mixing angle $\alpha$ and the heavy $CP$-even Higgs mass are presented 
in the left and right panels with red dotted lines, while the blue lines represent the relative and absolute differences 
of these quantities with the ones computed in the hMSSM approximation. 
We present our results for $M_S = 5$~TeV, for which the correct values of the Higgs mass, represented by black solid, dashed and dotted lines, 
may only be obtained for moderate to large values of $t_\b$ in this region of parameters. 
Differences in $\alpha$ of the order of 10\%--20\% are obtained for moderate values of $t_\b$ and values of the heavy $CP$-even Higgs bosons of the order of the weak scale. 
Since the mixing angle controls the coupling of the lightest $CP$-even Higgs boson to fermions and gauge bosons, 
relevant modifications of the Higgs phenomenology are expected in this region of parameters. 
Similarly, the heavy $CP$-even Higgs boson mass may be affected by values of a few to 10 GeV in this region of parameters. 

\begin{figure}[htb!]
\begin{center}
\includegraphics[width=0.45\linewidth]{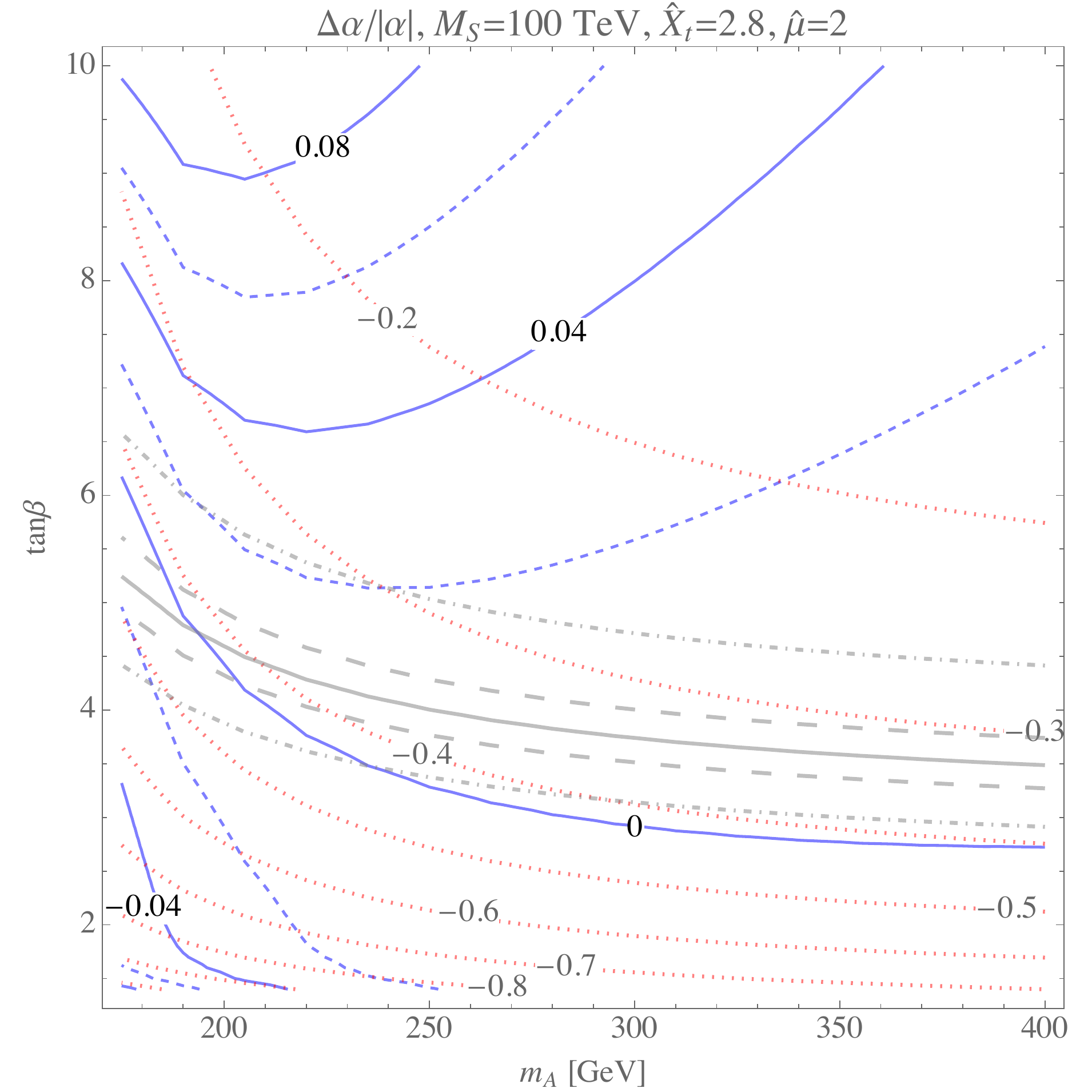} \quad
\includegraphics[width=0.45\linewidth]{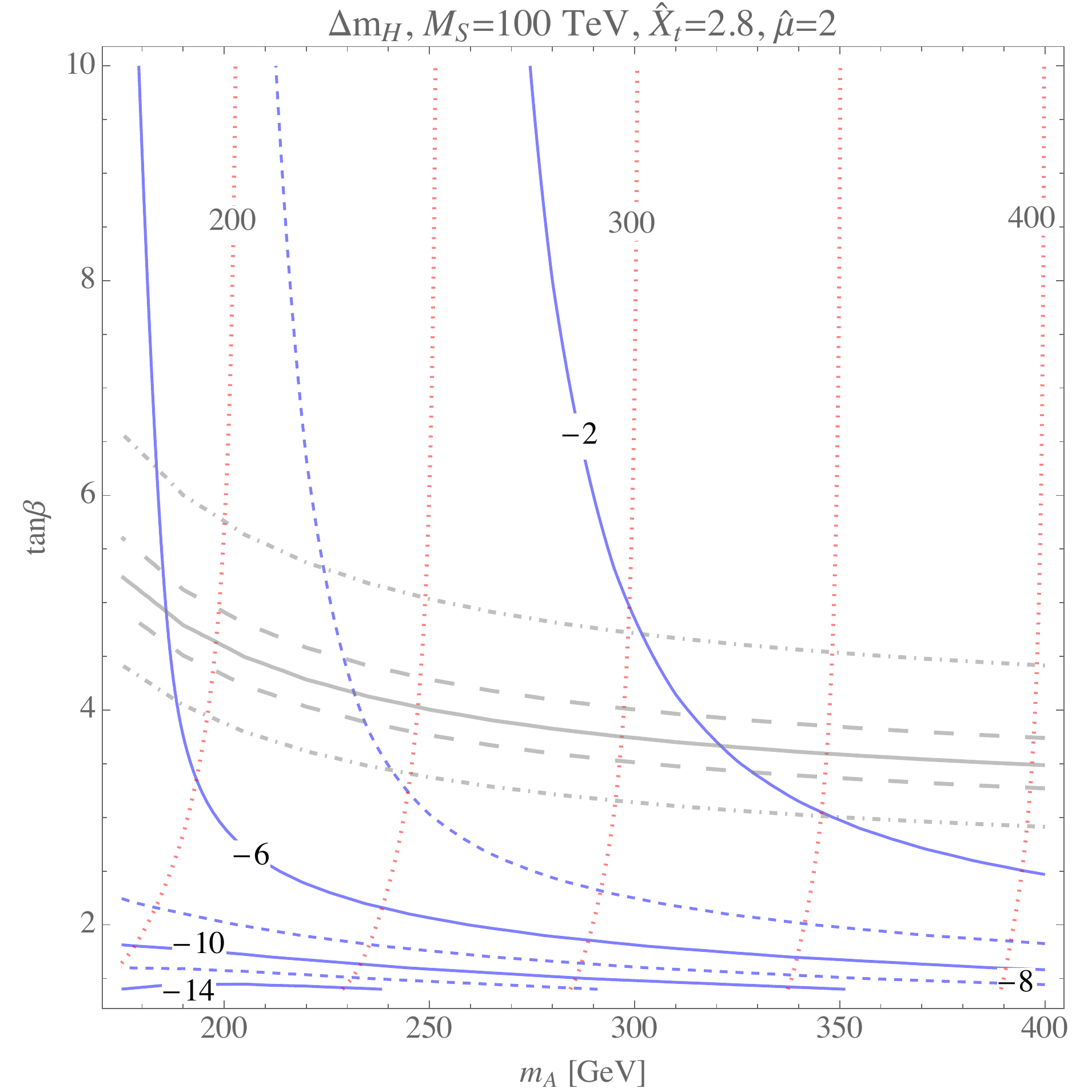}
\includegraphics[width=0.45\linewidth]{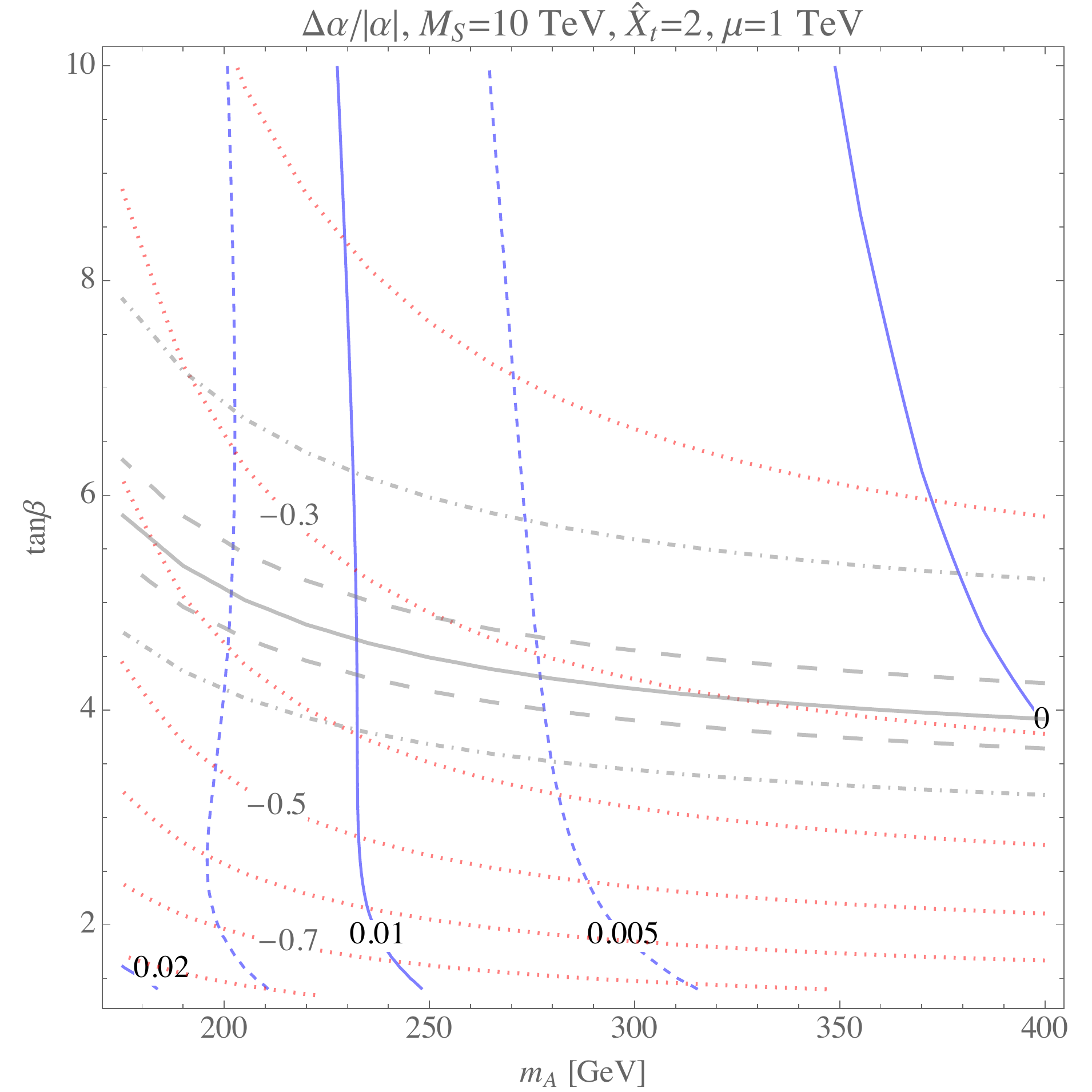} \quad
\includegraphics[width=0.45\linewidth]{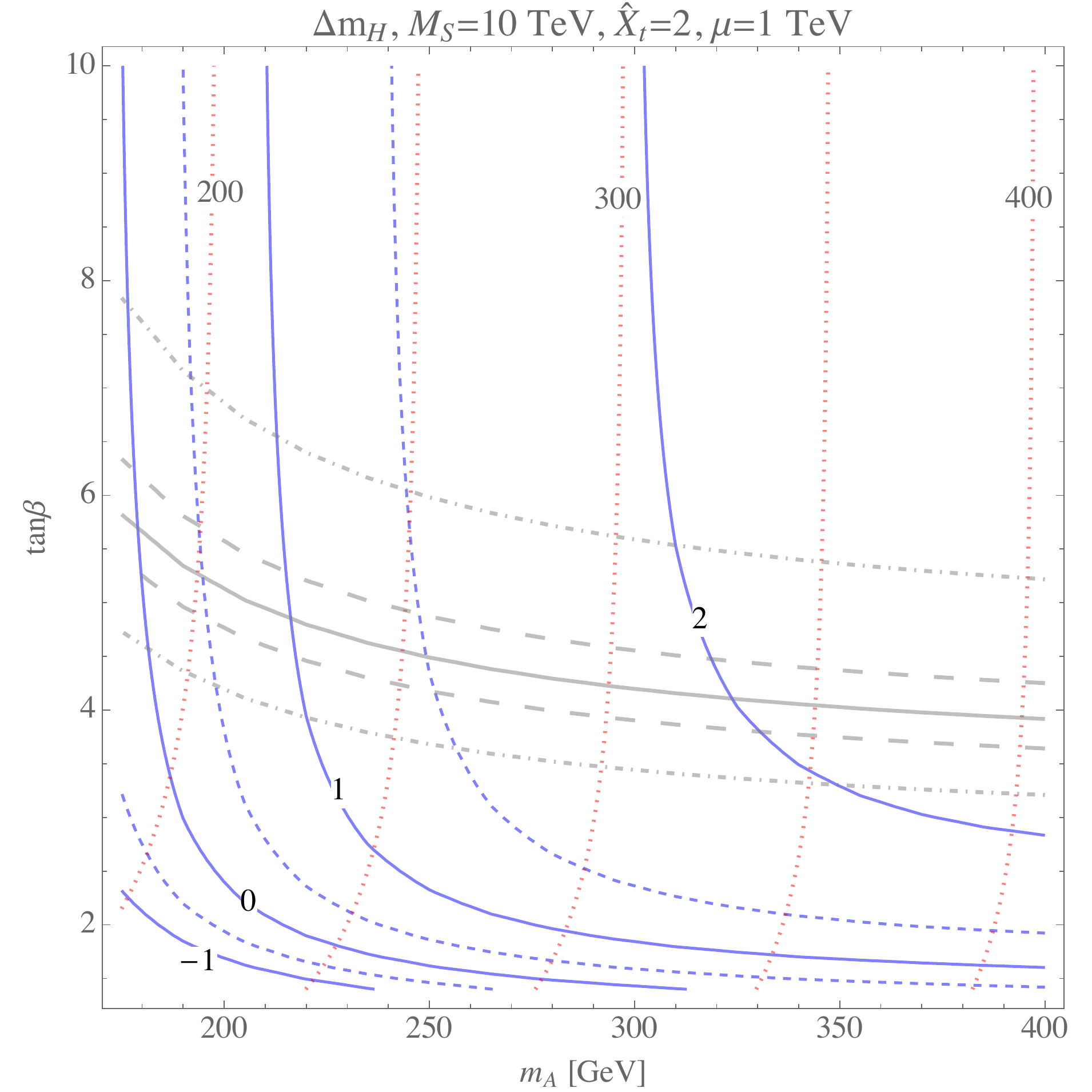}
\end{center}
\vspace{-1.5em}
\caption{As in Fig. \ref{fig:THDMvhMSSM_MS5_muMS2}, with $M_S = 100$ TeV, $\XtMS = 2.8, \muMS = 2$ [top] and $M_S = 10$ TeV, $\XtMS = 2, \mu = 1$ TeV [bottom].}
\label{fig:THDMvhMSSM_MS10-100}
\end{figure}

In Fig.~\ref{fig:THDMvhMSSM_MS10-100}, we present in the upper panels similar results but for $\XtMS = 2.8$ and large values of $M_S= 100$~TeV for which lower values of $t_\b \simeq 4$ are required to obtain the correct Higgs masses. 
We see that in this case, in the relevant region of parameters, the agreement is improved compared to the large $t_\b$ case, with differences in $\alpha$ of the order of a few percent and differences in $m_H$ of the order of a few GeV. 
In the lower panels, we present results for lower values of $\muMS$ and $M_S = 10$~TeV, for which values of $t_\b\simeq 5$ lead to the proper Higgs boson masses. We see that due to the smaller values of $\muMS$ and $t_\b$, the differences with the hMSSM reduce to values of at most 1\%--2\% in this case. 

A ``low-tan$\b$-high" scenario, in the region of $1 \lesssim t_\b \lesssim 10$, 150~GeV~$\lesssim m_A \lesssim$~500~GeV, has been presented by the LHC Cross Section Working Group \cite{Bagnaschi:2039911} with values for a subset of the MSSM parameters necessary to achieve $M_h = $ 122--128 GeV in \textsc{FeynHiggs}. 
In this scenario, a simplified, heavy MSSM spectrum above the scale $M_S$ is assumed; 
other MSSM parameters are chosen as $A_f = 2$ TeV ($f = b, \t, c, s, \mu, u, d, e$); $M_3 = M_S$;
$M_2 = 2$ TeV; $\mu = 1.5$ TeV; and $M_1 = M_2 \cdot \frac53 \tan^2\q_W \sim 950$ GeV fixed by the GUT relation.
$M_S^{\text{OS}}$ and $X_t^{\text{OS}}$, the values of the stop masses and mixing parameters in the on-shell scheme used in \textsc{FeynHiggs}, are then chosen to achieve $M_h$ in the desired range. 

We have used one-loop conversion formulas \cite{Espinosa:2000df, Draper:2013oza} to change $M_S^{\text{OS}}, X_t^{\text{OS}}$ in the OS scheme to $\oln{M}_S, \oln{X}_t$ in the $\MSbar$ scheme, which are the parameters used in our calculation.
The maximum $M_S^{\text{OS}}$ value specified is 100 TeV, which is used for points with low $m_A \lesssim $ 200--250 GeV and $t_\b \lesssim $ 1--3.
Maximal mixing is chosen for points in the region $t_\b \leq 2$. In \textsc{FeynHiggs}, this corresponds to the choice $X_t^{\text{OS}} = 2M_S^{\text{OS}}$;
in the $\MSbar$ scheme, the output value of $\oln{X}_t$ should be close to the maximal mixing value $\oln{X}_{t, \text{max}}$, for which $M_h$ as a function of $\oln{X_t}$ achieves its maximum (e.g., in Fig.~\ref{fig:MhvXt}, $\oln{X}_{t, \text{max}}$ lies close to the one-loop value $\oln{X}_{t, \text{max}}^{h_t^4} = \sqrt{6} \, \oln{M}_S$). 
For a selection of these points, we performed a similar scan and found that the output values of $\oln{X}_t$ yield $M_h(\oln{X}_t)$ values within 0.5 GeV of the maximal mixing values.

\begin{figure}[tb!]
\begin{center}
\includegraphics[width=0.415\linewidth]{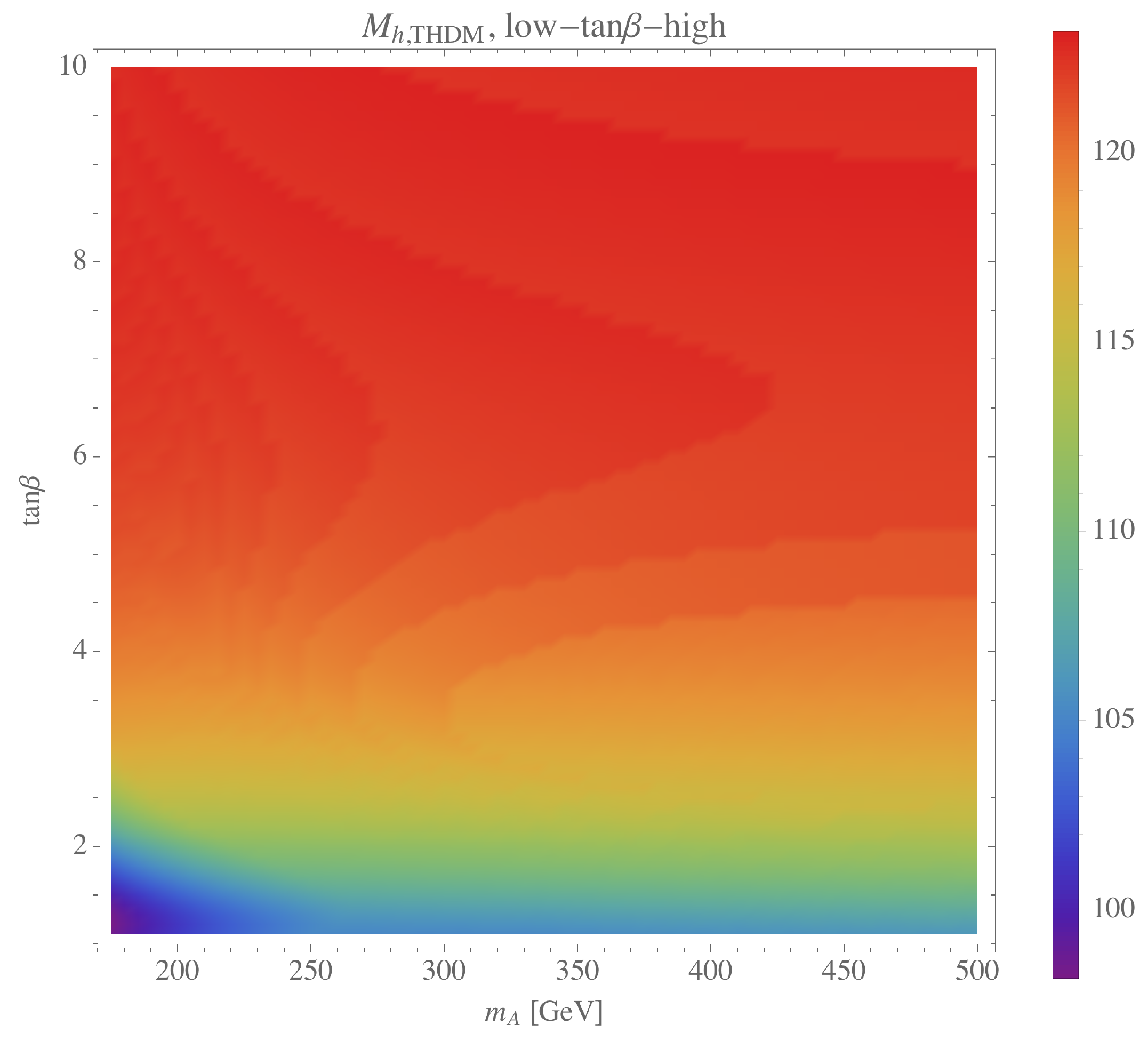} \quad
\includegraphics[width=0.425\linewidth]{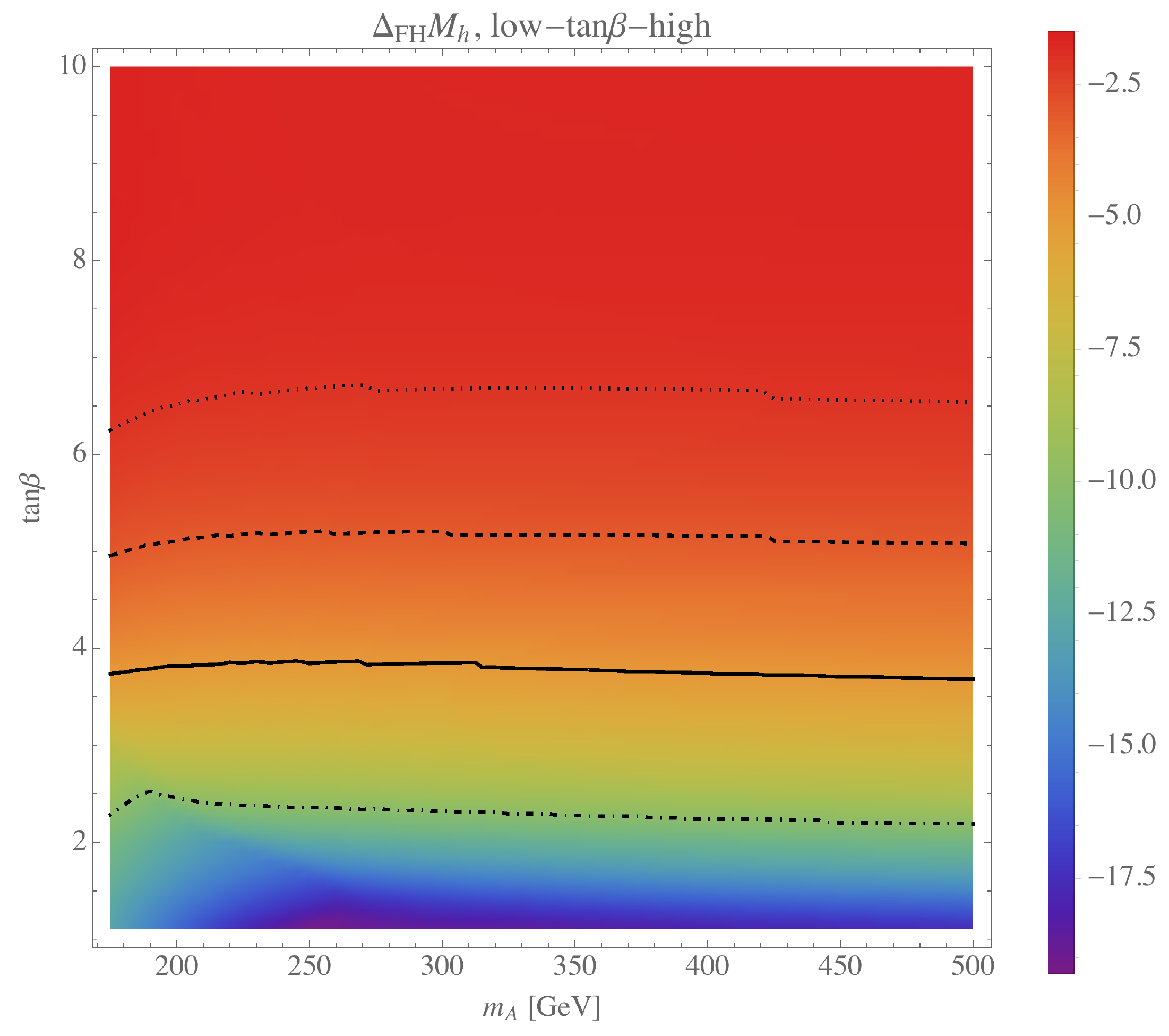}
\includegraphics[width=0.417\linewidth]{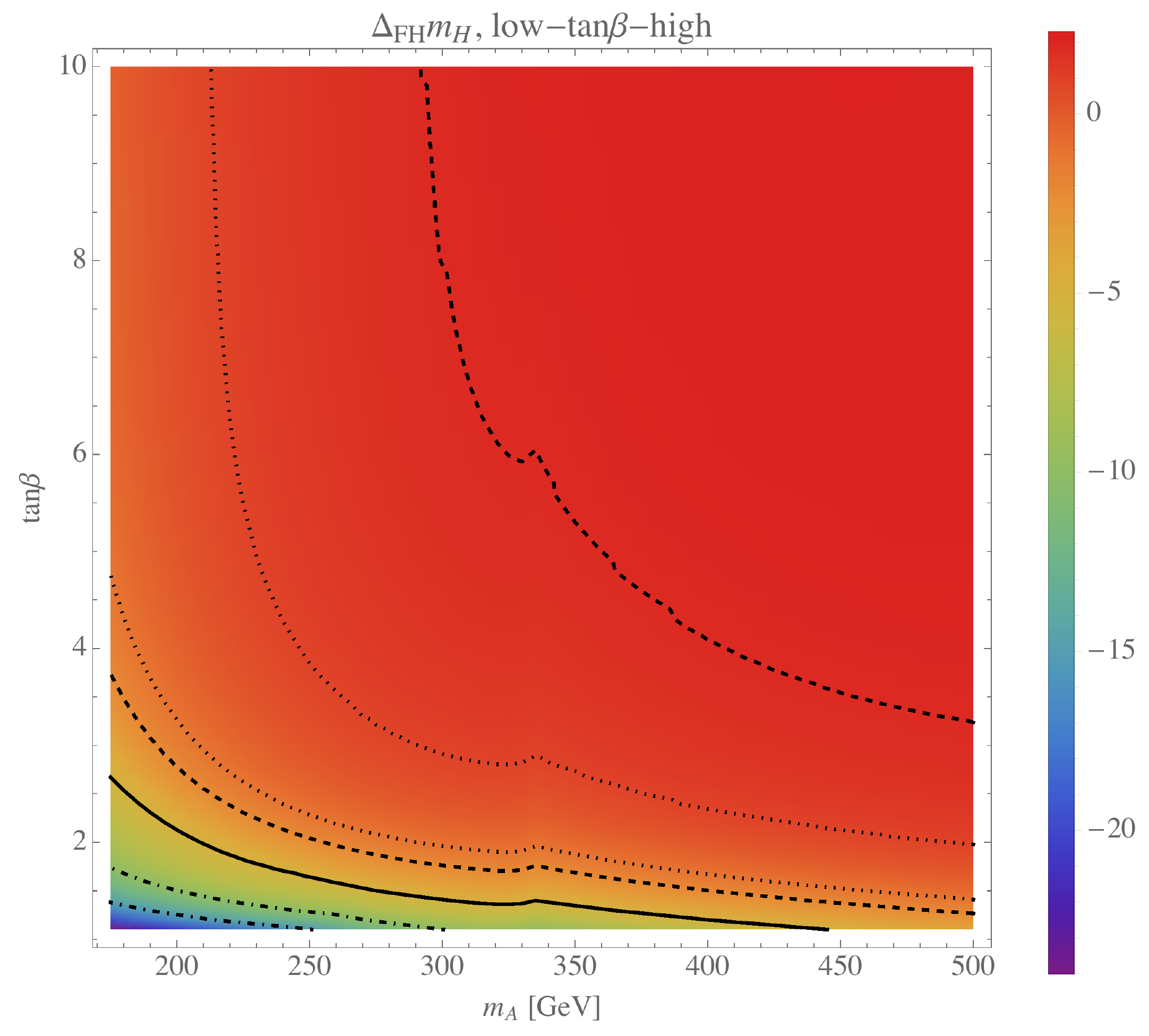} \quad
\includegraphics[width=0.424\linewidth]{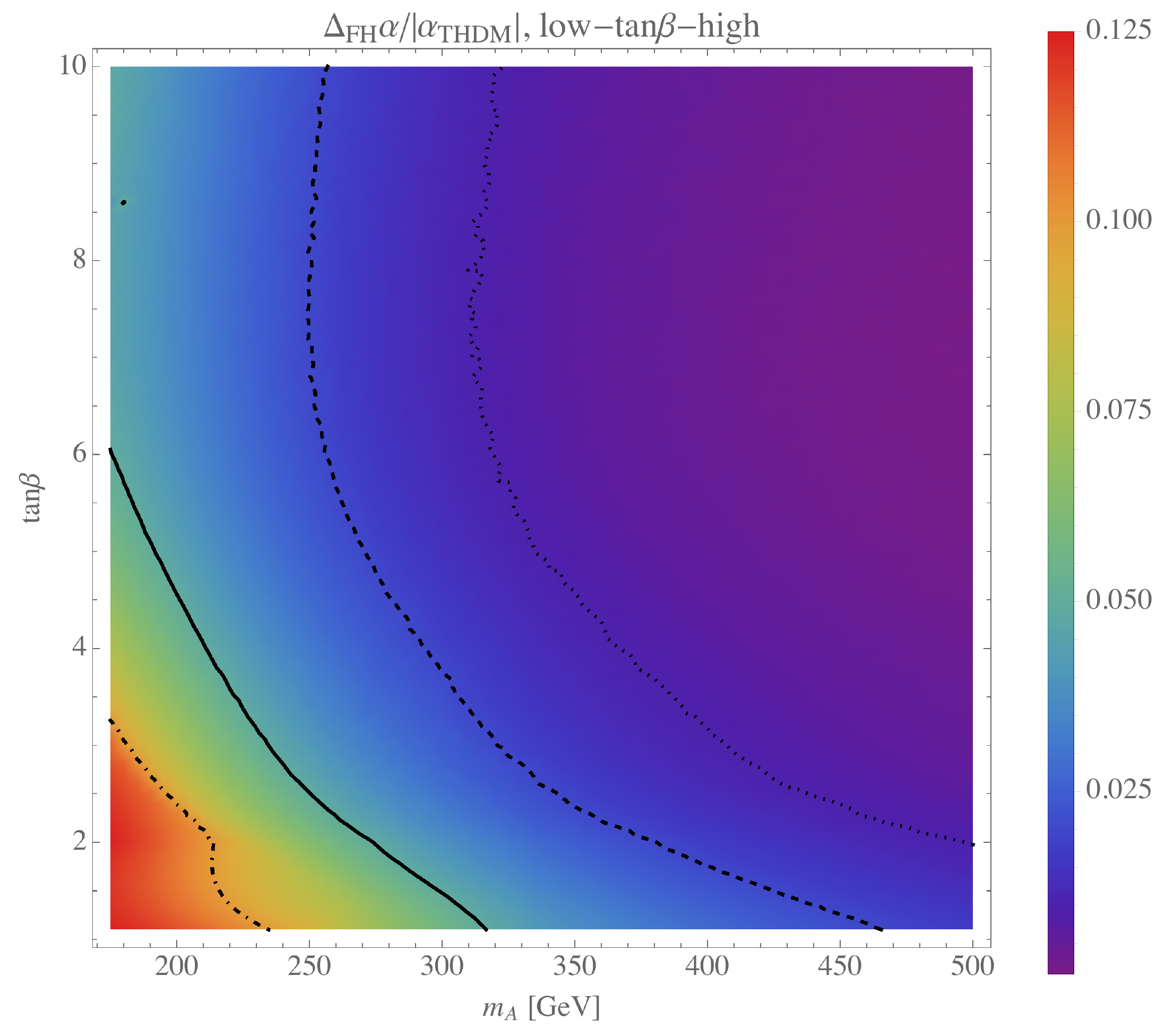}
\end{center}
\vspace{-1.5em}
\caption{\textbf{Top row:} Density plots for $M_h$ calculated using the effective THDM (left) and the difference between the left plot and the calculation of $M_h$ using \textsc{FeynHiggs}, for the low-$\tan\b$-high scenario (right). 
From top to bottom, the (dotted, dashed, solid, dot-dashed) black curves correspond to differences of $-(2, 3, 5, 10)$ GeV, respectively. 
\textbf{Bottom row:} The difference in $m_H$ (left) and fractional difference in $\alpha$ (right) calculated using the effective THDM and \textsc{FeynHiggs}. 
In the left plot, from the upper right to the lower left, the (dashed, dotted, dotted, dashed, solid, dot-dashed, dot-dashed) black curves 
correspond to differences of $(2, 1, -1, -2, -5, -10, -15)$ GeV, respectively.
In the right plot, from top to bottom, the (dotted, dashed, solid, dot-dashed) black curves correspond to differences of $(1, 2, 5, 10)\%$.}
\label{fig:low-tanb-high_DeltaFH}
\end{figure}

Our results in this scenario are shown in Fig.~\ref{fig:low-tanb-high_DeltaFH}. 
The top-left panel shows that across the range of parameter space using the tabulated values of $M_S$ and $X_t$, $M_h \lesssim 123$ GeV using the effective THDM calculation. 
In the top-right panel, the discrepancy between our calculated value of $M_h$ and that of \textsc{FeynHiggs} is clearly exhibited: for much of the parameter space above $t_\b \sim 6.5$, our calculation of $M_h$ is about 2 GeV lower. Between $t_\b \sim $ 4--5 ($t_\b \sim $ 2--4), this disagreement worsens to 3--5 (5--10) GeV.
This can also be seen in the lower-right panel of Fig. \ref{fig:MhGUT_tanbvMS_mA200}, where for $m_A = 200$ GeV and lower values of $\mu = M_1 = M_2 = 200$ GeV, $M_h = 122$ GeV is not achieved for $M_S = 100$ TeV until $t_\b \sim 3$.
The effective THDM calculation yields a higher value of the Higgs mixing angle $\alpha$ compared with \textsc{FeynHiggs}, but the two are in agreement at the level of $5\%$ except for a region $m_A \lesssim 300$ GeV and $t_\b \lesssim 6$. Below $t_\b \sim 3$ and $m_A \sim 225$ GeV, the fractional difference reaches 10\%--12\%. The values of the heavy Higgs mass $m_H$ are only significantly discrepant, more than 5 GeV, for low $t_\b \lesssim$ 2.5, although for $t_\b \lesssim 1.5$, $m_A \lesssim 250$ GeV, our calculation of $m_H$ is more than 10 GeV lower.

\begin{figure}[tb!]
\begin{center}
\includegraphics[width=0.42\linewidth]{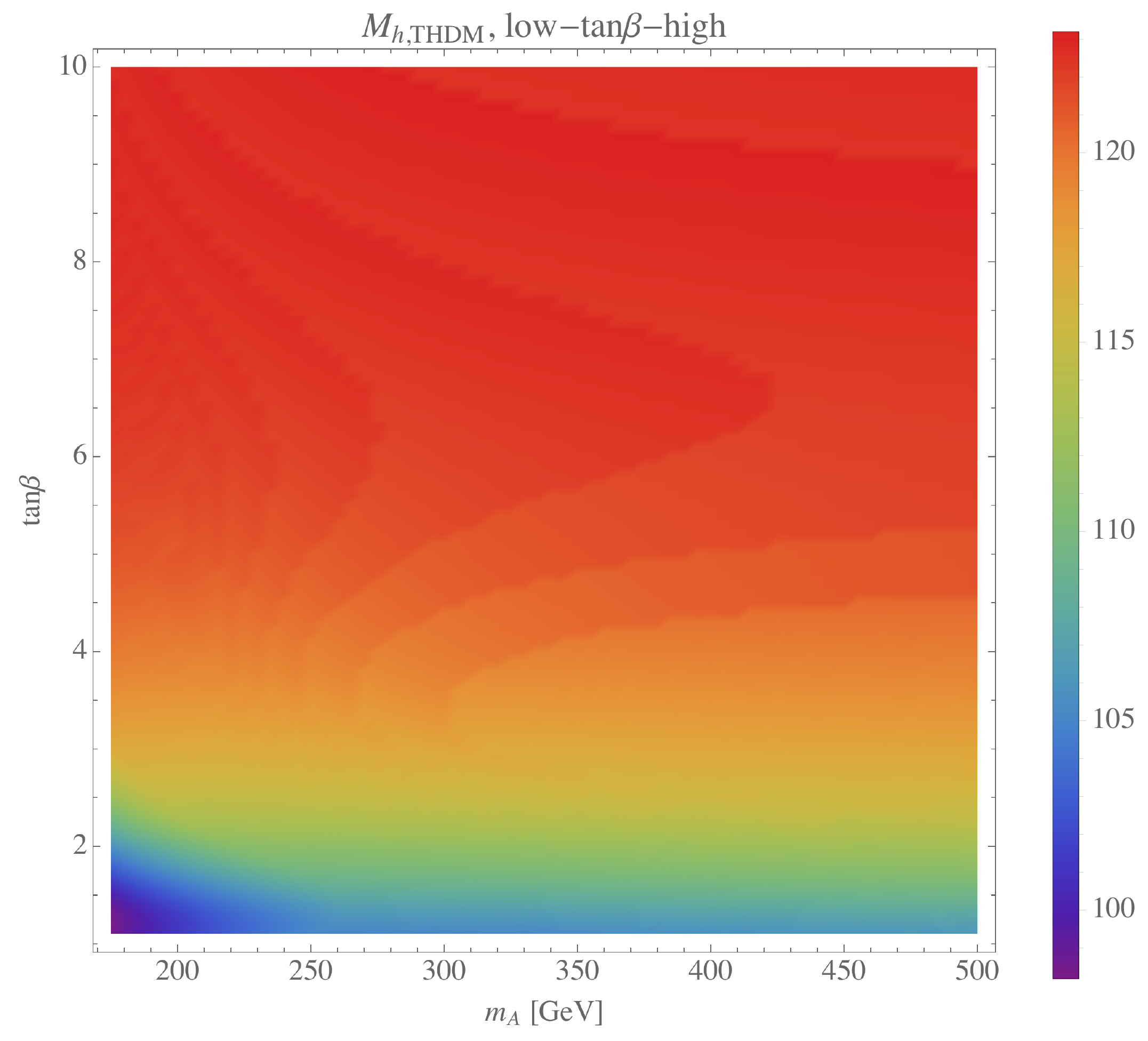} \quad
\includegraphics[width=0.423\linewidth]{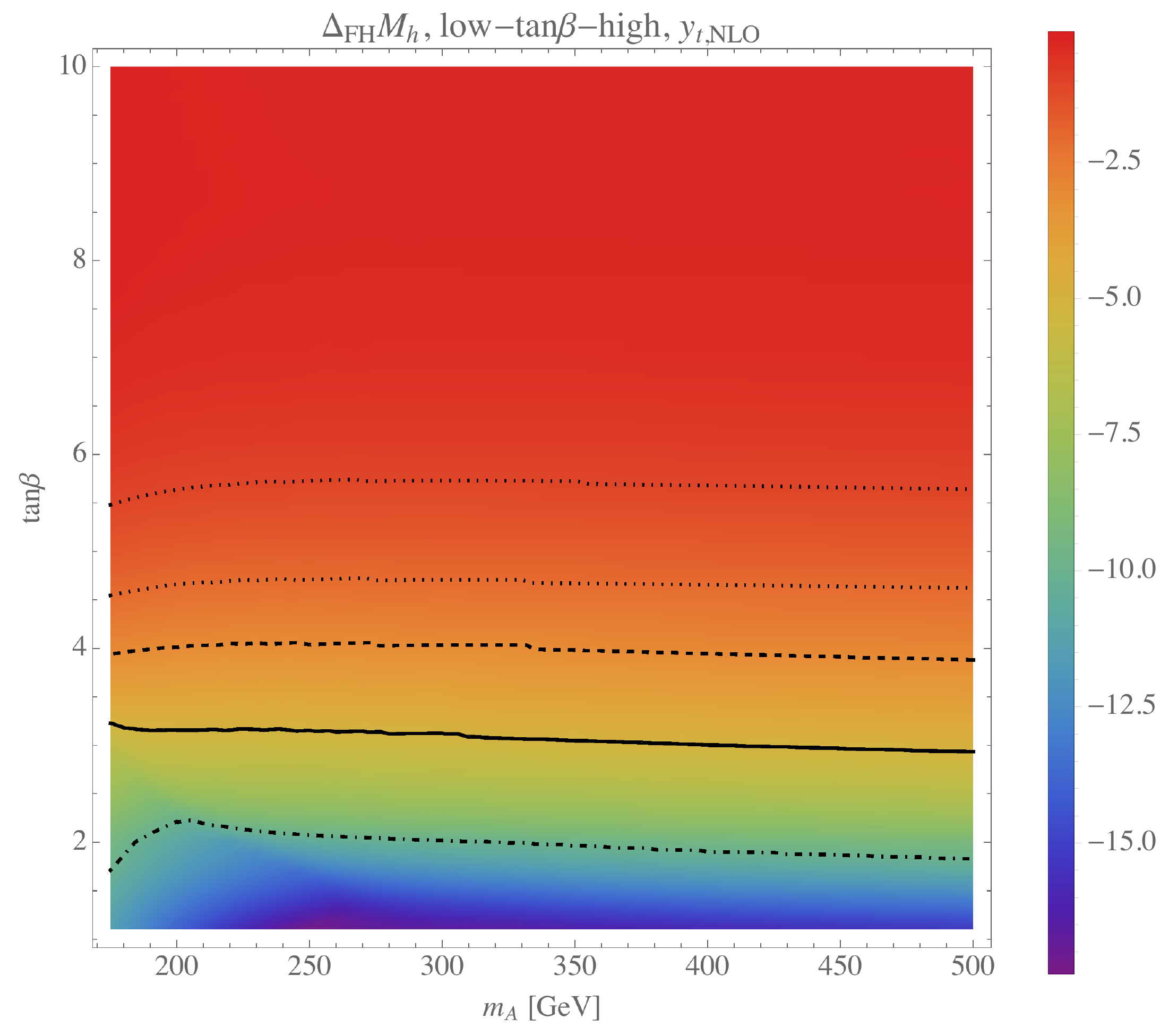}
\includegraphics[width=0.423\linewidth]{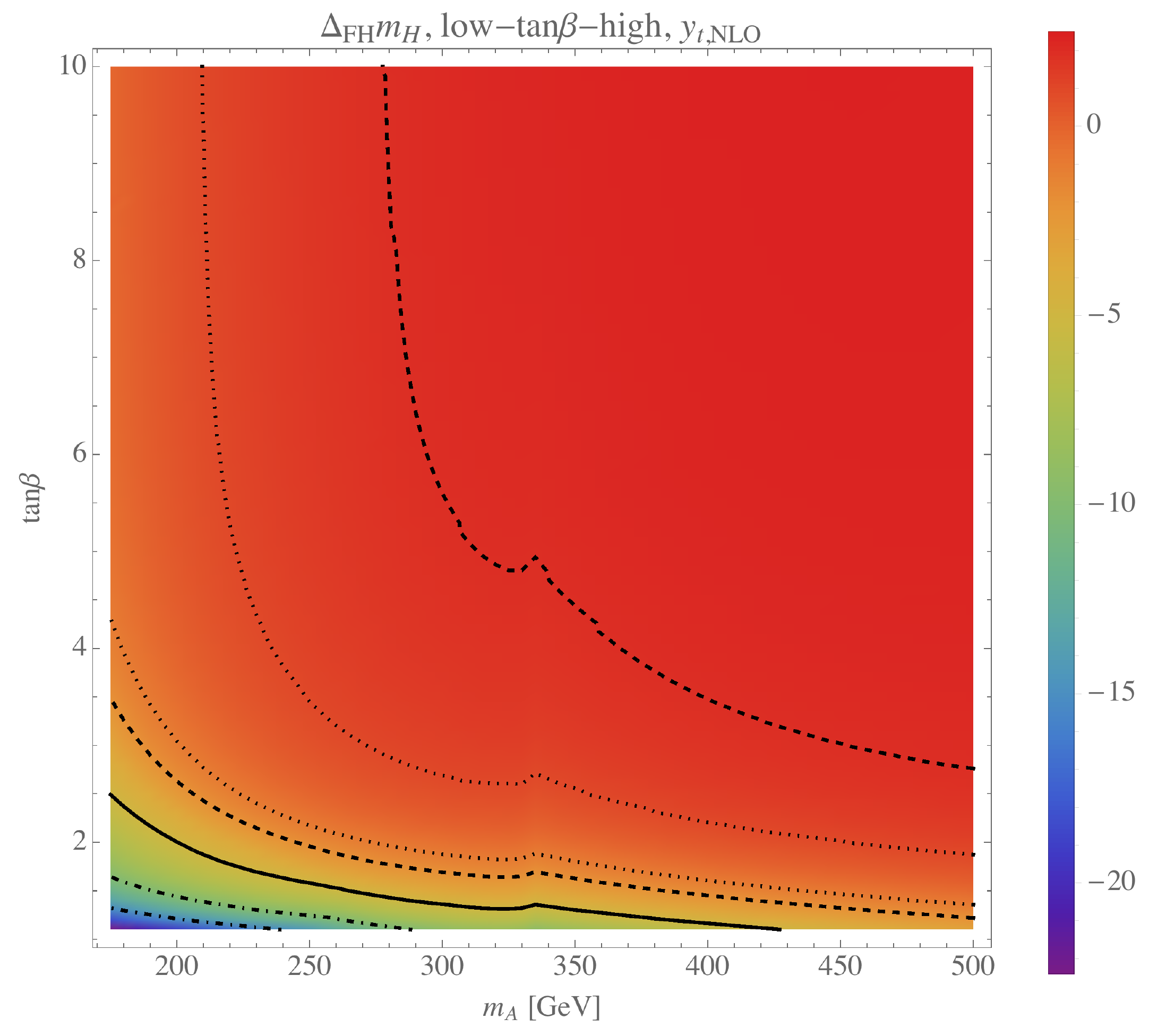} \quad
\includegraphics[width=0.423\linewidth]{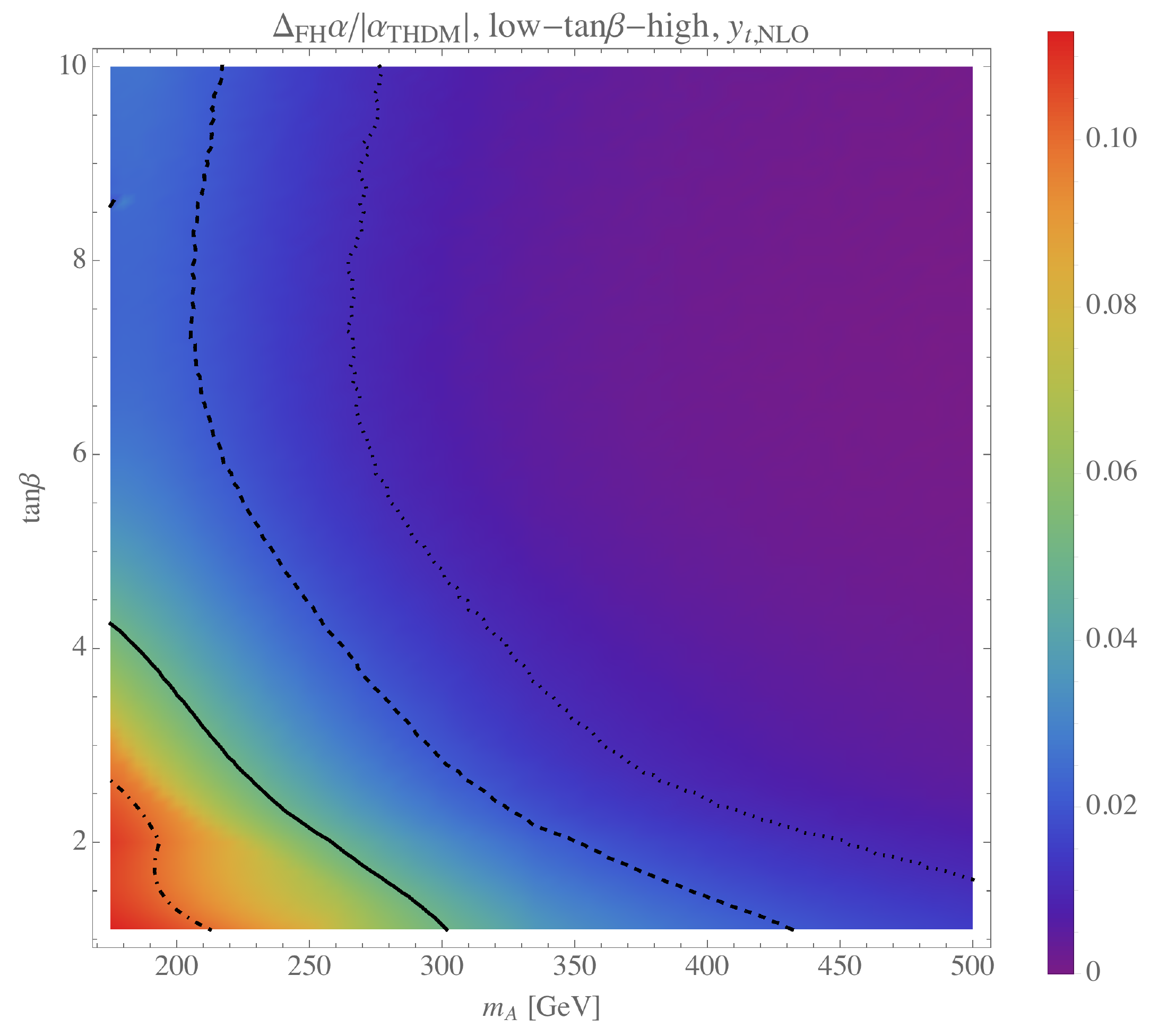}
\end{center}
\vspace{-1.5em}
\caption{As in Fig. \ref{fig:low-tanb-high_DeltaFH}, except using the boundary value $y_{t,\text{NLO}} (M_t) = 0.95113$ for the RG evolution.
In the top right plot, from top to bottom, the (dotted, dotted, dashed, solid, dot-dashed) black curves correspond here to differences of $-(1, 2, 3, 5, 10)$ GeV, respectively. 
}
\label{fig:low-tanb-high_DeltaFH_ytNLO}
\end{figure}

We can estimate how much of the differences in Fig.~\ref{fig:low-tanb-high_DeltaFH} are due to the use of a different boundary value for the top Yukawa $y_t (M_t)$, 
for which \textsc{FeynHiggs} uses the one-loop SM $\MSbar$ $\oln{m}_t$ running value.%
\footnote{For consistency at two-loop order, only the one-loop terms involving $g_3, y_t$ are employed in \textsc{FeynHiggs} to obtain $y_{t, \text{FH NLO}} (M_t) = 0.962$.}
In Fig.~\ref{fig:low-tanb-high_DeltaFH_ytNLO}, we reproduce the results in Fig.~\ref{fig:low-tanb-high_DeltaFH}, 
except that we use $y_{t, \text{NLO}}(M_t) = 0.95113$ as the boundary value for the RG running. 
In the top-right panel, $M_h$ in the region above $t_\b \sim 5.5$ (4.5) now agrees to within 1 (2) GeV; 
however, discrepancies larger than 5 (10) GeV still exist for $t_\b \lesssim 3$ (2). 
Likewise, there are modest reductions in the differences in $\alpha$ and $m_H$ across the parameter space. 
The remaining differences between the \textsc{FeynHiggs} results and our results could be explained 
by the different resummation method implemented in \textsc{FeynHiggs} in which the THDM effects are ignored.

\begin{figure}[tb!]
\begin{center}
\includegraphics[width=0.464\linewidth]{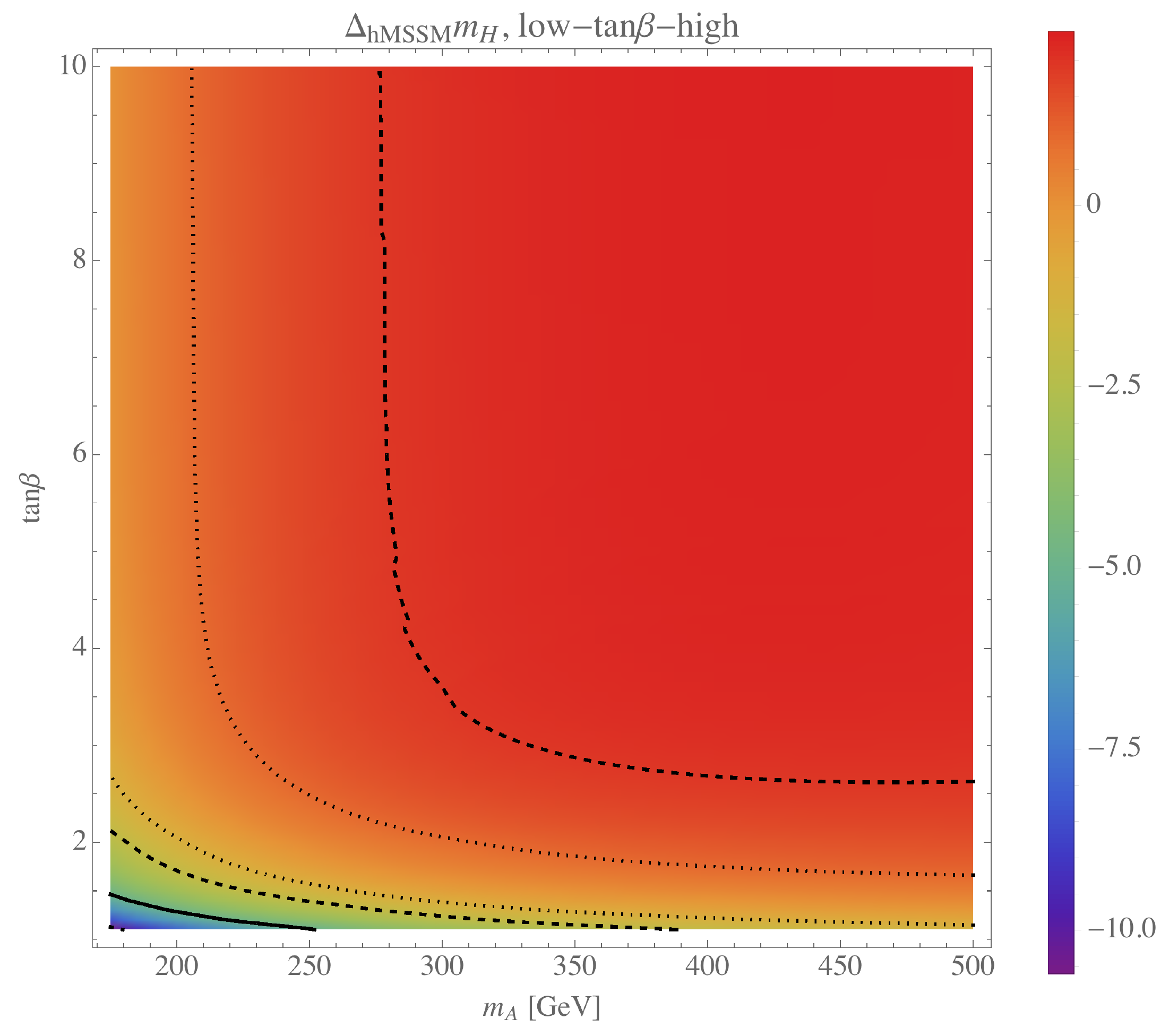} \quad
\includegraphics[width=0.456\linewidth]{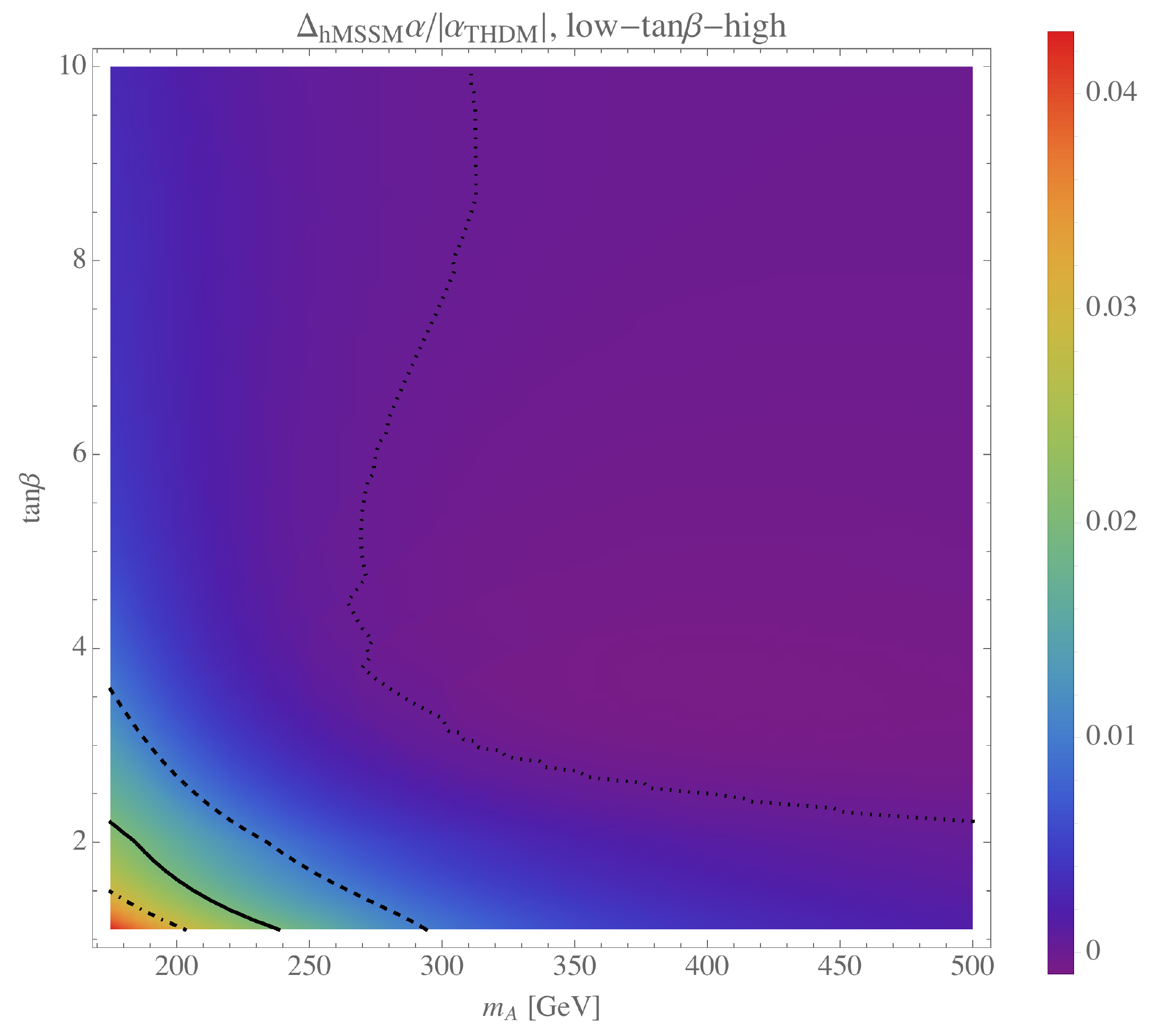}
\end{center}
\vspace{-1.5em}
\caption{Density plots for the difference in $m_H$ (left) and the fractional difference in $\alpha$ (right) 
calculated using the effective THDM and the hMSSM approximation, for the low-$\tan\b$-high scenario.
In the left plot, from the upper right to the lower left, the (dashed, dotted, dotted, dashed, solid, dot-dashed) 
black curves correspond to differences of $(2, 1, -1, -2, -5, -10)$ GeV, respectively.
In the right plot, from the upper right to the lower left, the (dotted, dashed, solid, dot-dashed) black curves 
correspond to differences of $(0, 1, 2, 3)\%$, respectively.}
\label{fig:low-tanb-high_DeltahMSSM}
\end{figure}

We turn now to the comparison with the hMSSM in this scenario, shown in Fig.~\ref{fig:low-tanb-high_DeltahMSSM}. We use Eqs.~(\ref{eqn:hMSSMMH2})--(\ref{eqn:hMSSMalpha}), inserting the value of $M_h$ obtained from the effective THDM calculation. The fractional difference in $\alpha$ between our calculation and the hMSSM is less than 4\% between the two calculations. Likewise, there is minimal deviation in $m_H$, except in the small corner of parameter space at $t_\b \sim 1$, $m_A \lesssim 200$ GeV, where the disagreement reaches the 5\% level. As was discussed in Sec.~\ref{sect:hMSSM}, sizable values of $\mu$ are needed for the hMSSM approximation to break down; however, throughout the parameter space of the low-tan$\b$-high scenario, $\mu \ll M_S$. Finally, we note that if instead the value of $M_h$ from \textsc{FeynHiggs} is used in the hMSSM equations, we see a similar level of disagreement between the hMSSM and our calculation as in Fig.~\ref{fig:low-tanb-high_DeltaFH}. 

\begin{figure}[htb!]
\begin{center}
\includegraphics[width=0.44\linewidth]{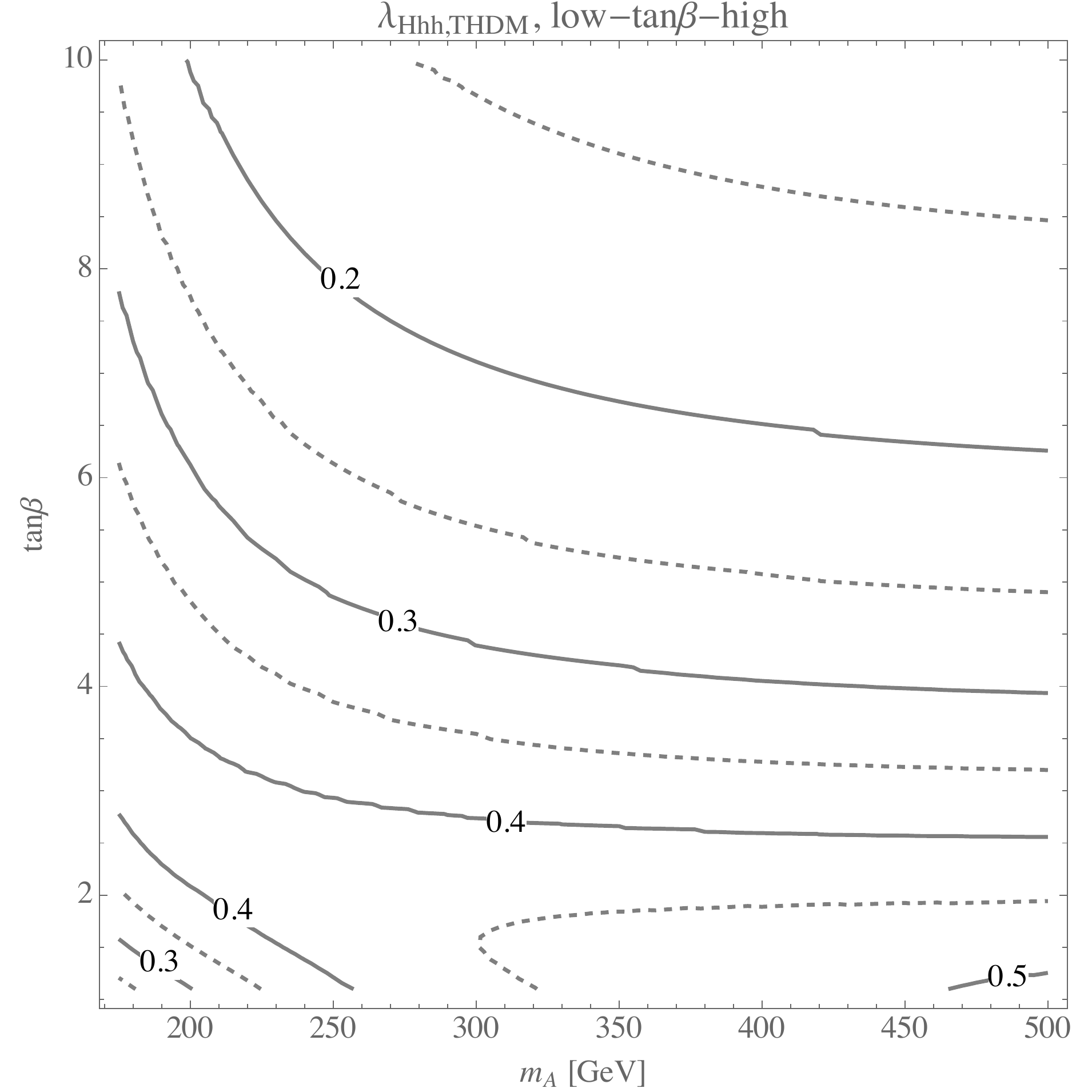} \quad
\includegraphics[width=0.44\linewidth]{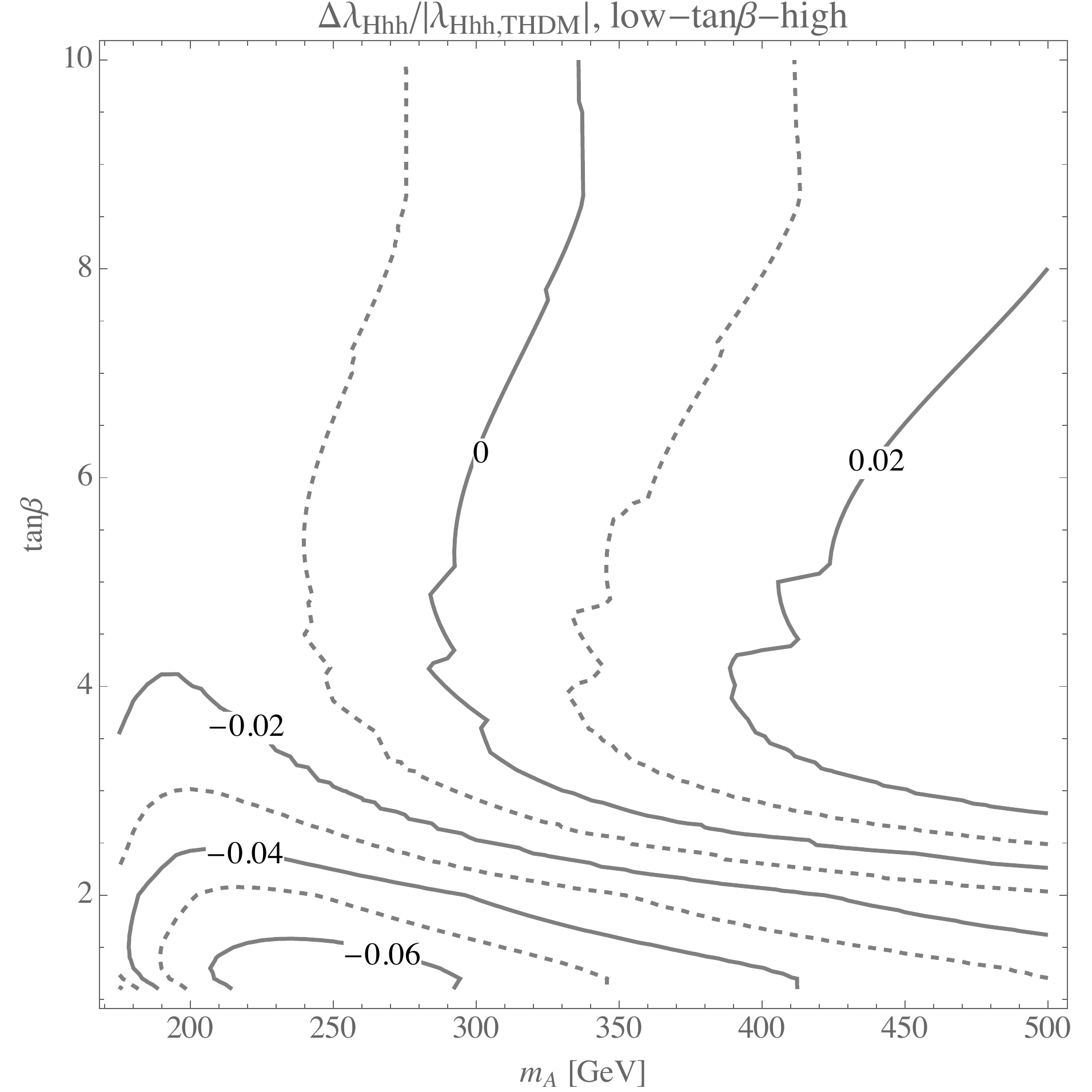}
\end{center}
\vspace{-1.5em}
\caption{Contour plots of $\l_{Hhh} = g_{Hhh}/v$ computed in the THDM [left] and the fractional difference between $\l_{Hhh}$ computed in the THDM and the hMSSM [right].}
\label{fig:low-tanb-high_gHhh}
\end{figure}

We can also test the formulas for the $g_{Hhh}$ coupling, Eqs.~(\ref{eqn:gHhh})--(\ref{eqn:hMSSMgHhh}), in the low-$\tan\b$-high scenario. 
In Fig.~\ref{fig:low-tanb-high_gHhh}, we show the results of our calculation and the fractional difference with the hMSSM using the effective THDM value of $M_h$. 
Fractional deviations of less than 6\%--7\% are observed. As above, differences between our calculation and the hMSSM when the \textsc{FeynHiggs} value of $M_h$ is used reach 30\% at low $\tan\b$ and larger values of $m_A$.

The dominant SM uncertainties come from the inputs $y_t, \a_s$ at $M_t$. 
The uncertainty from $\a_s(M_t)$ is subdominant as it enters at two-loop order for $M_h$ in both the RG running of $y_t, \l_i$ and in the threshold contributions. 
The uncertainty from $y_t$ has two sources: one from the experimental measurement of the top-quark pole mass $M_t$, 
and the other from the conversion of $M_t$ to the $\MSbar$ top Yukawa $y_t(M_t)$. 
An estimate of the uncertainty from the value of $M_t$ can be found in the $m_A \sim M_S$ case in \cite{Draper:2013oza}, 
where it was shown that using the $1\sigma$ high and low values of $M_t$ shift $M_h$ by about 1 GeV. 
As previously discussed, the use of the NLO, NNLO, or NNLO+N${}^3$LO QCD values of $y_t(M_t)$ can shift $M_h$ by 1--2 GeV.
There are also uncertainties from varying the renormalization scale $Q^2$ in the effective potential, 
from subleading two-loop threshold corrections to $\l_k$, and from higher-dimensional operators, 
but we expect these contributions are subdominant to those from the SM. 
For a more detailed discussion of uncertainties, see Ref.~\cite{Vega:2015fna}.


\section{Conclusion} \label{sect:concl}

In this article, we have computed the mass and couplings of the lightest $CP$-even Higgs boson in the MSSM, considering large values
of the masses of the scalar quarks, and intermediate values of the $CP$-odd Higgs mass. We performed these calculations using 
effective theory techniques and resumming the large logs appearing above and below the $CP$-odd Higgs mass scale. We worked in
the Higgs basis and showed that provided the threshold corrections to the off-diagonal $CP$-even Higgs mass matrix element are small,
all relevant radiative corrections may be effectively absorbed into the definition of the lightest $CP$-even Higgs mass.  This situation 
occurs for moderate or small values of the Higgsino mass parameter $\mu$ and/or of the trilinear stop mass parameter $A_t$, and the
resulting $CP$-even Higgs boson masses are well approximated by the hMSSM scenario. On the other hand, for sizable values
of $\mu$ and $A_t$, the alignment condition may be realized, in which case our results differ significantly from those in the hMSSM method.

The Higgs masses computed in our work tend to be lower than the results obtained by \textsc{FeynHiggs}, which
implements a different resummation method, and may differ by a few GeV or more. The difference may be traced to our use of an effective THDM
theory at scales above $m_A$ and also a higher-order computation of the relation between the running and the on-shell top-quark mass.  

Our calculation of $M_h$ leads to lower bounds on $t_\b$ for low values of $m_A$ in order to achieve $M_h = 125$ GeV: 
for $m_A = 200$ GeV, we find $t_\b \gtrsim 3.4$ (2.0) for $\mu = M_S$ (200 GeV). 
These bounds are due to the appearance of large mixing effects that push the lightest $CP$-even Higgs mass down and 
cannot be overcome by the positive contributions to $M_h$ from both the stop threshold corrections (even with stops as heavy as $M_{\text{GUT}}$) 
and from radiative corrections from light charginos and neutralinos. 
We also note that for values of $t_\b \sim 1$, a Landau pole of the top-quark Yukawa coupling may be induced at low values of $m_A$. 

Finally, we would like to stress that our work has been restricted to the computation of the neutral Higgs masses in the MSSM, 
without taking into account any experimental constraints beyond the measured value of the Higgs mass. While constraints from
precision measurements, new physics searches, Higgs, dark matter, and flavour physics will lead to relevant bounds on the values 
of the free parameters of the theory, in this article we have focused on the Higgs mass computation for arbitrary values of
those parameters.  Our results should be complemented with a careful analysis of the experimental constraints and can also
be used to determine in a more precise way the bounds on the free parameters of the model coming from those constraints. 
We reserve this analysis for a future publication. 
 

\begin{acknowledgments} \label{sect:acknow}
We are indebted to P.~Slavich for his careful reading of the manuscript and many suggestions.
We also thank S.~Heinemeyer and J.~P.~Vega for useful discussions, and F.~Staub and M.~Goodsell for help with \textsc{SARAH}.
Work at ANL is supported in part by the U.S. Department of Energy under Contract No.~DE-AC02-06CH11357. 
G.~L. acknowledges support from DOE Grant No.~DE-FG02-13ER41958, the ICORE Program of Planning and Budgeting Committee, and by ISF Grant No.~1937/12.
The authors acknowledge the hospitality of the Munich Institute for Astro- and Particle Physics (MIAPP) of the DFG cluster of excellence ``Origin and Structure of the Universe"
while this work was being completed.
This work was completed while C.~W. was at the Aspen Center for Physics, which is supported by National Science Foundation Grant No. PHY-1066293.
\end{acknowledgments}

\appendix


\section{Two-loop RGE's in the Type II THDM} \label{app:THDM2LRGEs}

The $\b$-functions for a coupling are
\be
\b_g (t) = \frac{d g}{d t} = \sum_{n=1}^{\infty} \k^n \b_g^{(n)} (t)
\ee
where $t = \log Q$ with $Q$ the renormalization scale, $\k = 1/(4\pi)^2$ is the loop factor, and $\b_g^{(n)}$ is the $n$th loop $\b$-function for $g$.
We have extracted these equations from the program \textsc{SARAH}, version 4.2. \cite{Goodsell:2014bna} 
Below, we list the two-loop RG equations for the type II THDM in the third generation approximation. 
$N_g$ is the number of fermion generations and $\q_X$ is the Heaviside function for the mass $X$.
These equations were also listed in Ref.~\cite{Dev:2014yca}, with which we find minor differences;
we use different conventions for three parameters 
$\l_1 = 2\tilde{\l}_1, \l_2 = 2\tilde{\l}_2, g_1^2 = 5g^{\prime 2}/3$,
where $g^\prime, \tilde{\l}_1, \tilde{\l}_2$ appear in Ref.~\cite{Dev:2014yca}.


\subsection{Gauge couplings}


Hypercharge coupling $g_1$ in the $SU(5)$ normalization, with $g_Y^2 = \frac35 g_1^2$:

\ba{
g_1^{-3} \b_{g_{1}}^{(1)} &= \frac15 + \frac43 N_g + \frac25 g_1^3 \q_\m \,, \\
g_1^{-3} \b_{g_{1}}^{(2)} &= \frac{44}5 g_3^2 + \frac{18}5 g_2^2 + \frac{104}{25} g_1^2 - \frac{17}{10} h_t^2 - \frac12 h_b^2 - \frac32 h_\tau^2 \,.
}


Weak gauge coupling $g_2$:

\ba{
g_2^{-3} \b_{g_2}^{(1)} &= - 7 + \frac43 N_g + \frac23 \Big[ \q_{\mu} + 2 \q_{M_2} \Big] \,, \\
g_2^{-3} \b_{g_2}^{(2)} &=  12 g_3^2 + 8 g_2^2 + \frac65 g_1^2 - \frac32 h_t^2 - \frac32 h_b^2 - \frac12 h_\tau^2 \,.
}


Strong gauge coupling $g_3$:

\ba{
g_3^{-3} \b_{g_3}^{(1)} &= - 11 + \frac43 N_g \,, \\
g_3^{-3} \b_{g_3}^{(2)} &= - 26 g_3^2 + \frac92 g_2^2 + \frac{11}{10} g_1^2 - 2 h_t^2 - 2 h_b^2 \,.
}


\subsection{Yukawa couplings}


Top Yukawa $h_t$:

\be
h_t^{-1} \b_{h_t}^{(1)} = - 8 g_3^2 - \frac94 g_2^2 - \frac{17}{20} g_1^2 + \frac92 h_t^2 + \frac12 h_b^2 + \frac32 \Big[ g_2^2 + \frac15 g_1^2 \q_{M_1} \Big] \q_{\mu} \,,
\ee

\be
\begin{split}
h_t^{-1} \b_{h_t}^{(2)} &= - 108 g_3^4 + 9 g_3^2 g_2^2 + \frac{19}{15} g_3^2 g_1^2 - \frac{21}4 g_2^4 - \frac9{20} g_2^2 g_1^2 + \frac{1267}{600} g_1^4 \\
& \quad + g_3^2 \bigg[ 36 h_t^2 + \frac{16}3 h_b^2 \bigg] 
+ \frac{3}{16} g_2^2 \Big[ 75 h_t^2 + 11 h_b^2 \Big] 
+ \frac1{240} g_1^2 \Big[ 1179 h_t^2 - 41 h_b^2 \Big] \\
& \quad - 12 h_t^4 - \frac52 h_t^2 h_b^2 - \frac52 h_b^4 - \frac34 h_b^2 h_\tau^2 
- 6 h_t^2 \l_2 + 2 h_b^2 \Big[ - \l_3 + \l_4 \Big] \\
& \quad + \frac32 \l_2^2 + \l_3^2 + \l_3 \l_4 + \l_4^2 + \frac32 \l_5^2 +\frac32 \l_6^2 + \frac92 \l_7^2 \,.
\end{split}
\ee


Bottom Yukawa $h_b$:

\be
h_b^{-1} \b_{h_b}^{(1)} = - 8 g_3^2 - \frac94 g_2^2 - \frac14 g_1^2 + \frac12 h_t^2 + \frac92 h_b^2 + h_\tau^2 \,,
\ee

\be
\begin{split}
h_b^{-1} \b_{h_b}^{(2)} &= - 108 g_3^4 + 9 g_3^2 g_2^2 + \frac{496}{240} g_3^2 g_1^2 - \frac{21}4 g_2^4 - \frac{27}{20} g_2^2 g_1^2 - \frac{113}{600} g_1^4 \\
& \quad + g_3^2 \bigg[ \frac{16}3 h_t^2 + 36 h_b^2 \bigg] 
+ \frac{3}{16} g_2^2 \Big[ 11 h_t^2 + 75 h_b^2 + 10 h_\tau^2 \Big]
- \frac{1}{240} g_1^2 \Big[ 53 h_t^2 - 711 h_b^2 - 450 h_\tau^2 \Big] \\
& \quad - \frac52 h_t^4 - \frac52 h_t^2 h_b^2 - 12 h_b^4 - \frac94 h_b^2 h_\tau^2  - \frac94 h_\tau^4
- 2 h_t^2 \l_3 + 2 h_t^2 \l_4 - 6 h_b^2 \l_1 \\
& \quad + \frac32 \l_1^2 + \l_3^2 + \l_3 \l_4 + \l_4^2 + \frac32 \l_5^2 + \frac92 \l_6^2 + \frac32 \l_7^2 \,.
\end{split}
\ee


Tau Yukawa $h_\tau$:

\be
h_\tau^{-1} \b_{h_\tau}^{(1)} = - \frac94 g_2^2 - \frac94 g_1^2 + 3 h_b^2 + \frac52 h_\tau^2 \,,
\ee

\be
\begin{split}
h_\tau^{-1} \b_{h_\tau}^{(2)} &= 20 g_3^2 h_b^2 - \frac{21}4 g_2^4 + \frac{27}{20} g_2^2 g_1^2 + \frac{1449}{200} g_1^4 \\
& \quad + \frac{15}{16} g_2^2 \Big[ 6 h_b^2 + 11 h_\tau^2 \Big] 
+ \frac{1}{80} g_1^2 \Big[ 50 h_b^2 + 537 h_\tau^2 \Big]  \\
& \quad - \frac94 h_t^2 h_b^2 - \frac{27}4 h_b^4 - \frac{27}4 h_b^2 h_\tau^2 - 3 h_\tau^4 - 6 h_\tau^2 \l_1 \\
& \quad + \frac32 \l_1^2 + \l_3^2 + \l_3 \l_4 + \l_4^2 + \frac32 \l_5^2 + \frac92 \l_6^2 + \frac32 \l_7^2 \,.
\end{split}
\ee


\subsection{Anomalous dimensions}


Down-type Higgs $\g_1 = d\log v_1/dt$:

\be
\g_{1}^{(1)} = \frac94 \bigg( g_2^2 + \frac15 g_1^2 \bigg) - 3 h_b^2 - h_\t^2 
\underbrace{ - \frac32 \bigg( g_2^2 \q_{M_2} + \frac15 g_1^2 \q_{M_1} \bigg) \q_{\mu} }_{\g_{1,\c}^{(1)}}
\ee

\be
\begin{split}
\g_{1}^{(2)} &= \frac{435}{32} g_2^4 - \frac9{80} g_2^2 g_1^2 - \frac{1341}{800} g_1^4
- 20 g_3^2 h_b^2 
- \frac{15}8 g_2^2 \Big[ 3 h_b^2 + h_\tau^2 \Big]
- \frac58 g_1^2 \Big[ h_b^2 + 3 h_\tau^2 \Big] \\
& \quad + \frac94 h_t^2 h_b^2 + \frac{27}4 h_b^4 + \frac94 h_\tau^4 
- \frac32 \l_1^2 - \l_3^2 - \l_3 \l_4 - \l_4^2 - \frac32 \l_5^2 - \frac92 \l_6^2 - \frac32 \l_7^2 \\
& \quad - \frac32 t_\b \Big[ \l_1 \l_6 + \l_2 \l_7 + \l_{345} (\l_6 + \l_7) \Big]
\end{split}
\ee


Up-type Higgs $\g_1 = d\log v_2/dt$:

\be
\g_{2}^{(1)} = \frac94 \bigg( g_2^2 + \frac15 g_1^2 \bigg) - 3 h_t^2 
\underbrace{ - \frac32 \bigg( g_2^2 \q_{M_2} + \frac15 g_1^2 \q_{M_1} \bigg) \q_{\mu} }_{\g_{2,\c}^{(1)}}
\ee

\be
\begin{split}
\g_{2}^{(2)} &=\frac{435}{32} g_2^4 - \frac9{80} g_1^2 g_2^2 - \frac{1341}{800} g_1^4
- h_t^2 \bigg[ 20 g_3^2 + \frac{45}8 g_2^2 + \frac{17}8 g_1^2 \bigg] \\
& \quad + \frac{27}4 h_t^4 + \frac94 h_b^2 h_t^2 
- \frac32 \l_2^2 - \l_3^2 - \l_3 \l_4 - \l_4^2 - \frac32 \l_5^2 - \frac32 \l_6^2 -\frac92 \l_7^2 \\
& \quad - \frac32 t_\b^{-1} \Big[ \l_1 \l_6 + \l_2 \l_7 + \l_{345} (\l_6 + \l_7) \Big]
\end{split}
\ee

\newpage

\subsection{Quartic couplings}


$\l_1$:

\be
\begin{split}
\b_{\l_1}^{(1)} &= \frac34 \bigg[ 2 g_2^4 + \bigg( g_2^2 + \frac35 g_1^2 \bigg)^2 \bigg] - 12 h_b^4 - 4 h_\tau^4  \\
& \quad + 12 \l_{1}^2+2 (\l_{3}+\l_{4})^2+2 \l_{3}^2+2 \l_{5}^2+24 \l_{6}^2 - 4 \l_{1} \g_1 \\
& \quad - \Big[ 5 g_2^4 \q_{M_2} + \frac65 g_2^2 g_1^2 \q_{M_2} \q_{M_1} + \frac9{25} g_1^4 \q_{M_1} \Big] \q_{\m} - 4 \l_1 \g_{1,\c}^{(1)}
\end{split}
\ee

\be
\begin{split}
\b_{\l_1}^{(2)} &= \frac{291}8 g_2^6 - \frac{303}{40} g_2^4 g_1^2 - \frac{1719}{200} g_2^2 g_1^4 - \frac{3537}{1000} g_1^6 \\
& \quad - \frac38 g_2^4 \Big[ 12 h_b^2 + 4 h_\tau^2 + 17 \l_{1} - 20 \Big( 2 \l_{3} + \l_{4} \Big) \Big]
+ \frac3{20} g_2^2 g_1^2 \Big[ 36 h_b^2 + 44 h_\tau^2 + 39 \l_{1} + 20 \l_{4} \Big] \\
& \quad + \frac{9}{200} g_1^4 \Big[ 20 h_b^2 - 100 h_\tau^2 + 217 \l_{1} + 20 \Big( 2 \l_{3} + \l_{4} \Big) \Big] \\
& \quad - 16 g_3^2 h_b^2 \Big[ 4 h_b^2 - 5 \l_{1} \Big]
+ \frac32 g_2^2 \Big[ 5 \l_1 \Big( 3 h_b^2 + h_\tau^2 \Big) + 4 \Big( 9 \l_{1}^2 + 4 \l_{3}^2 + 4 \l_{3} \l_{4} + \l_{4}^2 + 18 \l_{6}^2 \Big) \Big] \\
& \quad + \frac1{10} g_1^2 \bigg[ h_b^2 \Big( 16 h_b^2 + 25 \l_{1} \Big) + 3 h_\tau^2 \Big( - 16 h_\tau^2 + 25 \l_{1} \Big) \\
& \qquad \qquad + 12 \Big( 9 \l_{1}^2 + 4 \l_{3}^2 + 4 \l_{3} \l_{4} + 2 \l_{4}^2 - \l_{5}^2 + 18 \l_{6}^2 \Big) \bigg] \\
& \quad - 12 h_t^2 \Big[ 2 \l_{3}^2 + 2 \l_{3} \l_{4} + \l_{4}^2 + \l_{5}^2 + 6 \l_{6}^2 \Big]
+ 3 h_t^2 h_b^2 \Big[ 4 h_b^2 - 3 \l_{1} \Big] \\
& \quad - 3 h_b^2 \Big[ - 20 h_b^4 + h_b^2 \l_{1} + 24 \l_{1}^2 + 24 \l_{6}^2 \Big]
- h_\tau^2 \Big[ - 20 h_\tau^4 + h_\tau^2 \l_{1} + 24 \l_{1}^2  + 24 \l_{6}^2 \Big] \\
& \quad - 2 \l_1 \Big[ 39 \l_{1}^2 + 10 \l_{3}^2 + 10 \l_{3} \l_{4} + 6 \l_{4}^2 + 7 \l_{5}^2 + 159 \l_{6}^2 - 3 \l_{7}^2 \Big] \\
& \quad - 4 \l_3 \Big[ 4 \l_{3}^2 + 6 \l_{3} \l_{4} + 8 \l_{4}^2 + 10 \l_{5}^2 + 33 \l_{6}^2 + 18 \l_{6} \l_{7} + 9 \l_{7}^2 \Big] \\
& \quad - 4 \l_4 \Big[ 3 \l_{4}^2 + 11 \l_{5}^2 + 35 \l_{6}^2 + 14 \l_{6} \l_{7} + 7 \l_{7}^2 \Big]
- 4 \l_5 \Big[ 37 \l_{6}^2 + 10 \l_{6} \l_{7} + 5 \l_{7}^2 \Big]
\end{split}
\ee

\newpage

$\l_2$:

\be
\begin{split}
\b_{\l_2}^{(1)} &= \frac34 \bigg[ 2 g_2^4 + \bigg( g_2^2 + \frac35 g_1^2 \bigg)^2 \bigg] - 12 h_t^4 \\
& \quad + 12 \l_2^2 + 2 \l_3^2  + 2 (\l_3+\l_4)^2 + 2 \l_5^2 + 24 \l_6^2 - 4 \l_2 \g_2 \\
& \quad - \Big[ 5 g_2^4 \q_{M_2} + \frac65 g_2^2 g_1^2 \q_{M_2} \q_{M_1} + \frac9{25} g_1^4 \q_{M_1} \Big] \q_{\m} - 4 \l_2 \g_{2,\c}^{(1)}
\end{split}
\ee

\be
\begin{split}
\b_{\l_2}^{(2)} &= \frac{291}8 g_2^6 - \frac{303}{40} g_2^4 g_1^2 - \frac{1719}{200} g_2^2 g_1^4 - \frac{3537}{1000} g_1^6 \\
& \quad - \frac38 g_2^4 \Big[12 h_t^2 + 17 \l_2 - 20 \Big( 2 \l_3 + \l_4 \Big) \Big]
+ \frac3{20} g_2^2 g_1^2 \Big[ 84 h_t^2 + 39 \l_2 + 20 \l_4 \Big] \\
& \quad - \frac{9}{200} g_1^4 \Big[ 76 h_t^2 - 217 \l_2 - 20 \Big( 2 \l_3 + \l_4 \Big) \Big] \\
& \quad - 16 g_3^2 h_t^2 \Big[ 4 h_t^2 - 5 \l_2 \Big]
+ \frac32 g_2^2 \Big[ 15 h_t^2 \l_2 + 4 \Big( 9 \l_2^2 + 4 \l_3^2 + 4 \l_3 \l_4 + \l_4^2 +18 \l_7^2 \Big) \Big] \\
& \quad + \frac1{10} g_1^2 \Big[ - h_t^2 \Big( 32 h_t^2 - 85 \l_2 \Big) 
+ 12 \Big( 9 \l_2^2 + 4 \l_3^2 + 4 \l_3 \l_4 + 2 \l_4^2 - \l_5^2 + 18 \l_7^2 \Big) \Big] \\
& \quad + 3 h_t^2 \Big[ 20 h_t^4 - h_t^2 \l_2 - 24 \l_2^2 - 24 \l_7^2 \Big]
+ 3 h_t^2 h_b^2 \Big[ 4 h_t^2 - 3 \l_2 \Big] \\
& \quad - \Big( 12 h_b^2 + 4 h_\tau^2 \Big) \Big[  2 \l_3^2 + 2 \l_3 \l_4 + \l_4^2 + \l_5^2 + 6 \l_7^2 \Big] \\
& \quad - 2 \l_2 \Big[ 39 \l_2^2 + 10 \l_3^2 + 10 \l_3 \l_4 + 6 \l_4^2 + 7 \l_5^2 - 3 \l_6^2 + 159 \l_7^2 \Big] \\
& \quad - 4 \l_3 \Big[ 4 \l_3^2 + 6 \l_3 \l_4 + 8 \l_4^2 + 10 \l_5^2 + 9 \l_6^2 + 18 \l_6 \l_7 + 33 \l_7^2 \Big] \\
& \quad - 4 \l_4 \Big[ 3 \l_4^2 + 11 \l_5^2 + 7 \l_6^2 + 14 \l_6 \l_7 + 35 \l_7^2 \Big]
- 4 \l_5 \Big[ 5 \l_6^2 + 10 \l_6 \l_7 + 37 \l_7^2 \Big]
\end{split}
\ee

\newpage

$\l_3$:

\be
\begin{split}
\b_{\l_3}^{(1)} &= \frac34 \bigg[ 2 g_2^4 + \bigg( g_2^2 - \frac35 g_1^2 \bigg)^2 \bigg] - 12 h_t^2 h_b^2 \\
& \quad + 2 (\l_1 + \l_2) (3 \l_3 + \l_4) + 4 \l_3^3 + 2 \l_4^2 + 2 \l_5^2 + 4 \l_6^2 + 16 \l_6 \l_7 + 4 \l_7^2 - 2 \l_3 \Big[ \g_1^{(1)} + \g_2^{(1)} \Big] \\
& \quad - \bigg[ 5 g_2^4 \q_{M_2} - \frac65 g_2^2 g_1^2 \q_{M_2} \q_{M_1} + \frac9{25} g_1^4 \q_{M_1} \bigg] \q_\m - 2 \l_3 \Big[ \g_{1,\c}^{(1)} + \g_{2,\c}^{(1)} \Big]
\end{split}
\ee

\be
\begin{split}
\b_{\l_3}^{(2)} &= \frac{291}8 g_2^6 + \frac{33}{40} g_2^4 g_1^2 + \frac{909}{200} g_2^2 g_1^4 - \frac{3537}{1000} g_1^6 \\
& \quad - \frac34 g_2^4 \bigg[ 3 h_t^3 + 3 h_b^2 + h_\tau^2 - 15 \Big( \l_1 + \l_2 \Big) + \frac{37}2 \l_3 - 10 \l_4 \bigg] \\
& \quad - \frac3{10} g_2^2 g_1^2 \bigg[ 21 h_t^2 + 9 h_b^2 + 11 h_\tau^2 + 5 \Big( \l_1 + \l_2 \Big) - \frac{11}2 \l_3 + 6 \l_4 \bigg] \\
& \quad + \frac9{200} g_1^4 \Big[ - 38 h_t^2 + 10 h_b^2 - 50 h_\tau^2 + 30 \Big( \l_1 + \l_2 \Big) + 197 \l_3 + 20 \l_4 \Big] \\
& \quad - 8 g_3^2 \Big[ 8 h_t^2 h_b^2 - 5 \l_3 \Big( h_t^2 + h_b^2 \Big) \Big] \\
& \quad + 6 g_2^2 \bigg[ \frac58 \l_3 \Big( 3 h_t^2 + 3 h_b^2 + h_\tau^2 \Big) + 3 \Big( \l_1 + \l_2 \Big) \Big( 2 \l_3 + \l_4 \Big) + \Big( \l_3 - \l_4 \Big)^2 + 18 \l_6 \l_7 \bigg] \\
& \quad + \frac15 g_1^2 \bigg[ - 4 h_t^2 h_b^2 + \frac54 \l_3 \Big( 17 h_t^2 + 5 h_b^2 + 15 h_\tau^2 \Big) \\
& \qquad \qquad + 12 \Big( \l_1 + \l_2 \Big) \Big( 3 \l_3 + \l_4 \Big) + 6 \Big( \l_3^2 - \l_4^2 + 2 \l_5^2 + \l_6^2 + 16 \l_6 \l_7 + \l_7^2 \Big) \bigg] \\
& \quad - \frac92 \l_3 \Big[ 3 h_t^4 + 3 h_b^4 + h_\tau^4 \Big] + h_t^2 h_b^2 \Big[ 36 \Big( h_t^2 + h_b^2 \Big) + 15 \l_3 \Big] \\
& \quad - 6 h_t^2 \Big[ 6 \l_2 \l_3 + 2 \l_2 \l_4 + 2 \l_3^2 + \l_4^2 + \l_5^2 + 8 \l_6 \l_7 + 4 \l_7^2 \Big] \\
& \quad - \Big( 6 h_b^2 + 2 h_\tau^2 \Big) \Big[ 6 \l_1 \l_3 + 2 \l_1 \l_4 + 2 \l_3^2 + \l_4^2 + \l_5^2 + 4 \l_6^2 + 8 \l_6 \l_7 \Big] \\
& \quad - \Big( \l_1^2 + \l_2^2 \Big) \Big[ 15 \l_3 + 4 \l_4 \Big] 
- 2 \Big( \l_1 + \l_2 \Big) \Big[ 18 \l_3^2 + 8 \l_3 \l_4 + 7 \l_4^2 + 9 \l_5^2 \Big] \\
& \quad - 2 \l_1 \Big[ 31 \l_6^2 + 22 \l_6 \l_7 + 11 \l_7^2 \Big] - 2 \l_2 \Big[ 11 \l_6^2 + 22 \l_6 \l_7 + 31 \l_7^2 \Big] \\
& \quad - \l_3 \Big[ 12 \l_3^2 + 4 \l_3 \l_4 + 16 \l_4^2 + 18 \l_5^2 + 60 \l_6^2 + 176 \l_6 \l_7+ 60 \l_7^2 \Big] \\
& \quad - \l_4 \Big[ 12 \l_4^2 + 44 \l_5^2 + 68 \l_6^2 + 88 \l_6 \l_7 + 68 \l_7^2 \Big] - \l_5 \Big[ 68 \l_6^2 + 72 \l_6 \l_7 + 68 \l_7^2 \Big]
\end{split}
\ee

\newpage

$\l_4$:

\be
\begin{split}
\b_{\l_4}^{(1)} &= \frac95 g_2^2 g_1^2 + 12 h_t^2 h_b^2 + 2 \l_4 (\l_1 + \l_2 + 4 \l_3 + 2 \l_4) + 8 \l_5^2 + 10 \l_6^2 + 4 \l_6 \l_7 + 10 \l_7^2 \\
& \quad - 2 \l_4 \Big[ \g_1^{(1)} + \g_2^{(1)} \Big] - \bigg[ 4 g_2^4 \q_{M_2} - \frac{12}5 g_2^2 g_1^2 \q_{M_2} \q_{M_1} \bigg] \q_{\m} - 2 \l_4 \Big[ \g_{1,\c}^{(1)} + \g_{2,\c}^{(1)} \Big]
\end{split}
\ee

\be
\begin{split}
\b_{\l_4}^{(2)} &= - \frac{42}5 g_2^4 g_1^2 - \frac{657}{50} g_2^2 g_1^4 
+ \l_4 \Big[ - \frac{231}8 g_2^4 + \frac{1413}{200} g_1^4 \Big] \\
& \quad + \frac3{10} g_2^2 g_1^2 \bigg[ 42 h_t^2 + 18 h_b^2 + 22 h_\tau^2 + 10 \l_1 + 10 \l_2 + 4 \l_3 + \frac{51}{2} \l_4 \bigg] \\
& \quad + 8 g_3^2 \Big[ 8 h_t^2 h_b^2 + 5 \l_4 \Big( h_t^2 + h_b^2 \Big) \Big] \\
& \quad + g_2^2 \bigg[ \frac{15}4 \l_4 \Big( 3 h_t^2 + 3 h_b^2 + h_\tau^2 \Big) + 18 \Big( 2 \l_3 \l_4 + \l_4^2 + 3 \l_5^2 + 3 \l_6^2 + 3 \l_7^2 \Big) \bigg] \\
& \quad + \frac15 g_1^2 \bigg[ 4 h_t^2 h_b^2 + \frac54 \l_4 \Big( 17 h_t^2 + 5 h_b^2 + 15 h_\tau^2 \Big) 
+ 12 \l_4 \Big( \l_1 + \l_2 + \l_3 + 2 \l_4 \Big) \\
& \qquad \qquad + 6 \Big( 8 \l_5^2 + 7 \l_6^2 + 4 \l_6 \l_7 + 7 \l_7^2 \Big) \bigg] \\
& \quad - \frac92 \l_4 \Big[ 3 h_t^4 + 3 h_b^4 + h_\tau^4 \Big]
- h_t^2 h_b^2 \Big[ 24 \Big( h_t^2 + h_b^2 \Big) + 24 \l_3 + 33 \l_4 \Big] \\
& \quad - 12 h_t^2 \Big[ \l_2 \l_4 + 2 \l_3 \l_4 + \l_4^2 + 2 \l_5^2 + \l_6 \l_7 + 5 \l_7^2 \Big] \\
& \quad - \Big( 12 h_b^2 + 4 h_\tau^2 \Big) \Big[ \l_1 \l_4 + 2 \l_3 \l_4 + \l_4^2 + 2 \l_5^2 + 5 \l_6^2 + \l_6 \l_7 \Big] \\
& \quad - 7 \l_4 \Big[ \l_1^2 + \l_2^2 \Big] - 4 \Big( \l_1 + \l_2 \Big) \Big[ 10 \l_3 \l_4 + 5 \l_4^2 + 6 \l_5^2 \Big] \\
& \quad - 2 \l_1 \Big[ 37 \l_6^2 + 10 \l_6 \l_7 + 5 \l_7^2 \Big] - 2 \l_2 \Big[ 5 \l_6^2 + 10 \l_6 \l_7 + 37 \l_7^2 \Big] \\
& \quad - 4 \l_3 \Big[ 7 \l_3 \l_4 + 7 \l_4^2 + 12 \l_5^2 + 18 \l_6^2 + 20 \l_6 \l_7 + 18 \l_7^2 \Big] \\
& \quad - 2 \l_4 \Big[ 13 \l_5^2 + 34 \l_6^2 + 80 \l_6 \l_7 + 34 \l_7^2 \Big] 
- 16 \l_5 \Big[ 5 \l_6^2 + 6 \l_6 \l_7 + 5 \l_7^2 \Big]
\end{split}
\ee

\newpage

$\l_5$:

\be
\begin{split}
\b_{\l_5}^{(1)} &= 2 \l_5 \Big[ \l_1 + \l_2 + 4 \l_3 + 6 \l_4 \Big] + 10 \l_6^2 + 4 \l_6 \l_7 + 10 \l_7^2 
- 2 \l_5 \Big[ \g_{1}^{(1)} + \g_{2}^{(1)} \Big] 
- 2 \l_5 \Big[ \g_{1,\c}^{(1)} + \g_{2,\c}^{(1)} \Big]
\end{split}
\ee

\be
\begin{split}
\b_{\l_5}^{(2)} &= \l_5 \Big[ - \frac{231}8 g_2^4 + \frac{57}{20} g_2^2 g_1^2 + \frac{1413}{200} g_1^4 \Big] 
+ 40 g_3^2 \l_5 \Big[ h_t^2 + h_b^2 \Big] \\
& \quad + \frac34 g_2^2 \Big[  5 \l_5 \Big( 3 h_t^2 + 3 h_b^2 + h_\tau^2 \Big) + 48 \l_5 \Big( \l_3 + 2 \l_4 \Big) + 72 \Big( \l_6^2 + \l_7^2 \Big) \Big] \\
& \quad + \frac1{20} g_1^2 \Big[ 5 \l_5  \Big( 17 h_t^2 + 5 h_b^2 + 15 h_\tau^2 \Big) 
- 24 \l_5 \Big( \l_1 + \l_2 - 8 \l_3 - 12 \l_4 \Big) + 48 \Big( 5 \l_6^2 - \l_6 \l_7 + 5 \l_7^2 \Big) \Big] \\
& \quad - \frac12 \l_5 \Big[ 3 h_t^4 + 3 h_b^4 + h_\tau^4 \Big] 
- h_t^2 \Big[ 33 h_b^2 \l_5 + 12  \l_5 \Big( \l_2 + 2 \l_3 + 3 \l_4 \Big) + 12 \l_7 \Big( \l_6 + 5 \l_7 \Big) \Big] \\
& \quad - \Big( 12 h_b^2 + 4 h_\tau^2 \Big) \Big[ \l_5 \Big( \l_1 + 2 \l_3 + 3 \l_4 \Big) + \l_6 \Big( 5 \l_6 + \l_7 \Big) \Big] \\
& \quad - 7 \l_5 \Big[ \l_1^2 + \l_2^2 \Big] - 4 \l_5 \Big[ \l_1 + \l_2 \Big] \Big[ 10 \l_3 + 11 \l_4 \Big] \\
& \quad - 2 \l_1 \Big[ 37 \l_6^2 + 10 \l_6 \l_7 + 5 \l_7^2 \Big] - 2 \l_2 \Big[ 5 \l_6^2 + 10 \l_6 \l_7 + 37 \l_7^2 \Big] \\
& \quad - 4 \l_3 \Big[ 7 \l_3 \l_5 + 19 \l_4 \l_5 + 18 \l_6^2 + 20 \l_6 \l_7 + 18 \l_7^2 \Big] \\
& \quad - 4 \l_4 \Big[ 8 \l_4 \l_5 + 19 \l_6^2 + 22 \l_6 \l_7 + 19 \l_7^2 \Big] + 6 \l_5^2 - 8 \l_5 \Big[ 9 \l_6^2 + 21 \l_6 \l_7 + 9 \l_7^2 \Big] 
\end{split}
\ee

\newpage

$\l_6$:

\be
\begin{split}
\b_{\l_6}^{(1)} &= 2 \l_6 \Big[ 6 \l_1 + 3 \l_3 + 4 \l_4 + 5 \l_5 \Big] + 2 \l_7 \Big[ 3 \l_3 + 2 \l_4 + \l_5 \Big] - \l_6 \Big[ 3 \g_{1}^{(1)} + \g_{2}^{(1)} \Big] 
- \l_6 \Big[ 3 \g_{1,\c}^{(1)} + \g_{2,\c}^{(1)} \Big] 
\end{split}
\ee

\be
\begin{split}
\b_{\l_6}^{(2)} &= \frac18 g_2^4 \Big[ - 141 \l_6 + 90 \l_7 \Big] 
+ \frac3{20} g_2^2 g_1^2 \Big[ 29 \l_6 + 10 \l_7 \Big] 
+ \frac9{200} g_1^4 \Big[ 187 \l_6 + 30 \l_7 \Big]
+ 20 g_3^2 \l_6 \Big[ h_t^2 + 3 h_b^2 \Big] \\
& \quad + \frac98 g_2^2 \Big[ 5 \l_6 \Big( h_t^2 + 3 h_b^2 + h_\tau^2 \Big) 
+ 48 \l_6 \Big( \l_1 + \l_5 \Big) + 16 \l_6 \Big( \l_3 + 2 \l_4 \Big) + 16 \l_7 \Big( 2 \l_3 + \l_4 \Big) \Big] \\
& \quad + \frac1{40} g_1^2 \Big[ 5 \l_6 \Big( 17 h_t^2 + 15 h_b^2 + 45 h_\tau^2 \Big) 
+ 48 \l_6 \Big( 9 \l_1 + 3 \l_3 + 5 \l_4 + 10 \l_5 \Big) + 48 \l_7 \Big( 4 \l_4 - \l_5 \Big) \Big] \\
& \quad - \frac14 \l_6 \Big[ 27 h_t^4 + 84 h_t^2 h_b^2 + 33 h_b^4 + 11 h_\tau^4 \Big] \\
& \quad - 6 h_t^2 \Big[ \l_6 \Big( 3 \l_3 + 4 \l_4 + 5 \l_5 \Big) + \l_7 \Big( 6 \l_3 + 4 \l_4 + 2 \l_5 \Big) \Big] \\
& \quad - \Big( 6 h_b^2 + 2 h_\tau^2 \Big) \l_6 \Big[ 12 \l_1 + 3 \l_3 + 4 \l_4 + 5 \l_5 \Big] \\
& \quad - \frac32 \l_6 \Big[ 53 \l_1^2 - \l_2^2 \Big] 
- 2 \l_1 \l_6 \Big[ 33 \l_3 + 35 \l_4 + 37 \l_5 \Big] 
- 2 \l_2 \l_6 \Big[ 9 \l_3 + 7 \l_4 + 5 \l_5 \Big] \\
& \quad - 2 \l_6 \Big[ 16 \l_3^2 + 34 \l_3 \l_4 + 36 \l_3 \l_5 + 17 \l_4^2 + 38 \l_4 \l_5 + 18 \l_5^2 \Big] \\
& \quad - 2 \Big( \l_1 + \l_2 \Big) \l_7 \Big[ 9 \l_3 + 7 \l_4 + 5 \l_5 \Big]
- 2 \l_7 \Big[ 18 \l_3^2 + 28 \l_3 \l_4 + 20 \l_3 \l_5 + 17 \l_4^2 + 22 \l_4 \l_5 + 21 \l_5^2 \Big] \\
& \quad - 3 \Big[ 37 \l_6^3 + 42 \l_6^2 \l_7 + 11 \l_6 \l_7^2 + 14 \l_7^3 \Big]
\end{split}
\ee

\newpage

$\l_7$:

\be
\begin{split}
\b_{\l_7}^{(1)} &= 2 \l_7 \Big[ 6 \l_2 + 3 \l_3 + 4 \l_4 + 5 \l_5 \Big] + 2 \l_6 \Big[ 3 \l_3 + 2 \l_4 + \l_5 \Big] - \l_7 \Big[ \g_{1}^{(1)} + 3 \g_{2}^{(1)} \Big] 
- \l_7 \Big[ \g_{1,\c}^{(1)} + 3 \g_{2,\c}^{(1)} \Big] 
\end{split}
\ee

\be
\begin{split}
\b_{\l_7}^{(2)} &= \frac18 g_2^4 \Big[ 90 \l_6 - 141 \l_7 \Big] 
+ \frac3{20} g_2^2 g_1^2 \Big[ 10 \l_6 + 29 \l_7 \Big]
+ \frac9{200} g_1^4 \Big[ 30 \l_6 + 187 \l_7 \Big]
+ 20 g_3^2 \l_6 \Big[ 3 h_t^2 + h_b^2 \Big] \\
& \quad + \frac38 g_2^2 \Big[ 5 \l_7 \Big( 9 h_t^2 + 3 h_b^2 + h_\tau^2 \Big) 
+ 144 \l_7 \Big( \l_2 + \l_5 \Big) + 48 \l_3 \Big( 2 \l_6 + \l_7 \Big) + 48 \l_4 \Big( \l_6 + 2 \l_7 \Big) \Big] \\
& \quad + \frac1{40} g_1^2 \Big[ 5 \l_7 \Big( 51 h_t^2 + 5 h_b^2 + 15 h_\tau^2 \Big)
+ 48 \l_6 \Big( 6 \l_3 + 4 \l_4 - \l_5 \Big) + 48 \l_7 \Big( 9 \l_2 + 3 \l_3 + 5 \l_4 + 10 \l_5 \Big) \Big] \\
& \quad - \frac34 \l_7 \Big[ 11 h_t^4 + 28 h_t^2 h_b^2 + 9 h_b^4 + 3 h_\tau^4 \Big] \\
& \quad - 6 h_t^2 \l_7 \Big[ 12 \l_2 + 3 \l_3 + 4 \l_4 + 5 \l_5 \Big] \\
& \quad - \Big( 6 h_b^2 + 2 h_\tau^2 \Big) \Big[ \l_6 \Big( 6 \l_3 + 4 \l_4 + 2 \l_5 \Big) + \l_7 \Big( 3 \l_3 + 4 \l_4 + 5 \l_5 \Big) \Big] \\
& \quad - \frac32 \l_7 \Big[ - \l_1^2 + 53 \l_2^2 \Big] 
- 2 \l_1 \l_7 \Big[ 9 \l_3 + 7 \l_4 + 5 \l_5 \Big]
- 2 \l_2 \l_7 \Big[ 33 \l_3 + 35 \l_4 + 37 \l_5 \Big]  \\
& \quad - 2 \l_7 \Big[ 16 \l_3^2 + 34 \l_3 \l_4 + 36 \l_3 \l_5 + 17 \l_4^2 + 38 \l_4 \l_5 + 18 \l_5^2 \Big] \\
& \quad - 2 \Big( \l_1 + \l_2 \Big) \l_6 \Big[ 9 \l_3 + 7 \l_4 + 5 \l_5 \Big]
- 2 \l_6 \Big[ 18 \l_3^2 + 28 \l_3 \l_4 + 20 \l_3 \l_5 + 17 \l_4^2 + 22 \l_4 \l_5 + 21 \l_5^2 \Big] \\
& \quad - 3 \Big[ 14 \l_6^3 + 11 \l_6^2 \l_7 + 42 \l_6 \l_7^2 + 37 \l_7^3 \Big]
\end{split}
\ee



\end{document}